\documentclass[manuscript,nonacm]{acmart}

\AtBeginDocument{%
  }

\usepackage{hyperref}
\usepackage{setspace} 
\usepackage{amsmath} 
\usepackage{mathtools}  
\usepackage{url} 
\usepackage{xurl}
\usepackage{color} 
\usepackage{booktabs} 
\usepackage{textcomp} 
\usepackage{graphicx} 
\usepackage{subcaption} 
\usepackage{multirow} 
\usepackage{algpseudocode}
\usepackage{algorithm}
\usepackage{enumitem}
\usepackage[switch]{lineno}
\usepackage{array}
\newcommand\pmunit{\mu g/m^3}

\author{Ankit Bhardwaj}
\affiliation{%
    \institution{New York University}
    \city{New York}
    \state{New York}
    \country{USA}
}

\author{Ananth Balashankar}
\authornote{Work done while at NYU.}
\affiliation{%
  \institution{Google Research}
  \city{New York}
  \state{New York}
  \country{USA}
}

\author{Shiva Iyer}
\authornotemark[1]
\affiliation{%
  \institution{Toyota InfoTechnology Center}
  \city{Mountain View}
  \country{USA}
}

\author{Nita Soans}
\affiliation{%
  \institution{Kaiterra Inc}
  \city{New Delhi}
  \country{India}
}

\author{Anant Sudarshan}
\affiliation{%
 \institution{University of Warwick}
 \city{Coventry}
 \country{United Kingdom}
 }

\author{Rohini Pande}
\affiliation{%
  \institution{Yale University}
  \city{New Haven}
  \state{Connecticut}
  \country{USA}
}

\author{Lakshminarayanan Subramanian}
\affiliation{%
  \institution{New York University}
  \city{New York}
  \state{New York}
  \country{USA}
}

\renewcommand{\shortauthors}{Bhardwaj et al.}

\begin{document}
\title{Comprehensive Monitoring of Air Pollution Hotspots Using Sparse Sensor Networks}

\begin{abstract}
Urban air pollution hotspots pose significant health risks, yet their detection and analysis remain limited by the sparsity of public sensor networks. This paper addresses this challenge by combining predictive modeling and mechanistic approaches to comprehensively monitor pollution hotspots. We enhanced New Delhi's existing sensor network with 28 low-cost sensors, collecting PM$_2.5$ data over 30 months from May 1, 2018, to Nov 1, 2020. Applying established definitions of hotspots to this data, we found the existence of additional 189 hidden hotspots apart from confirming 660 hotspots detected by the public network. Using predictive techniques like Space-Time Kriging, we identified hidden hotspots with 95\% precision and 88\% recall with 50\% sensor failure rate, and with 98\% precision and 95\% recall with 50\% missing sensors. The projected results of our predictive models were further compiled into policy recommendations for public authorities. Additionally, we developed a Gaussian Plume Dispersion Model to understand the mechanistic underpinnings of hotspot formation, incorporating an emissions inventory derived from local sources. Our mechanistic model is able to explain 65\% of observed transient hotspots. Our findings underscore the importance of integrating data-driven predictive models with physics-based mechanistic models for scalable and robust air pollution management in resource-constrained settings.
\end{abstract}




\maketitle

\section{Introduction}
\label{sec:intro}

\textbf{Problem Context:} Across the developing world, urbanization has led to the emergence of heavily polluted regions that are at the focus of local economic growth and are ground zero for one of the most urgent public health challenges the world faces today \citep{Alpert2012}. Over 4.5 billion people live in parts of the world where the average ambient concentration of fine particulate matter is above twice the maximum level the World Health Organization (WHO) considers safe \citep{world2021air}. Most urban centers in the world can only deploy sparse coarse-resolution sensor networks for air pollution monitoring due to the high costs of establishing industrial-grade gold-standard air quality monitoring stations. The cost of one fixed automated station is dependent on the negotiation between government officials and contractors, but can be as high as 200,000 USD to set up with a 10\% yearly operational and maintenance cost \citep{hindustantimes2017,ncap18,downtoearth2014,hindustantimes2020}. 

On the other hand, prior analysis of city-scale pollution data \citep{APH_paper} shows the existence of ``air pollution hotspots'', which loosely defined, are areas with elevated pollution levels. Such hotspots can cause disproportionate exposure, potentially hurting a vulnerable population like those in urban slums \citep{kulshreshta2009, lueker2020indoor, sharma1998indoor}. Even if air pollution levels do not vary over time, levels of human exposure may vary widely depending on lifestyle choices, localities, and travel routes frequented \citep{Pant2017PM25}. Thus, identifying fine-grained hotspots is accepted to be very valuable for targeted policy interventions \citep{MAHATO2020109835}. In fact, many existing mitigation strategies in some of the most polluted cities are hotspot-centric \cite{dpcc_hotspot,PriyangiAgarwal_2020,Shrangi_Pillai_2019,toi2024,doedelhi,it2023}. In this paper, we show that the space-time resolution of the public sensor networks is insufficient to detect and analyze these hotspots effectively, and present the predictive and mechanistic modeling strategies to complement sparse sensor networks for comprehensive monitoring of air pollution hotspots. \\

\noindent\textbf{Study Area:} We chose the city of New Delhi, India for the target of our study. In 2016, the annual average level of airborne fine particulate matter (PM$_{2.5}$) in New Delhi, the capital of India, was 142 $\pmunit$, 4 times higher than the least stringent interim target of 35 $\pmunit$ \citep{world2021air}, making it one of the most polluted cities in the world. 
The pollution in New Delhi is caused due to a combination of factors. One of the most cited macro sources is the burning of crop residue in the farmlands of the neighboring states \citep{Bikkina2019Air,Cusworth2018Quantifying}, which contributes 17-26\% of the total pollution in the winter, and about 7-12\% in the summer \citep{iitkanpur}. The remaining is attributed to a myriad of localized sources within the city, such as road dust, vehicular traffic, domestic emissions (from domestic activities like waste burning and cooking), construction and demolition activities, etc. \citep{Apte2017High,iitkanpur}. 
New Delhi has a network of air quality monitoring stations operated by three different public bodies -- Central Pollution Control Board (CPCB), Delhi Pollution Control Committee (DPCC), and the Indian Meteorological Department (IMD), that provide quality-controlled public data. The spatial monitoring resolution of this network with 32 sensors spread over a ground area of about 858 km$^2$ is the densest such network in any Indian city. Based on official financial and policy reports \cite{pib_funds,caqm2022}, while the government has the funds to expand their sensor network, the focus in the near future will be on establishing new monitoring stations in the National Capital Region as these areas have a much sparser coverage. Thus, for the foreseeable future, the public sensor network resolution in the urban areas of New Delhi is unlikely to increase. Some experts believe that the current resolution is not fine enough to capture the more localized causes of high pollution that inform us about citizen exposure \citep{thunis2019source}. \\

\noindent \textbf{Summary of Contributions:} In this paper, we show that the spatial coverage of the public sensor network in New Delhi is insufficient for monitoring air pollution hotspots, and present complementary predictive and mechanistic modeling techniques to enable comprehensive monitoring of such hotspots. We augmented New Delhi's public network by deploying and maintaining our own network of low-cost air quality sensors from May 1, 2018 to November 1, 2020. Our deployment, in collaboration with Kaiterra \citep{kaiterra}, incurred a total cost of 8000 USD for 28 sensors, only 0.2\% of the estimated cost of using high-end reference monitors. Applying the definitions of air pollution hotspots formalized in prior literature \citep{APH_paper} to the sensor measurements, we show the existence of additional hotspots that were not detected by the public network, which we call \textit{hidden hotspots}. Given that the public sensor network size is likely to remain constant in the future, we developed predictive models based on Space-Time Kriging to uncover hidden hotspots effectively with \textbf{95\% precision} and \textbf{88\% recall}. Our choice of predictive model leads to increased robustness and generalization with minimal dependence on additional data collection infrastructure that might not be present in many cities, as compared to approaches that depend on multi-dimensional urban features. Further, using the projections from our predictive models, we compile a set of policy recommendations for public authorities. To deeply understand the mechanism of hotspot formation and identify the emission sources that are largely responsible for hotspots in New Delhi, we developed a city-scale Gaussian-Plume Dispersion model based on an approximate 0.01$^o$x0.01$^o$ emissions inventory that was collated using alternative online data sources. Our mechanistic model is implemented as matrix convolution operations, leveraging GPUs for accelerated computations. Using our mechanistic model, we are able to explain \textbf{65\%} of transient hotspots observed in sensor measurements. This dual predictive-mechanistic modeling approach enables comprehensive monitoring of air pollution hotspots in the city despite the sparsity of public sensor networks.

\section{Sensing Hotspots}
\label{sec:proof}

\begin{figure}[t]
  \centering
  \begin{subfigure}{0.7\columnwidth}
    \centering
      \includegraphics[width=\columnwidth]{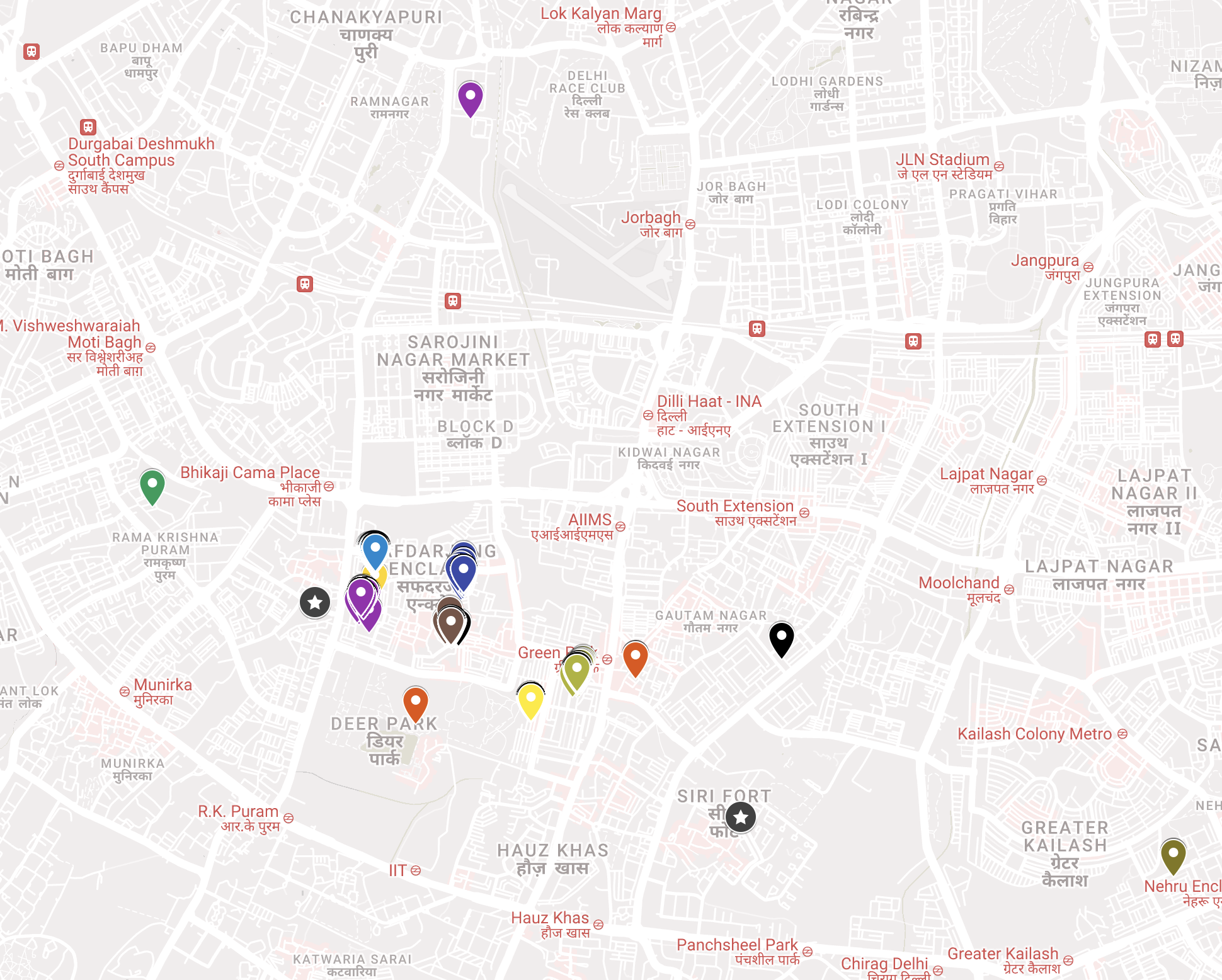}
      \caption{The map of locations for the experiment including two government sensors in, namely at R.K.Puram and Sirifort (represented by black stars). The two locations are separated by a 3.14 km straight line distance. Our deployed sensors, on average, are about 1.5 km from a reference monitor.}
  \end{subfigure}
  \begin{subfigure}{0.25\columnwidth}
      \begin{subfigure}{\columnwidth}
        \centering
        \includegraphics[width=\columnwidth]{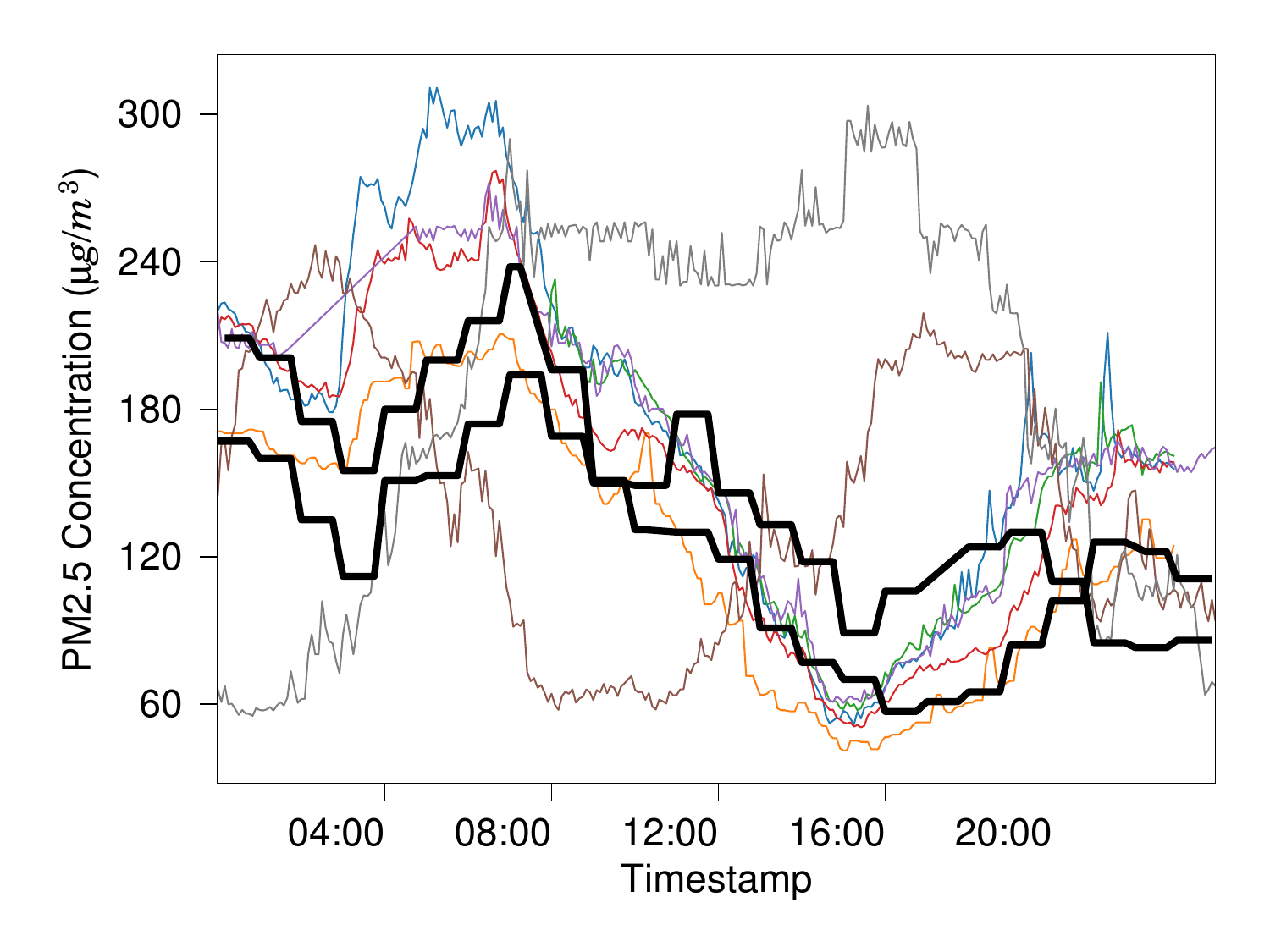}
        \caption{Jan 30, 2018}
        \label{fig:variation-1}
      \end{subfigure}
      
      \begin{subfigure}{\columnwidth}
        \centering
        \includegraphics[width=\columnwidth]{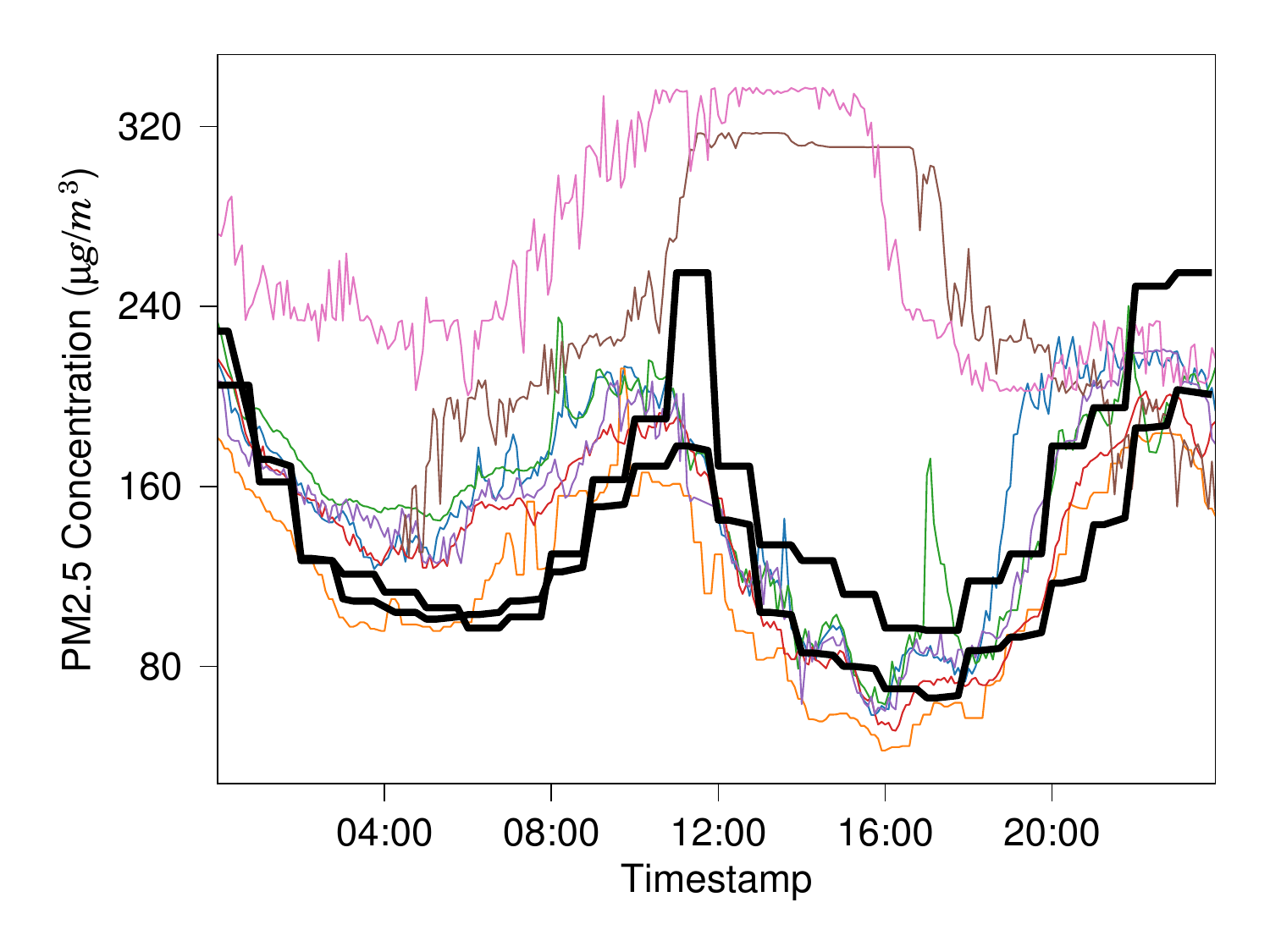}
        \caption{Feb 1, 2018}
        \label{fig:variation-2}
      \end{subfigure}
      
      \begin{subfigure}{\columnwidth}
        \centering
        \includegraphics[width=\columnwidth]{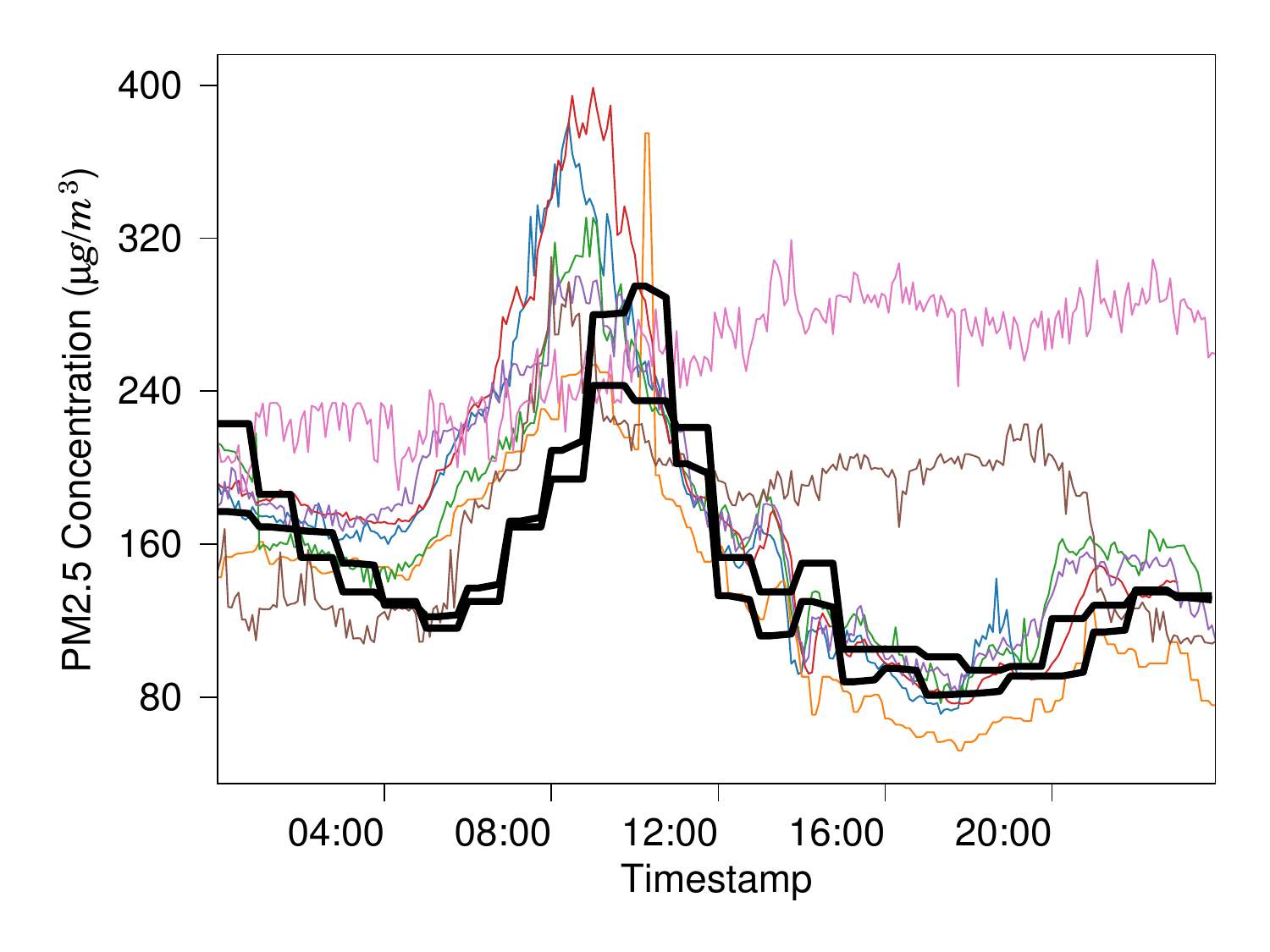}
        \caption{Feb 2, 2018}
        \label{fig:variation-3}
      \end{subfigure}
  \end{subfigure}
      
  \caption{The adjoining plots show a sample of readings in a small neighborhood containing 7 of our low-cost sensors and 2 public stations on three consecutive days. The black lines show the readings from the closest public monitors, and the other lines show the readings from the rest of the sensors. While most of our sensors are in agreement with the public monitors, note that some of our sensors have recorded spikes in pollution levels that are missed by the public monitors.}
  \label{fig:spatial-variation}
\end{figure}

We conducted an experiment in 2018, in which we traversed repeatedly through the roads of two different neighborhoods in New Delhi (area 4-5 km$^2$ each) over a period of four hours while carrying portable handheld air quality monitors. We were able to attribute some of these local hotspots to various local activities such as open food carts, open ironing shops (with the traditional coal iron press), incense burning, roadside construction, and earth-moving activity. Some of these sources generated PM$_{2.5}$ concentration in the air upwards of $100\pmunit$ while being present up to 10 feet away from those sources. Neither of these neighborhoods contained a government monitoring station in its vicinity. In Figure \ref{fig:spatial-variation}, we show the readings of 2 government sensors and 7 of our sensors in a small neighborhood, along with the locations placed on a map. Some of our sensors are able to record high concentrations of PM$_{2.5}$ when nearby government sensors do not record this phenomenon, thus validating our hypothesis that localized and transient, but significant sources of pollution do not get captured by the existing public network of monitors.

This experiment showed that at least on a transient time scale, the public sensor network misses significantly large fluctuations in the surrounding air pollution. The natural extension this observation is to investigate if the public network misses hotspots on longer time scales, where it is relevant for policy. For this, let us first define air pollution hotspots in a way that would be relevant for policy-makers. \\

\noindent\textbf{Defining Hotspots:} Borrowing the definitions of air pollution hotspots from \citep{APH_paper,dpcc_hotspot} that cover a similar study area in New Delhi, we consider the following definitions of hotspots.

\begin{itemize}
\item \textbf{\textit{Frequency Hotspots}}: As defined in \citep{APH_paper},if the daily average of a sensor's readings exceeds the threshold of $60 \pmunit$ on more than 60\% of the days on which the sensor was active during a month, we categorize the location as a frequency hotspot for that month.
\item \textbf{\textit{Scale Hotspots}}: As defined in \citep{APH_paper}, if the monthly average of readings of a sensor's readings exceeds $90 \pmunit$, we categorize the location as a scale hotspot for that month.
\item \textbf{\textit{Consistency Hotspots}}: As defined in \citep{APH_paper}, if the sensor's readings exceed $90 \pmunit$ on three consecutive days of a month, we categorize the location as a consistency hotspot for that month.  
\item \textbf{\textit{Annual Hotspots}}: If the annual average of sensor readings is more than $100 \pmunit$, the location is categorized as a hotspot as stated in the hotspot definition in \citep{dpcc_hotspot}.
\end{itemize}

Beyond the notion of annual hotspots defined on a yearly basis, any location is considered a hotspot for a given  month, if it shows hotspot behavior in any of the first three hotspot criteria listed above for frequency, scale or consistency hotspots.
It should be noted that these definitions are not the only possible ways to define a hotspot. We chose these definitions since they capture the notion of pollution hotspots in the context of New Delhi relevant to our study area. To answer the question posed earlier concretely, we deployed our private sensor network and collect data for longer durations.

\subsection{Sensor Network Deployment}
\label{sec:sensor_specs}

\begin{figure}
    \centering
    \includegraphics[width=0.8\columnwidth]{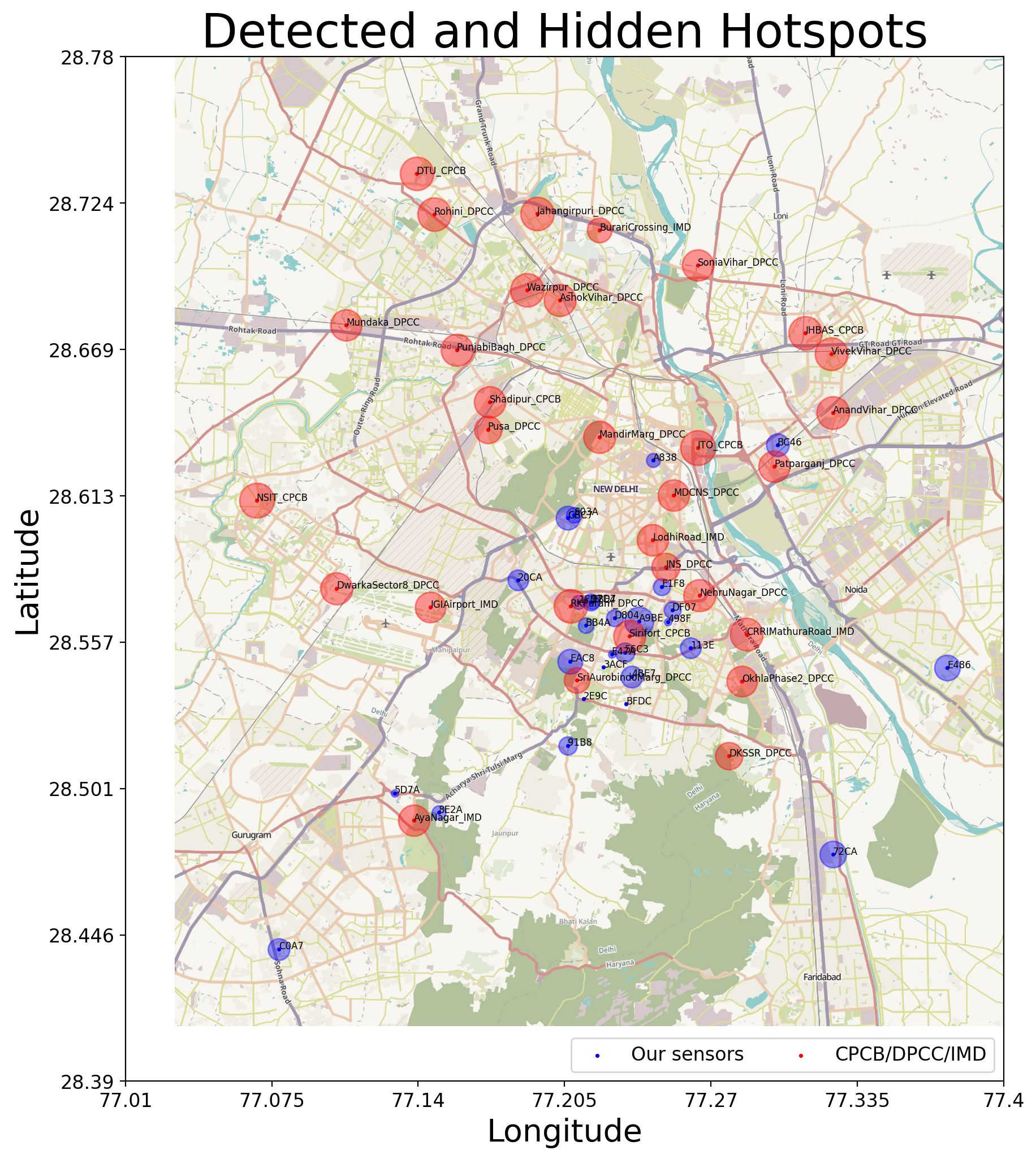}
    \caption{The map in Figure \ref{fig:monitor-locations} shows the placement of all the sensors for our study, 32 public (red), and the 28 low-cost sensors that we have deployed (blue). The size of individual points is proportional to the number of months they act as hotspots according to the definition presented in Section \ref{sec:proof} \citep{APH_paper}.}
    \label{fig:monitor-locations}
\end{figure}

\noindent\textbf{Sensor Placement:} We augmented the existing pollution monitoring sensor network of the government by deploying 28 additional sensors in the city, in collaboration with Kaiterra \citep{kaiterra}, a company that manufactures air quality monitors and air filters, bringing the total number of air pollution monitors up to 61. Most of our sensors were deployed in the region of south Delhi, with some spread out farther out in the city. The deployment of our sensors was primarily governed by the availability of volunteers willing to host our sensors. The sensor height was kept between 3 to 10 meters based on domain experts' recommendation \cite{chem_transport}. Figure \ref{fig:monitor-locations} shows the placement of both the government and our deployed sensors on the map of New Delhi. The final dataset was obtained by combining this data with publicly available data from government sensors on the CPCB portal \citep{cpcb_portal}. \\

\noindent\textbf{Sensor Specifications:} Each low-cost sensor is equipped with a Reliance Jio 4G device that enables connection over 4G network. It periodically sends the sensor readings to a database in China, from where the data was available to be pulled using a regular RESTful API. The sensors reported data approximately every 5 minutes or so. Our deployed sensors have the ability to report PM$_{2.5}$ and PM$_{10}$ values every 5 minutes, but most government sensors only report data once every hour. Since the calibration was also done using government sensors, we average our readings for the hour and the lowest temporal resolution for our data is one reading per hour.\\

\begin{figure}[t]
    \centering
    \begin{subfigure}{0.45\columnwidth}
        \centering
        \includegraphics[width=\columnwidth]{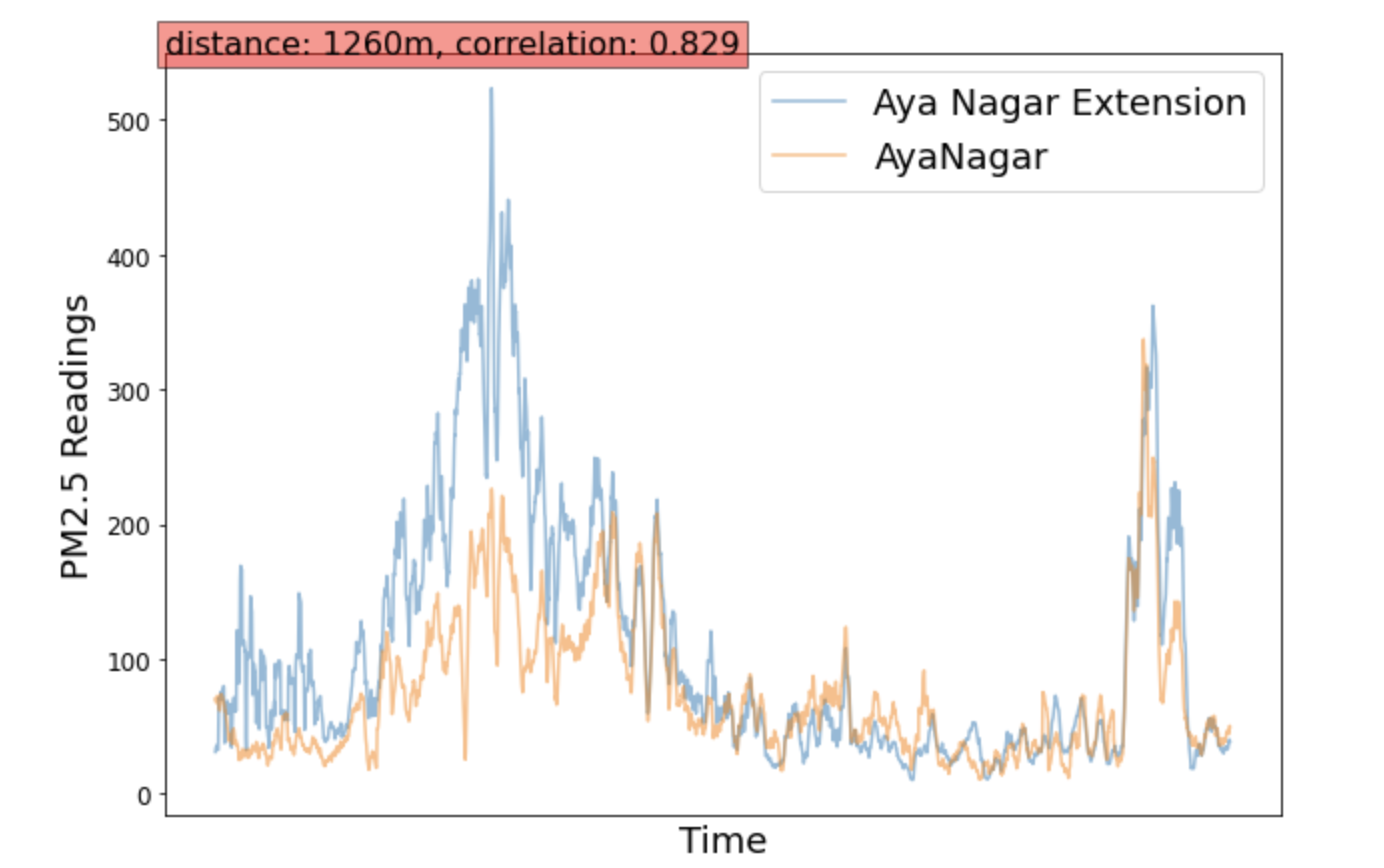}
        \caption{Aya Nagar Extension and Aya Nagar}
        \label{fig:ayanagar}
    \end{subfigure}
  ~
    \begin{subfigure}{0.45\columnwidth}
        \centering
        \includegraphics[width=\columnwidth]{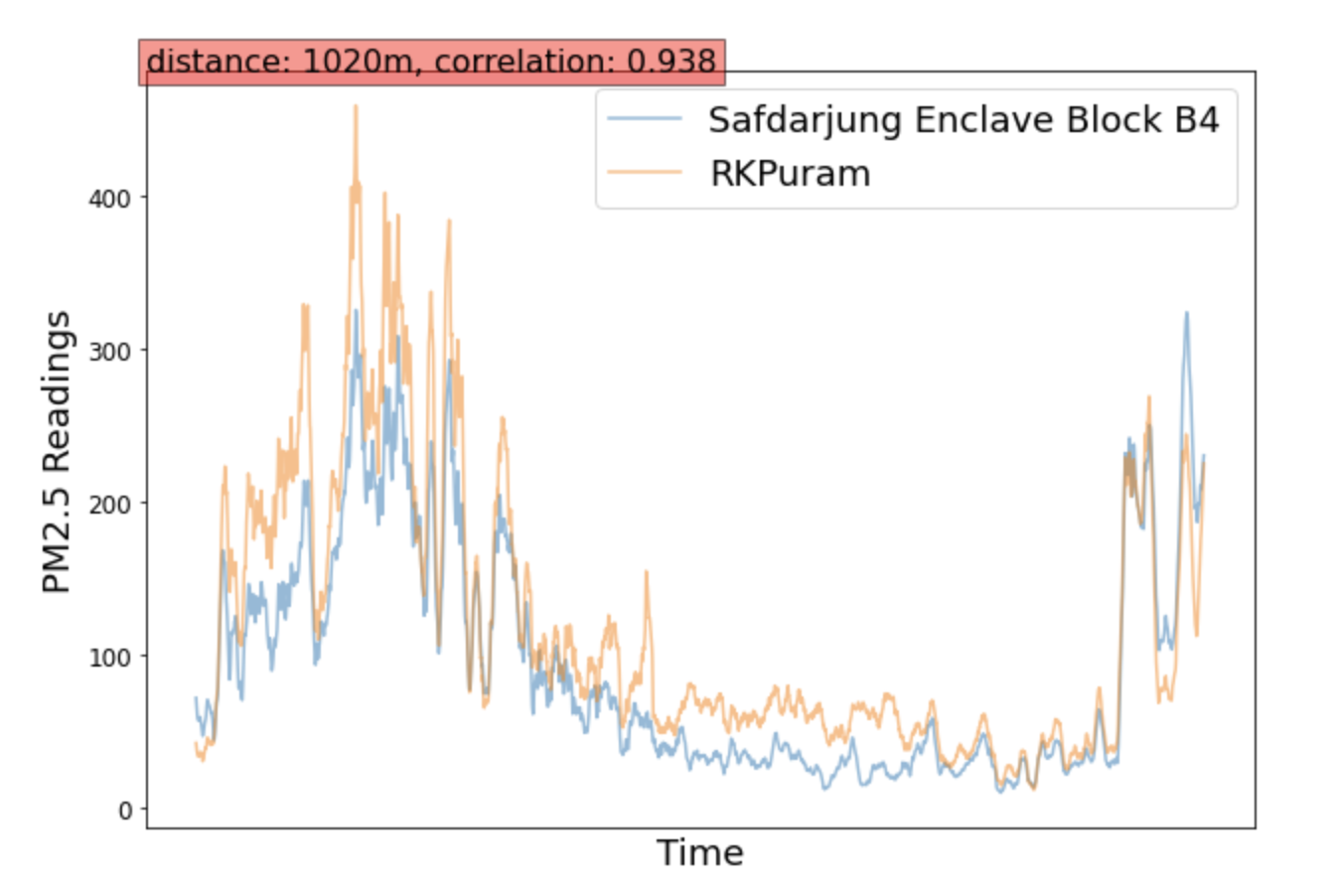}
        \caption{Safdurjung Enclave and RK Puram}
        \label{fig:safdurjung}
    \end{subfigure}
  
    \caption{
    Figures \ref{fig:ayanagar} and \ref{fig:safdurjung} show the temporal correlation of readings of our deployed sensors (in blue) and nearest government sensors (in orange). These charts are representative of most of our deployed sensors.}
  \label{fig:time-corr}
\end{figure}

\noindent\textbf{Sensor Calibration:} Kaiterra sensors are extensively calibrated using over 6,000 individual measurements in a testing environment with specifications much higher than the required industry standards. The sensors are calibrated both before and after the assembly process. The calibration is done by placing multiple sensors in a testing chamber that ensures uniform pollution concentration and temperature by carefully controlling the airflow in the chamber using ventilation equipment. The chamber can be set to known values of pollution concentration, which is used for sensor calibration during the manufacturing process. After assembly, every sensor is comprehensively tested for different particulate matter concentrations using more than six thousand measurements. Only sensors with a magnitude of deviation from the reference within specification for all readings leave the production line.

Our deployed sensors were further calibrated against the government sensors, by conducting a longitudinal comparison study by measuring in proximity to the location of the government monitoring centers. After calibration, we noted the variation of raw values of our sensor and the corresponding government sensor with time and observed strong correlation between the two sensor values with a high linear regression agreement and an $R^2$ of 0.996. The sensors were also regularly serviced and re-calibrated. In Figure \ref{fig:time-corr}, we show the readings of our deployed sensors (in blue) and the nearest government monitors (in orange), along with the distance between the sensors and the correlation between the readings for two locations. We observe strong temporal correlation between our sensors and nearest government sensors for the entire duration of the study. This indicates that our sensors maintained calibration with government sensors for the full duration.\\

\noindent\textbf{Admissibility of Low-Cost Sensors:} CPCB provides an official specification manual \cite{cpcb_specs} for the establishment of CAAQM stations. Clearly, such specifications can not be met by low-cost sensors, and the corresponding error in measurement is expected to be higher. But, Kaiterra sensors that we deployed have undergone quality control checks and periodic maintenance and calibration, ensuring that the PM2.5 data used in the study is as accurate as possible. The CAAQM sensors work on the beta-ray attenuation principle, which is considered the gold-standard approach. The kaiterra sensors work on the laser diffraction principle. Based on a 2022 comparison study \cite{Dinh2022}, light scattering sensors and beta ray attenuation sensors have insignificant differences when compared over a 24 hour averaged resolution. Given that the hotspot definitions we use in our study take averages over at least 24 hours, we are confident that the data from our deployed low-cost sensors is admissible for modeling.

\subsection{Results}
\label{sec:prelim_results}

\begin{figure}[t]
  \centering
  \begin{subfigure}{0.43\columnwidth}
    \centering
    \includegraphics[width=\columnwidth]{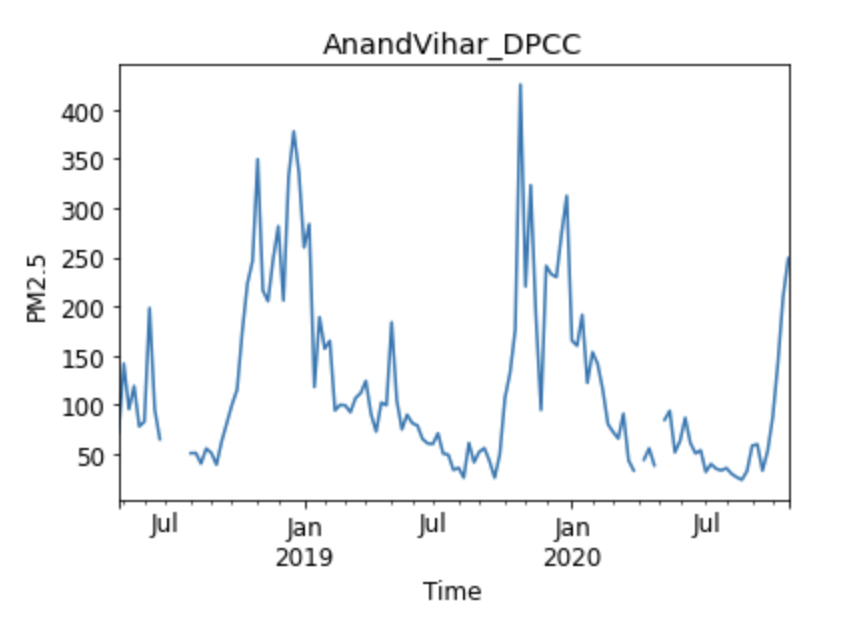}
    \caption{}
    \label{fig:long_av}
  \end{subfigure}
  ~
  \begin{subfigure}{0.52\columnwidth}
      \centering
    \includegraphics[width=\columnwidth]{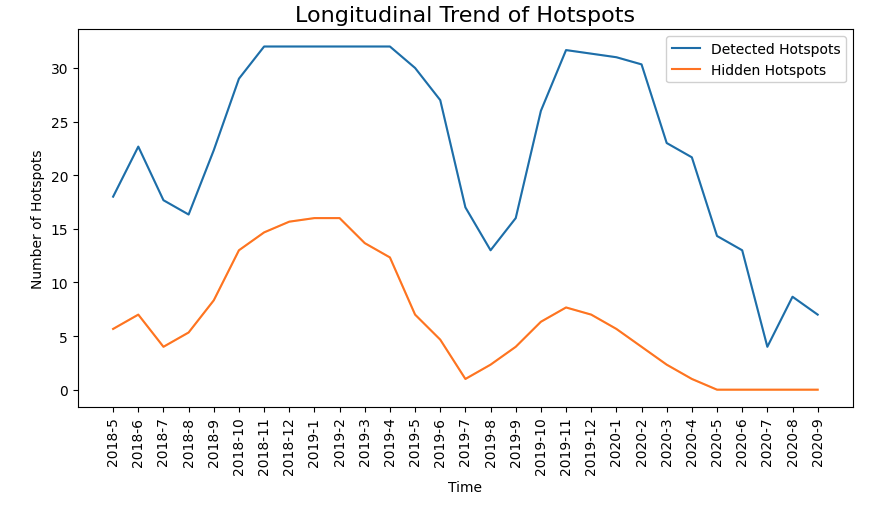}
    \caption{}
    \label{fig:long_hsps}
  \end{subfigure}
  \begin{subfigure}{0.32\columnwidth}
    \centering
    \includegraphics[width=\columnwidth]{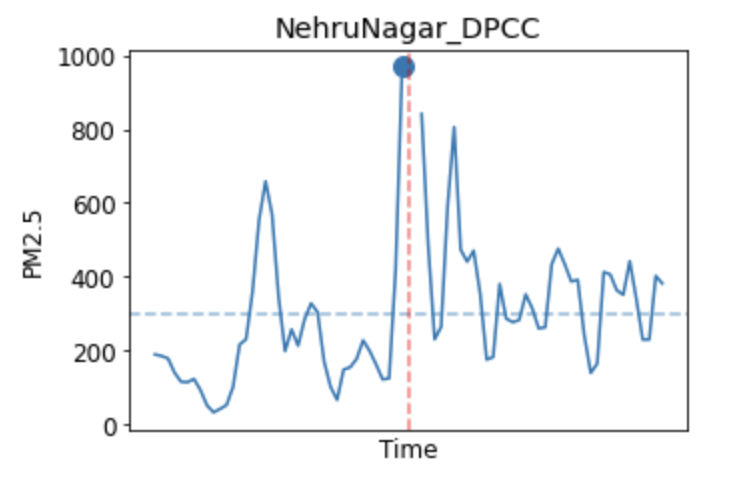}
    \caption{}
    \label{fig:dd_nn}
  \end{subfigure}
  ~
  \begin{subfigure}{0.32\columnwidth}
    \centering
    \includegraphics[width=\columnwidth]{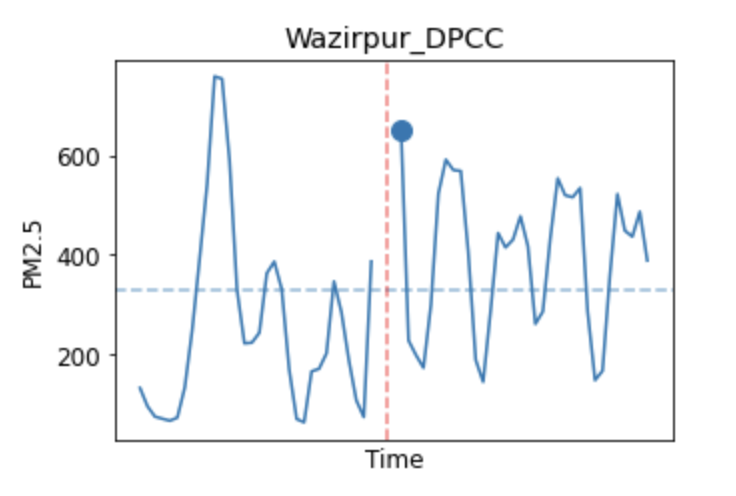}
    \caption{}
    \label{fig:dd_wz}
  \end{subfigure}
  ~
  \begin{subfigure}{0.32\columnwidth}
    \centering
    \includegraphics[width=\columnwidth]{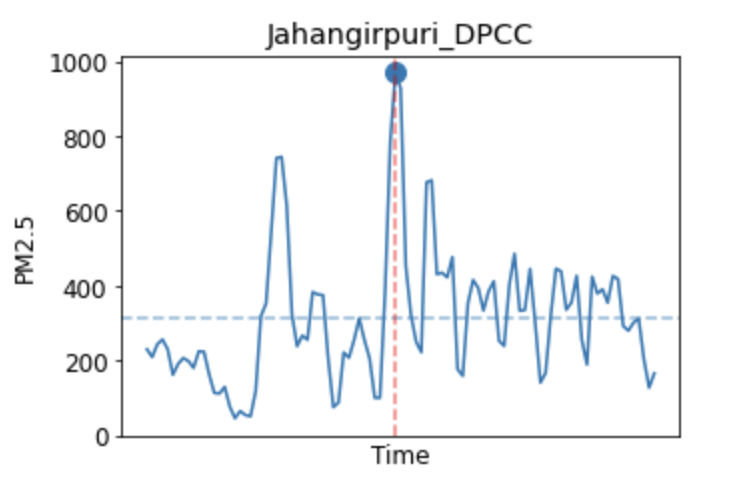}
    \caption{}
    \label{fig:dd_jh}
  \end{subfigure}
  \caption{In the above figure, we first show the longitudinal analysis of hotspots (Figure \ref{fig:long_hsps}) and corresponding pollution data plot for one of the monitoring stations (Anand Vihar) over two years at 1 week resolution (Figure \ref{fig:long_av}). There is a clear elevation in both pollution levels and detected as well as hidden hotspots during the winter months. We observe the same trend in the remaining monitoring stations as well. During the months of winter, on specific occasions like Diwali (Figures \ref{fig:dd_nn}-\ref{fig:dd_jh}), the total pollution level goes beyond sensor's measurement capacity of 1000 $\pmunit$.}
  \label{fig:longitudnal-1W}
\end{figure}

\noindent\textbf{Hotspot Statistics:} We applied the definitions of different types of hotspots (based on frequency, scale, and consistency) to the longitudinal data generated by the public and our low-cost sensor networks. We call the hotspots detected by the government network ``detected hotspots'' and the hotspots detected by our low-cost network ``hidden hotspots''. From Figure \ref{fig:monitor-locations}, we can see that the two sensor networks are non-overlapping, so any hotspots detected by our network are almost guaranteed to be ``hidden'' from the public network, and hence the name. The results of the analysis are presented in Figure \ref{fig:long_hsps}, where plot the number of hotspots for each month for Detected and Hidden categories. In total, we found 660 detected and 189 hidden hotspots in our study duration. In addition, for 2019, we found 4 additional annual hotspots. The additional hotspots discovered by our low-cost network are proof that our hypothesis is correct.\\

\begin{figure}[t]
    \centering
    \includegraphics[width=0.7\columnwidth]{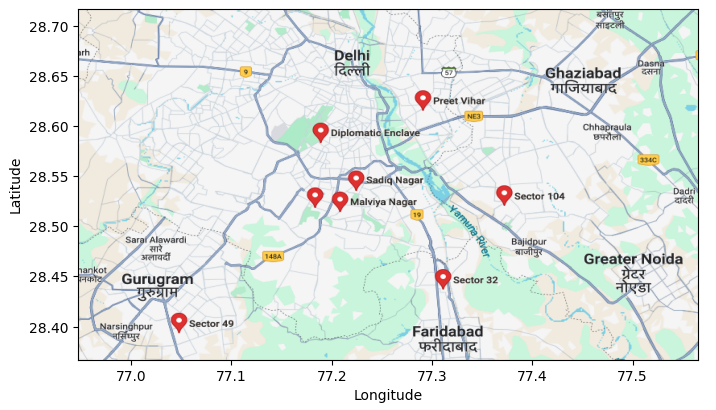}
    \begin{tabular}{|c|c|c|c|}
    \hline
        \textbf{Location} & \textbf{Number of months} & \textbf{Population} & \textbf{Area} (km$^2$)\\
    \hline
        Sadiq Nagar & 18 & 13,036 & 0.88 \\
        Faridabad Sector 32 & 15 & 22,330 & 1.43 \\
        Noida Sector 104 & 14 & 12,471 & 2.15 \\
        Qutab Institutional Area (ISI Delhi) & 13 & 14,307 & 1.62\\
        Diplomatic Enclave (US Embassy) & 12 & 9,887 & 1.25 \\
        Preet Vihar & 11 & 36,583 & 1.46 \\
        Gurugram Sector 49 & 10 & 25,500 & 3.12 \\
        Malviya Nagar & 10 & 63,866 & 2.13 \\
    \hline
    \end{tabular}
    \caption{Top hotspot locations covered by our sensor network in New Delhi, their demographic and geometric information, and the number of months they behave as hotspots. The population and areas were sourced from \cite{geoiq}. This is direct evidence that more than 150,000 people live in ``pollution hotspot'' neighborhoods that are not on the city government's map.}
    \label{tab:top_locations}
\end{figure}

\noindent\textbf{Local Context and Seasonal Trends:} As seen before in section \ref{sec:proof}, further definitions of hotspots have been considered in prior work, that are defined over the time duration of a month rather than a year. Looking at the results in Figure \ref{fig:long_hsps}, the number of hidden and detected hotspots peaks during the winter months. To observe this seasonality in pollution levels, we can look at the behavior of Anand Vihar monitoring stations' data over the entire study duration in Figure \ref{fig:longitudnal-1W}. This one-week resolution view of the data shows clear elevations in pollution levels during the months of winter. According to \cite{urbanem_winter}, this is largely due to meteorological reasons such as low mixing height, low amount of precipitation, stagnant wind conditions, and increased emissions due to open fires, heating, and agricultural burning. This trend is visible in other monitoring stations as well. In fact, on specific days of Diwali (which occurs in November), the burning of firecrackers contributes further to the elevated pollution. The resulting pollution level is so high that most sensors fail to capture readings beyond their maximum limit of $\sim 1000 \pmunit$ (see Figures \ref{fig:dd_nn}-\ref{fig:dd_jh} and Appendix \ref{sec:diwali} for details).

Considering the seasonal trends in pollution, we see that the number of hotspot locations increase significantly during the winter months (see Figure \ref{fig:long_hsps}). In fact, for majority of winter months, we find that \textit{all} government monitoring locations can be categorised as hotspots! The readings from our low-cost network also show many hidden hotspots during the winter months. We list these locations along with the number of months they behave as hotspots, their population, and the locality area in Figure \ref{tab:top_locations}. 

\section{Predictive Modeling of Hotspots}
\label{sec:predictive}
Given that we have shown the existence of hidden hotspots that go unnoticed by the public sensor network in New Delhi, the logical next step is to devise solutions to discover these hidden hotspots. The trivial and most effective solution would consist of scaling up our sensor network and collecting the pollution data at a higher spatial resolution. But, this solution comes at the significant cost of deploying and maintaining the larger sensor network. Assuming a simple grid network, increasing the spatial resolution of the network by a factor of two requires four times as many sensors. That is, the number of sensors and consequently the cost grows quadratically with the spatial resolution. Even with low-cost sensors, this would add up to a significant cost for obtaining sufficient resolution capable of detecting all hidden hotspots. In this section, we first present our algorithmic solution that works complimentarily to increasing the scale of sensor network, by interpolating the measured pollution values into a pollution field based on different modeling strategies. Then, we present a thorough evaluation of our predictive models, investigating the models from the perspective of performance, robustness, and generalization. Finally, using the projections of our predictive models, we provide policy recommendations for New Delhi public authorities.

\subsection{Models}
\noindent\textbf{Model Selection:} In related literature, we have come across four broad predictive modeling strategies that have been used in the context of air pollution modeling, namely spatial interpolation, multidimensional urban feature models, temporal graphs, and land use regression. Land-use regression requires auxiliary meteorological data that is available on coarse spatio-temporal granularity, it is not applicable to city-scale settings. A city is generally represented by a point in maps of such scale. Multidimensional urban feature models \cite{zheng2013uair,ge2021multi,zareba2023big,hofman2021spatiotemporal} utilize diverse urban datasets as input features and generally employ neural networks for processing. These models have demonstrated efficacy in cities such as Beijing and Shanghai. However, applying this approach in New Delhi poses significant challenges due to the unavailability or unreliability of key data (e.g., detailed emissions data and wind speed) at the required granular level, leading to substantial estimation errors, as shown in our results in Appendix \ref{sec:inventory} and supported by prior research \cite{GUTTIKUNDA2013101}. Furthermore, this multi-modal approach introduces difficulties in terms of result interpretability and explainability. Temporal graphs have been shown to perform incredibly well on temporal forecasting tasks \cite{our_npj}, but the task of short-term temporal forecasting is of limited use from a policy perspective. The sensor network in our settings generates data that is spatially sparse and temporally dense (with readings every hour). Spatial prediction and hence extensibility, is more valuable in our setting, and thus, spatial interpolation is the appropriate modeling strategy for our purpose. Temporal graph approaches lack this ability, and hence, are not the right modeling choice. Other temporal state models like Kriged Kalman Filters \cite{Mardia1998}, etc, have similar problems with spatial extensibility. One of the challenges of our sparse sensor network setting is that it renders deep learning approaches inapplicable, as any temporal snapshot of the network only has less than 60 data points. Thus, in our work, we have applied the conventional Kriging and simplistic neural network modeling approaches for spatial interpolation. We have not found any similar applications in the literature for our study area and in the context of hotspot analysis.\\

\noindent\textbf{Kriging:} Gaussian Process regression (also called Kriging) \cite{cressie2011,stein1999interpolation} is one of the standard methods of spatial interpolation in geostatistics. Kriging is an interpolation method that aims to estimate the value of a random variable $Z(u_0)$ at an unsampled location $u$ based on observations $z_1, z_2, \ldots, z_n$ at sampled locations $u_1, u_2, \ldots, u_n$. The Kriging estimate $Z(u_0)$ is typically given by a linear combination of observed values
\begin{align}
    \hat{Z}(u_0) = \sum_{i=1}^n \lambda_i z_i
\end{align}
Where $\lambda_i$ are the Kriging weights that need to be determined. The Kriging weights $\lambda_i$ are chosen to minimize the estimation error variance, subject to unbiasedness
\begin{align}
    E[Z(u_0) - Z(u_i)] &= 0\\
    \implies \sum_{i=1}^n \lambda_i = 1
\end{align}
\begin{align}
    \text{Var}[\hat{Z}(u_0) - Z(u_0)] = - \sum_{i=1}^n \sum_{j=1}^n \lambda_i \lambda_j \gamma_{ij} + 2\sum_{i=1}^n \lambda_i \gamma_{0i}
\end{align}
Where $\gamma(h)$ is the semivariogram or covariance function, and $\gamma_{ij}$ represents the covariance between locations $u_i$ and $u_j$. The semivariogram $\gamma(h)$ describes the spatial correlation or covariance between observations as a function of their separation distance $h$. Common models include Spherical model ($\gamma(h) = C(1.5(h/a) - 0.5(h/a)^3)$), Exponential model ($\gamma(h) = C\exp(-3h/a)$), and Gaussian model ($\gamma(h) = C\exp(-3(h/a)^2)$).

Here, $C$ is the sill (variance of $Z(u)$) and $a$ is the range (distance at which spatial correlation becomes negligible). The covariance between pairs of observations can be represented in a variogram matrix $\Gamma$, where $\Gamma_{ij} = \gamma_{ij}$. The weights $\lambda_i$ are found by solving the system of linear equations
\begin{align}
    \Gamma \cdot \lambda = \gamma_{0i}
\end{align}
where $\gamma_{0i}$ represents the covariance vector between the target location $u_0$ and observed locations $u_i$. Kriging provides not only the best linear unbiased estimate (BLUE) of $Z(u_0)$ but also the estimation variance, which quantifies the uncertainty of the prediction. The detailed derivation can be found in Appendix \ref{sec:kriging_proof}.\\

\noindent\textbf{Space-Time Kriging:} Extending the kriging methodology to include time as a separate dimension, as done by \cite{graler2012spatio} with a small temporal window (from a few hours to a week) allows us to utilize the nearby temporal data points for the problem of spatial interpolation without building a detailed temporal state variable for specific locations. This approach is spatially extensible as there is no temporal state involved. It should be noted that large window sizes become computationally infeasible since the time complexity of Gaussian processes increases as a cubic function of the number of data points. \\

\noindent\textbf{Validating Stationarity Assumptions:} The stationarity assumption in ordinary kriging model is the wide-sense or weak sense stationarity, in which the mean and correlation function of samples does not change when the process is shifted in time. One approach for testing for time-series stationarity is the Augmented Dickey-Fuller (ADF) test \cite{adftest}. In this test, the null hypothesis is that the time series is non-stationary. We used the statsmodel python library \cite{seabold2010statsmodels} to test for stationarity in the time-series data of different sensors. For the 1 hour resolution data from both our and government sensors, we find that the raw data for most sensors (except 2 of our sensors) shows statistically significant stationarity behavior (with p value of the Null Hypothesis > 0.05). After our pre-processing steps (log-conversion followed by averaged temporal cubic spline subtraction) on the data, we found that the stationarity property of sensor readings did not change. This is surprising because we know that the raw data shows seasonal trends. When we repeated the test with daily average values instead of 1 hour resolution data, we found that only 17 sensors satisfy stationarity. Thus, we conclude that since we use kriging models on 1 hour resolution data, the stationarity modeling assumption holds on the data. Furthermore, since we only learn models with small time intervals (at most one week) of samples, longer time-scale variations like seasonal trends do not affect the hypothesis. It should be noted that spatial stationarity is not easy to estimate, and we have only validated the assumption along the temporal dimension. \\

\noindent\textbf{Neural Networks:} Another approach to spatial interpolation is to featurize the available data as $X = \{\text{lat, long, time}\}$ and $Y = \{PM_{2.5} \text{values}\}$, and use a neural network to learn the mapping between the input and output space. But, with the sparsity of sensors, it should be noted that non-linear models are likely to easily overfit the data, and hence, we do not delve into sophisticated architecture designs. It should also be noted that Kriging approaches essentially learn a new model for every time snapshot, the neural network model learns a single model for the entire dataset.

\subsection{Evaluation}
\label{sec:pred_results}

\begin{figure}[ht!]
    \begin{subfigure}{0.5\columnwidth}
        \centering
        \includegraphics[width=\columnwidth]{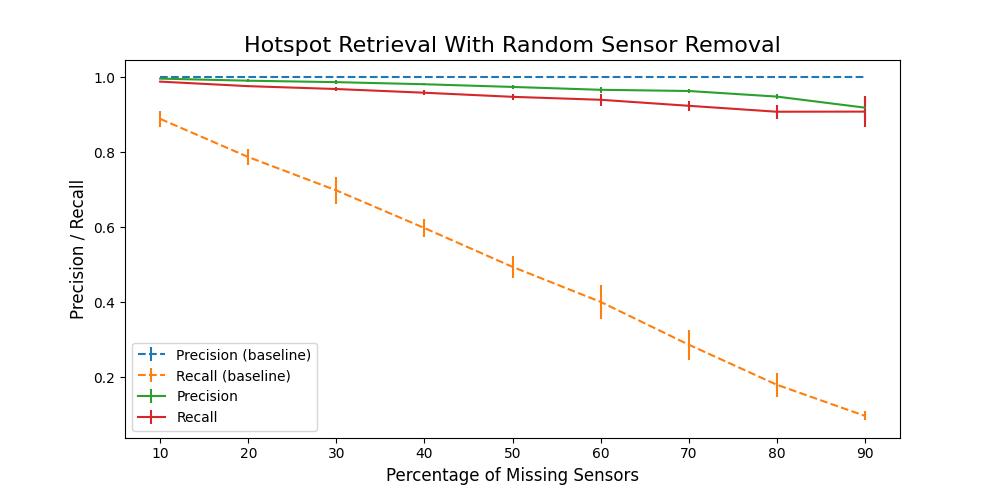}
        \caption{}
        \label{tab:missing_sensors}
    \end{subfigure}
~
    \begin{subfigure}{0.5\columnwidth}
        \centering
        \includegraphics[width=\columnwidth]{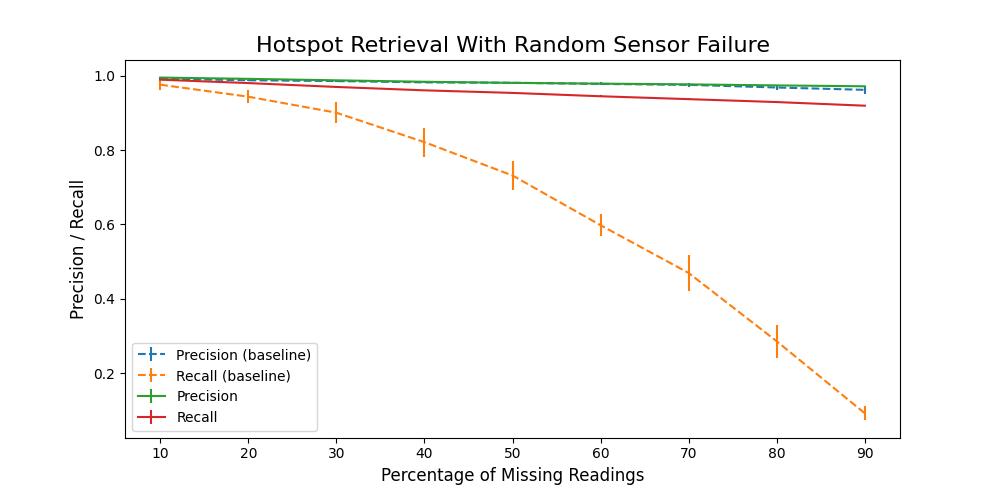}
        \caption{}
        \label{tab:missing_readings}
    \end{subfigure}

    \begin{subfigure}{\columnwidth}
        \centering
        \begin{tabular}{|c|c|c|}
            \hline
            \textbf{Method} & \textbf{MAPE} (\%) & \textbf{RMSE} ($\pmunit$) \\
            \hline
            Kriging & 35.06$\pm$ 0.57 & 42.37 $\pm$ 1.01 \\
            Space-Time Kriging & \textbf{30.24 $\pm$ 0.31\%} & \textbf{37.02 $\pm$ 0.90}\\
            Neural Network & 37.31 $\pm$ 0.78 & 45.94 $\pm$ 1.42\\
            \hline
        \end{tabular}
        \caption{}
        \label{tab:interpolation}
    \end{subfigure}

     \begin{subfigure}{\columnwidth}
     \centering
        \begin{subfigure}{0.2\columnwidth}
            \includegraphics[width=\columnwidth]{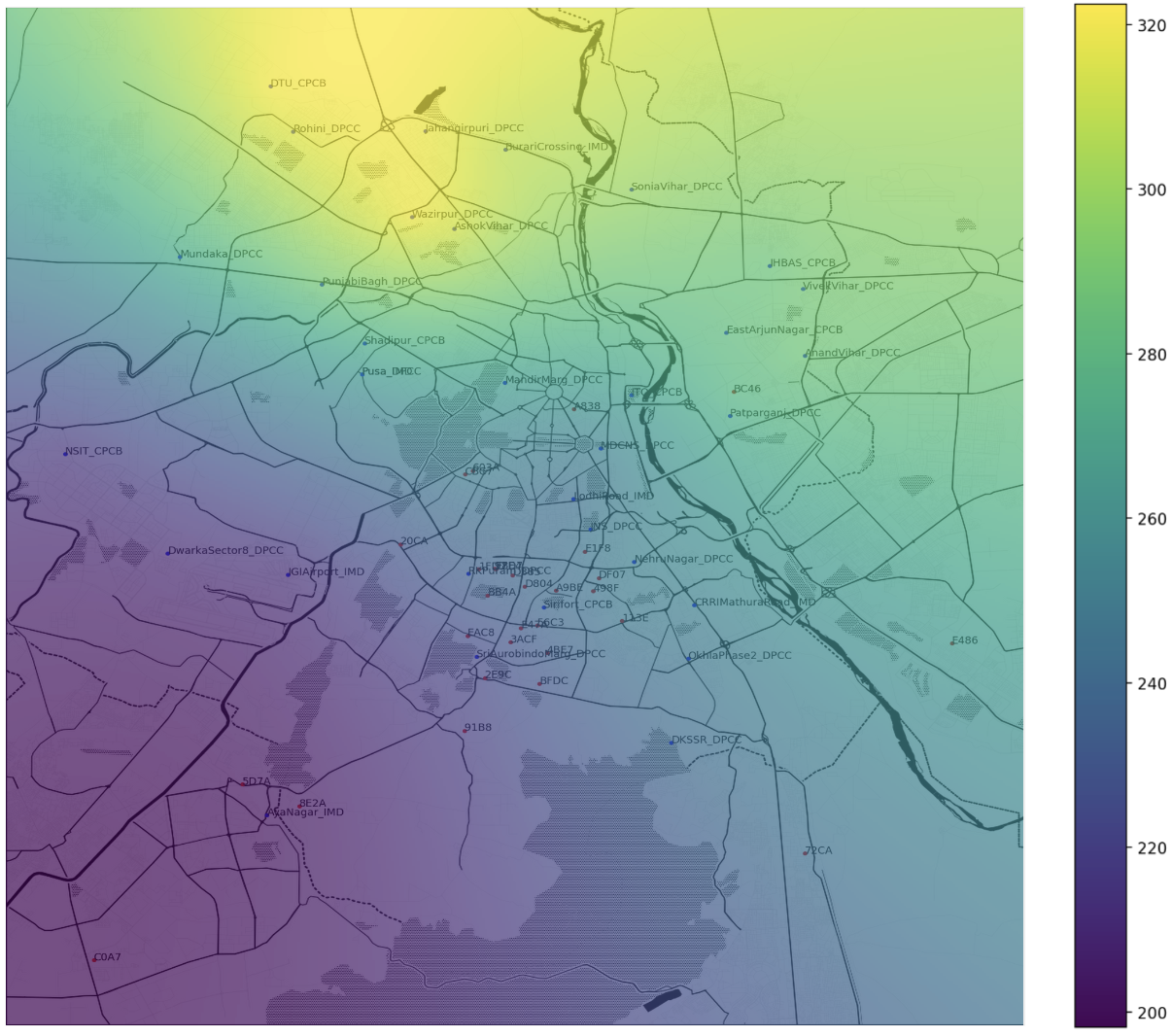}
            \caption{}
            \label{fig:heatmap1}
        \end{subfigure}
        ~
        \begin{subfigure}{0.2\columnwidth}
            \includegraphics[width=\columnwidth]{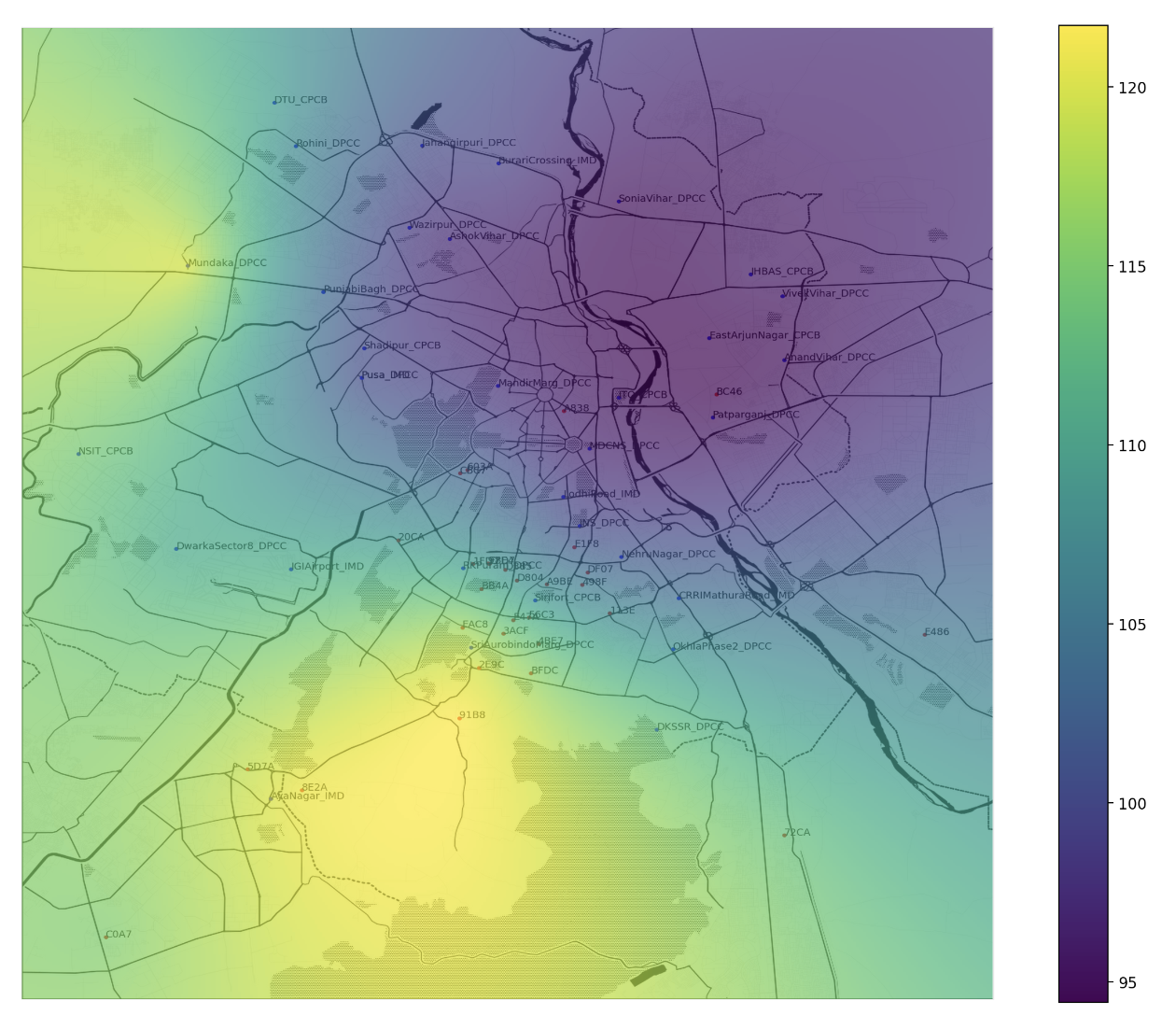}
            \caption{}
            \label{fig:heatmap2}
        \end{subfigure}
        ~
        \begin{subfigure}{0.2\columnwidth}
            \includegraphics[width=\columnwidth]{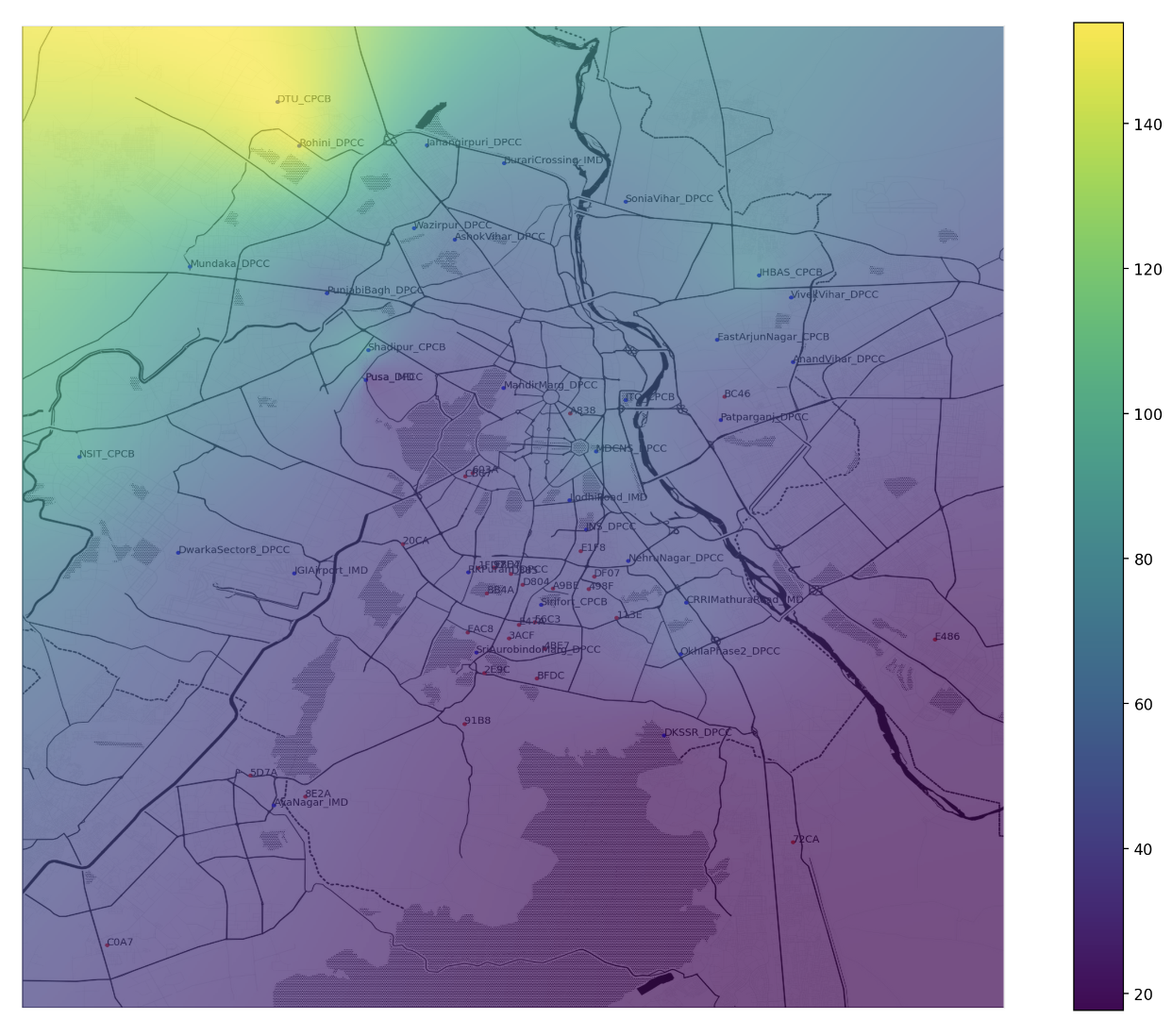}
            \caption{}
            \label{fig:heatmap3}
        \end{subfigure}
        ~
        \begin{subfigure}{0.2\columnwidth}
            \includegraphics[width=\columnwidth]{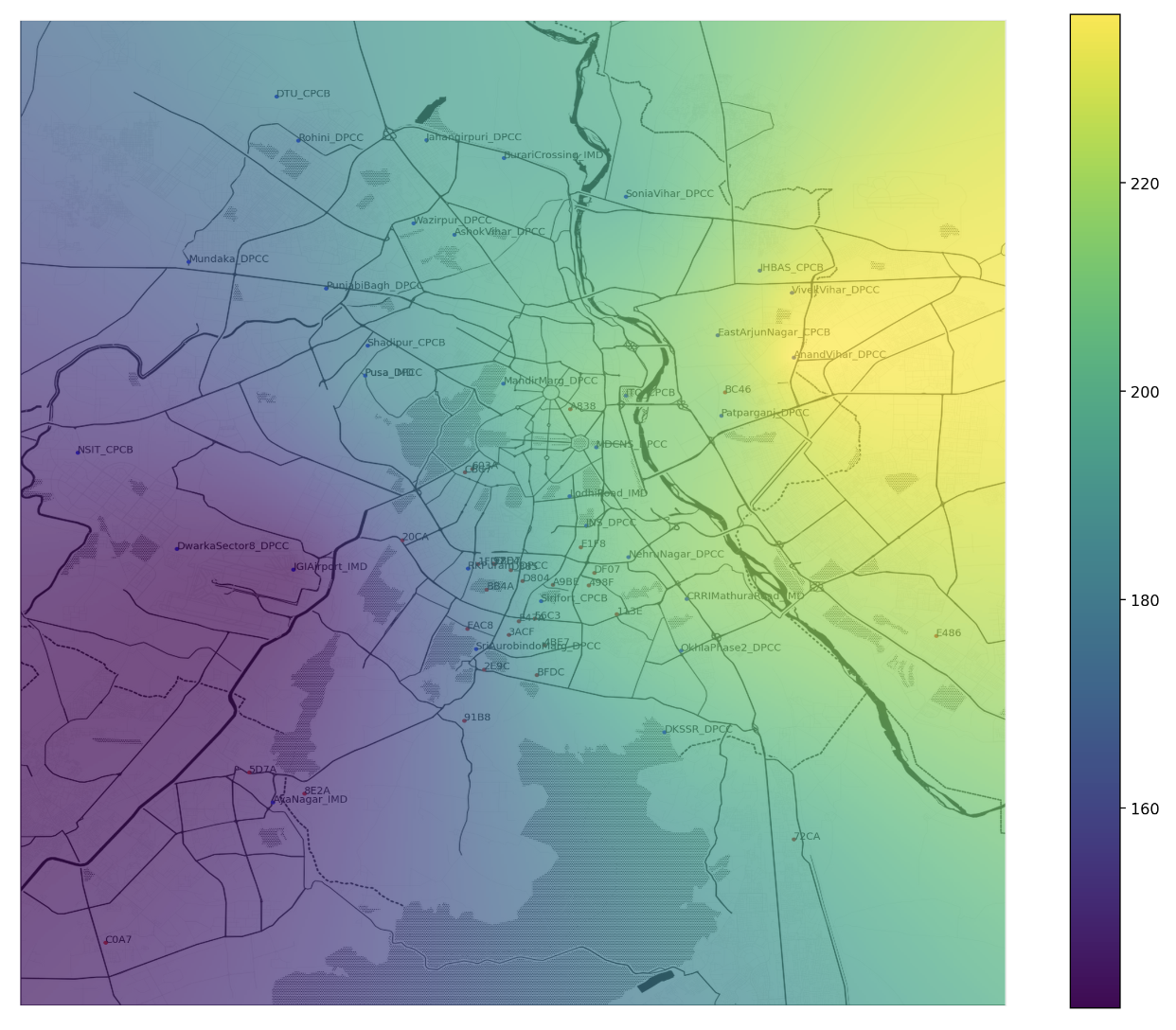}
            \caption{}
            \label{fig:heatmap4}
        \end{subfigure}
    \end{subfigure}
    \caption{Results of different modeling approaches. In Figure \ref{tab:missing_sensors}, we present the result of hotspot retrieval performance at new spatial locations using the Space-Time Kriging model. As can be seen, we are able to predict hotspots at completely new spatial locations with high precision and recall. In Figure \ref{tab:missing_readings}, we show similar numbers for hotspot retrieval performance under random sensor failure. Compared to no interpolation, which is the current baseline, our method gives much better recall numbers. When the percentage of missing values is extended beyond 90\%, the ST-Kriging model fails to run due to limited number of data points. In Figure \ref{tab:interpolation}, we show the spatial interpolation performance of different modeling strategies. We chose Space-Time Kriging for other experiments due to getting lowest error using this method. Finally, Figures \ref{fig:heatmap1}-\ref{fig:heatmap4} show the heatmaps of the examples of pollution gradient generated by the Space-Time Kriging method for qualitative analysis. Such heatmaps can be further used for delineating the spatial boundaries of hotspots.}
\end{figure}

\noindent\textbf{Interpolation Performance:} This evaluation of different modeling strategies focuses on the spatial interpolation performance of different models. For this purpose, we divide our data generated by the augmented network into a train and evaluation split in the following manner. For every timestep (a row in our dataframe), if the number of sensors active is more than 30 (assuming 50\% availability), we divide the sensors randomly into training and evaluation sets in an 80-20 split. Different models are then trained on the training set, and their predictions for the test set sensors are compared with actual readings. For Space-Time Kriging, we choose a window size of 7 around the timestep, which means that the model is given all available readings for 3 timesteps before and after the timestep on which interpolation is performed, along with the values of the training split of sensors for the corresponding interpolation timestep. Note that for Kriging and Space-Time Kriging, we learn a separate model at every interpolation timestep, and hence, there is no data leakage in this experimental procedure.

Figure \ref{tab:interpolation} presents the computed metrics, mean absolute percentage error (MAPE), and root mean squared error (RMSE) for the different modeling methods on five different random splits of the data. As shown in the table, the Space-Time Kriging approach provides the best performance with the lowest errors, while ordinary spatial Kriging follows that. The neural network approach lags behind these methods further, but it does provide us with a single model that predicts the spatio-temporal field, while kriging models are essentially different models for different timestamps, and can not be generalized in the time dimension. \\

\noindent\textbf{Hotspot Identification at New Spatial Locations:} For this goal, we take a certain percentage of sensors from our sensor network randomly and remove all readings from these sensors. This is equivalent to assuming that these sensors do not exist. Then, using the Space-Time Kriging approach (see section \ref{sec:methods} for details), we interpolate the readings of these sensors. Following this, we apply the definitions of hotspots, and compute the retrieval performance of hotspots on the interpolated data, as compared to the original data for the missing sensors. This allows us to evaluate the performance of our interpolation model in retrieving hotspots at any new spatial location. Figure \ref{tab:missing_sensors} has the precision and recall numbers for the entire task, for different percentages of missing sensors. As we can see, even for 90\% missing sensors, we find that the algorithm has very high precision and recall values. Comparing to the baseline case, where no imputation was done, we see a consistent drop in the recall values. Thus, we show that using our methodology, we can expand the effective sensor network size and capture many of the hidden hotspots with high accuracy.

While the performance of our algorithm on identifying hotspots at new locations is remarkable, note that the interpolation method is likely to perform better for sensors that have more neighboring sensors. To obtain an idea of relative performance in different cases, we consider two sensors that have contrasting locations, one is surrounded by other nearby sensors (JNS\_DPCC), while the other is in an isolated location (NSIT\_CPCB). The Space-Time Kriging models are then used to predict the pollution values at these locations across different timestamps. The MAPE and RMSE values for JNS\_DPCC are 28.51\% and 27.53 $\pmunit$, while for NSIT\_CPCB, they are 40.72\% are 51.48 $\pmunit$ respectively. Using these numbers, we can somewhat estimate the variance of prediction error of interpolation models on new spatial locations. \\

\noindent\textbf{Mapping and Hotspot Boundary:} The interpolation models can be further used to generate pollution maps of the city at individual timesteps, which can be used for visualization as well as for finding other potential hotspot locations as well as parameters like the geometry and area of hotspots. We present one example of the pollution map of the city generated by the method of Space-Time Kriging in Figures \ref{fig:heatmap1}-\ref{fig:heatmap4}. For delineating the geometric boundary of hotspots, we use the saddle-point method as described in \cite{6588606}. The details of the method can be found in the Appendix \ref{sec:bounds}. Note that the map and the boundaries of the hotspots generated by this method can not be evaluated using the sensor data, hence we can only use these as tools for qualitative understanding.\\

\noindent\textbf{Robustness Against Random Sensor Failure:} Since we have a rough idea of the prediction error at new locations, another evaluation that we are interested in is the usefulness of these models in the context of hotspot identification. For this, we designed the following experiment. First, we randomly drop a percentage of readings, simulating random sensor failure. We then use the Space-Time Kriging model to interpolate on these readings and tabulate the effects of these methods in hotspot retrieval with an increasing percentage of sensor failures. For the baseline, we consider the case when the dropped sensor readings are left as it is, and hotspot definitions are applied to the incomplete data. The results of the experiments are presented in Figure \ref{tab:missing_readings} as precision and recall of hotspots based on the hotspots from the original data as the ground truth. Note that the hotspots are already defined over the time period of a month, and thus, are robust to random sensor failures to a large extent. As seen in the table, with appropriate imputation, the information of hotspots can be retrieved even if 90\% of the deployed sensors randomly fail. \\

\noindent\textbf{Generalization Across Environments:} Our methodology can be easily extended to other urban environments, irrespective of the diverse environmental, economic, and infrastructural landscapes which might result in a different pollution pattern. Due to budget constraints, we have not done a multi-city deployment of low-cost sensor networks, the effectiveness of this approach has been shown in other published research \citep{jiao2016community, lin2015evaluation, shusterman2016berkeley, moltchanov2015feasibility, sun2016development, tsujita2005gas, gao2015distributed, 10.1145/2668332.2668346, kumar_rise_2015, kortoci_air_2022, qin_fine-grained_2022,ijcai2022}. The comparisons between our deployment efforts and prior work has been presented in Appendix \ref{sec:related}. One such deployment has been done by Purple Air \cite{purpleair} in New York City. NYC has significant differences in the underlying factors (environmental, economic, and infrastructural) compared to New Delhi, and Purple Air’s network consists of low-cost sensors, making this data the ideal testbed to show generalization. Taking the data from a subset of pollution sensors for the full year of 2023 from the Purple Air data portal, we ran Space-Time Kriging model on this data. Since the scale of the two datasets and observed values of pollution differ significantly, we normalized the RMSE observed on the NYC dataset by the ratio of average spatial variance between the two locations. The normalized RMSE value for Space-Time Kriging comes out to $32.598\pm 4.898 \pmunit$, which is equivalent to what we observe for the New Delhi dataset. Thus, we show that the Space-Time Kriging model provides consistent performance across different urban environments. The map of the NYC sensors is given in Appendix \ref{sec:purpleair}.\\

\noindent\textbf{Generalization Across Definitions:} We conducted additional experiments and evaluated the performance of our model when the threshold in hotspot definition is adjusted. As the hotspot definition has multiple thresholds, we decided to vary the pollution level threshold for all three types of hotspots while keeping their ratio the same as in the original definition. We found that for lower thresholds, our model’s performance remains similar. For higher thresholds, we found both the precision and recall of our model to be lower than before for the random sensor removal experiment. The recall value is still significantly higher than the baseline, but the drop in precision shows that for increasing thresholds, the model’s accuracy deteriorates.

\subsection{Policy Recommendations}
\label{sec:recommendations}
\textbf{Current Policy:} The CPCB and DPCC have published a list of 13 pollution `hotspots' in the greater Delhi region \citep{Shrangi_Pillai_2019,PriyangiAgarwal_2020,dpcc_hotspot}. These regions have been declared hotspots by the virtue of exhibiting consistently high levels of pollution (poor air quality measured in terms of PM concentration values) compared to neighboring regions. Green norms set by the Supreme Court-mandated Environment Pollution (Prevention and Control) Authority were reported to have been violated at these locations. These regions are -- Anand Vihar, DTU, Dwarka, Jahangirpuri, Mundka, NSIT, Okhla Phase 2, Punjabi Bagh, Rama Krishna Puram, Rohini, Vivek Vihar and Wazirpur. All these locations have regulatory monitors installed in the area. According to the official 2019 report, the pollution levels were noted to be ``usually higher at these places round the year''. According to the action plan mentioned in \cite{dpcc_hotspot}, some emission sources that have been noted for high pollution in these areas are localized burning of domestic waste (plastic, rubber, etc.), illegal dumping of construction and demolition waste, traffic congestion, road dust, and industries. \\

\begin{figure}[t]
    \centering
    \begin{subfigure}{0.28\columnwidth}
        \includegraphics[width=\columnwidth]{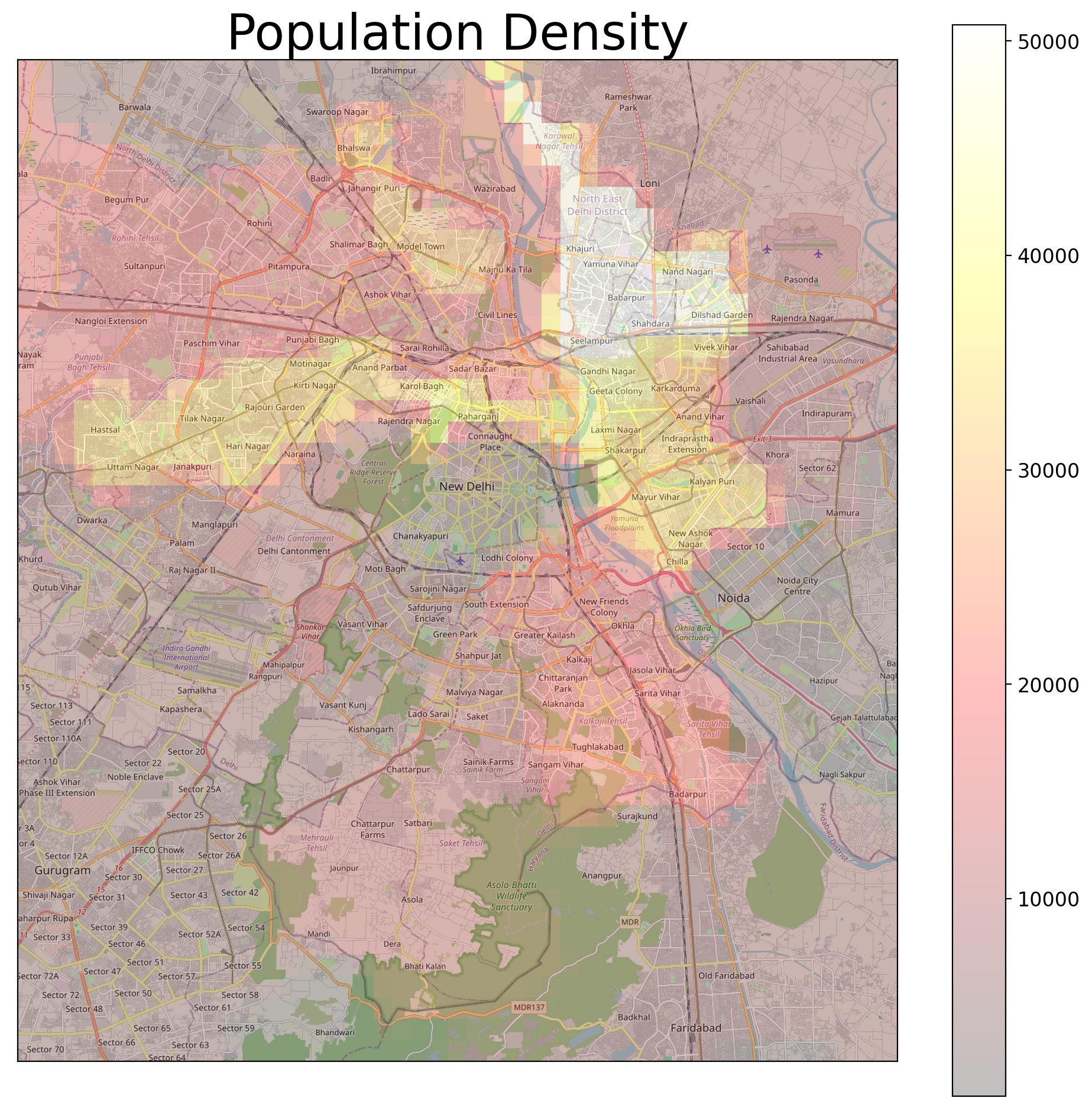}
        \caption{}
    \label{fig:population}
    \end{subfigure}
    ~
    \begin{subfigure}{0.28\columnwidth}
        \includegraphics[width=\columnwidth]{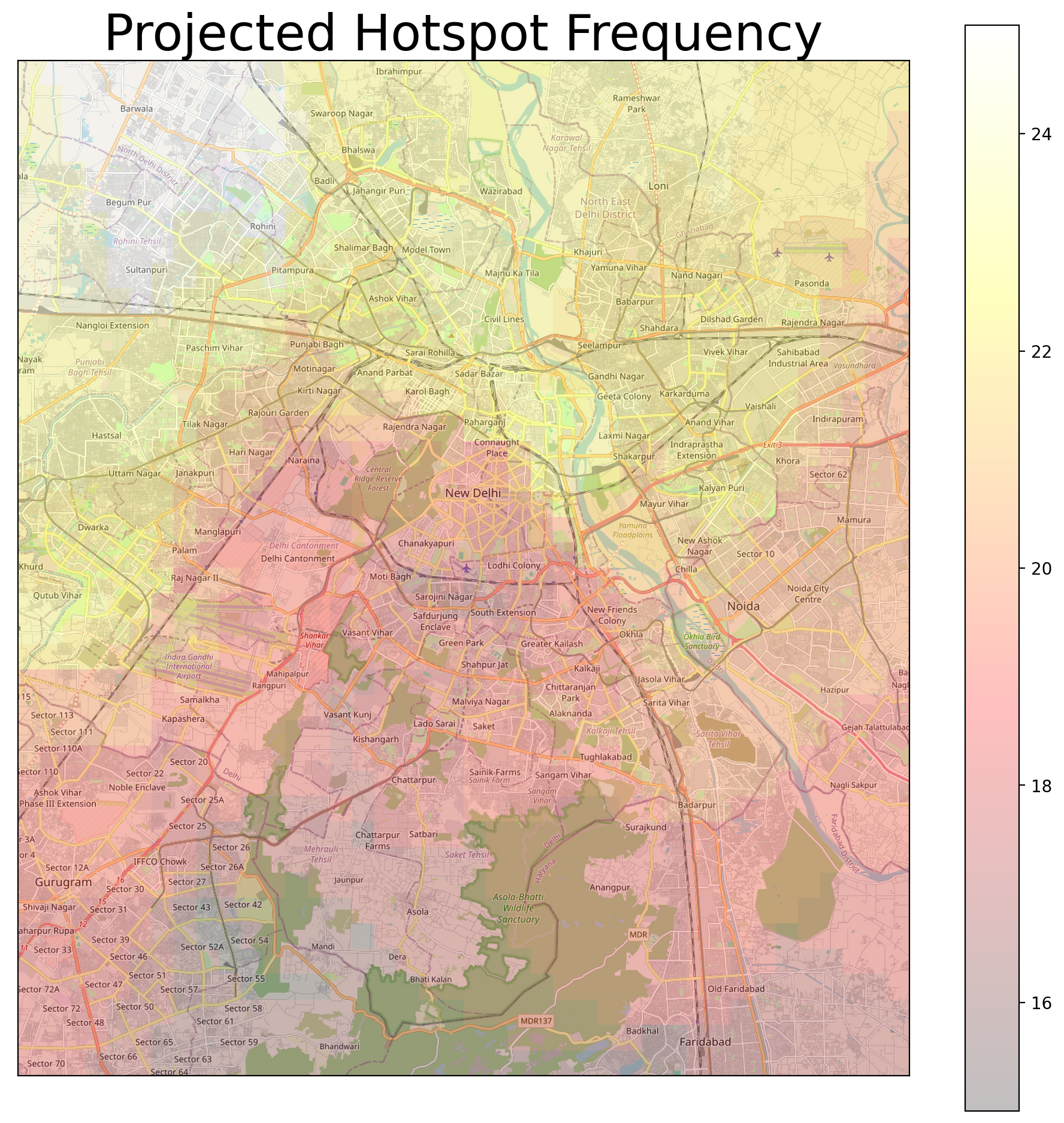}
        \caption{}
    \label{fig:hotspot_spread}
    \end{subfigure}
    ~
    \begin{subfigure}{0.28\columnwidth}
        \includegraphics[width=\columnwidth]{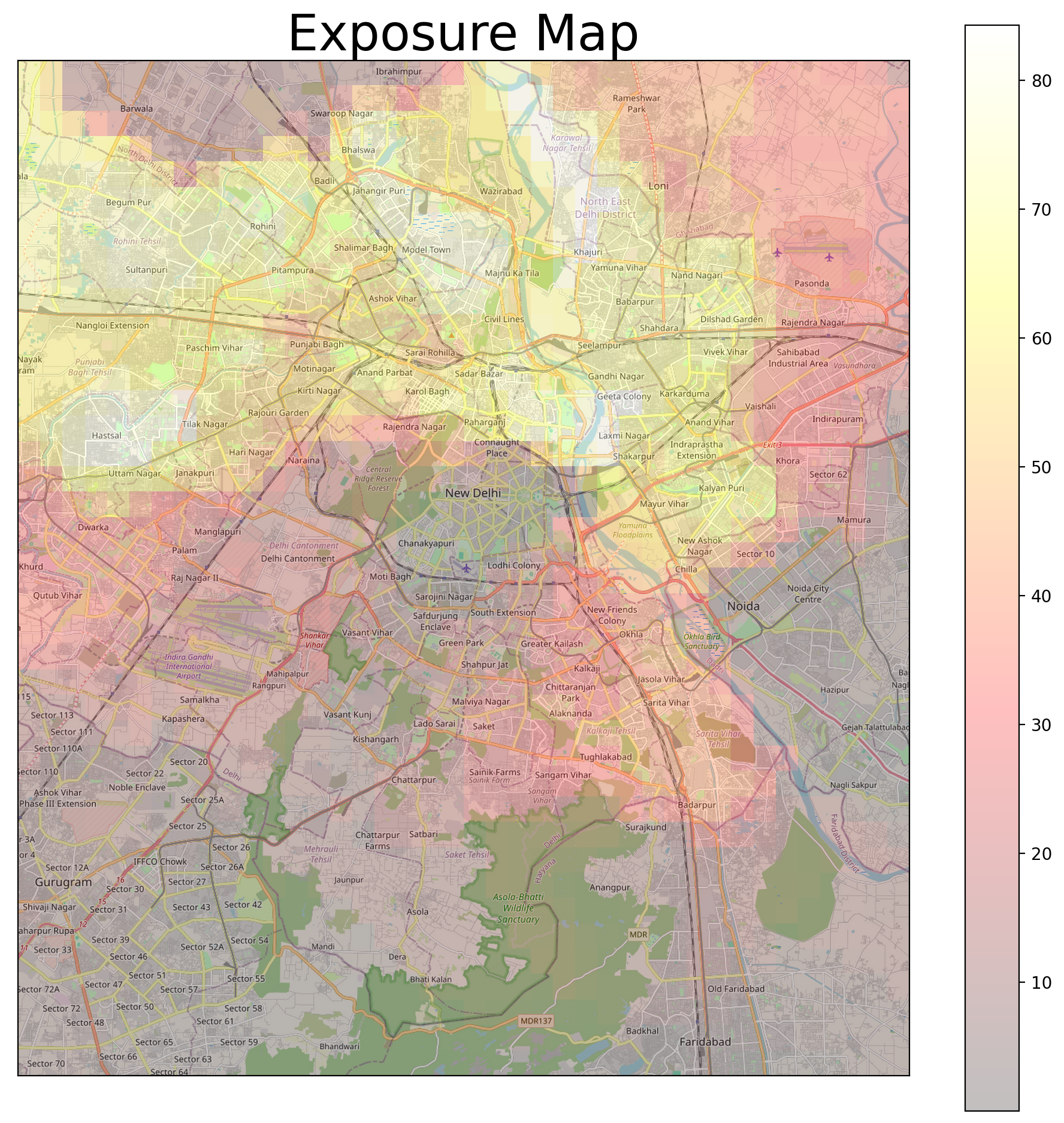}
        \caption{}
    \label{fig:exposure_map}
    \end{subfigure}
    \caption{In Figure \ref{fig:population}, we show the spread of population density in the city based on \citep{ColumbiaUniversity2018}. In Figure \ref{fig:hotspot_spread}, we show the frequency (in months) of projected hotspots on the map obtained by applying the hotspot definitions to our predictions. Finally, in Figure \ref{fig:exposure_map}, we show the population-hotspot exposure map, which is a normalized product of the previous two.}
    \label{fig:overview}
\end{figure}

\noindent\textbf{Recommendations:} We split the collected data over the study duration into two years, from 1 July, 2018 to 30 June, 2019, and from 1 July, 2019 to 30 June, 2020. We do not consider the remaining data to account for the seasonal variation (see Figure \ref{fig:longitudnal-1W}). Over the two years, we find that most locations mentioned in the \cite{dpcc_hotspot} report exhibit annual average value greater than $100 \pmunit$ in both years, with the exception of Jaghangirpuri, NSIT, and Okhla Phase 2, which exceed the threshold for only one year. Apart from these locations, we found \textbf{32 locations for year 1 and 7 locations for year 2 that can be considered as annual hotspots}. Thus, we can confirm that the official policy is correct, but not exhaustive. Out of these locations, \textbf{18 locations in year 1 hotspots and 4 locations in year 2 hotspots do not have a public monitoring station nearby}. Out of these, the newly uncovered locations of Fullbright House, Faridabad, and Preet Vihar have shown recurring hotspot behavior for both years. One clear trend observable in our findings is that the pollution levels for year 2 (average $87 \pmunit$) improved as compared to year 1 (average $109 \pmunit$).

Further, using interpolation methods described before, we computed the pollution heatmaps of the study area for each timestep in our collated dataset. By combining the 21,952 heatmaps, we can predict the occurrence of hotspots at a resolution of 1km x 1km grid cells. This projected hotspot map was presented in Figure \ref{fig:hotspot_spread} at the beginning. Based on this map, we can see that the northern part of New Delhi is much more heavily polluted and shows hotspot behavior as compared to the southern part. This can be explained in part by the presence of brick kilns in the northern parts of New Delhi. Combining percentile plots of this data and the population density data obtained from \cite{ColumbiaUniversity2018}, we can obtain an population-exposure map (see Figure \ref{fig:exposure_map}). Based on the calculated exposure map, \textbf{we project that areas of North-East Delhi district (Karawal Nagar Tehshil), East Delhi (Geeta Colony and Lakshmi Nagar), West Delhi (Punjabi Bagh Tehshil), North West Delhi (Rohini Tehshils), and some affluent residential areas in Old Delhi (Model Town) should be prioritized for maximal reduction in population pollution exposure}. In fact, while calculating the projected exposure of the city's population to hotspots, \textbf{we found that all locations in the city were predicted to have been hotspots for at least 50\% of the study duration}.

\section{Mechanistic Modeling of Hotspots}
\label{sec:mechanistic}

It is important that we understand the mechanics of the dispersion of pollution in the air and the parameters that affect this process. This first principle mechanistic approach to modeling hotspots is required to connect the hotspots to the possible sources of emission and generate proactive actions from the policy perspective, instead of the reactive nature of current policy. This bottom-up approach to source apportionment \cite{src_apportionment_primer} has been studied in prior works \cite{GUTTIKUNDA2013101,10.1007/978-94-007-2540-9_11} for New Delhi, but to our knowledge, we are the first to apply mechanistic models in context of air pollution hotspots. In this section, we first present our method of deriving an approximate emissions inventory consisting of different emission sources, followed by the Gaussian plume dispersion model setup and our accelerated approximate implementation of the model, and finally describe the framework for explaining hotspots.


\subsection{Deriving Approximate Emissions Inventory}
\label{sec:inventory}

The general bottom-up approach to source apportionment starts with building a detailed emissions inventory, which is a detailed spatiotemporal map of all pollution-causing activities in the city, along with estimates of their fuel consumption and emission \cite{src_apportionment_primer}. This emissions inventory is then used as input for pollution dispersion models to compute the pollution map of the city. The process of collating an accurate and detailed emissions inventory is a significant undertaking, requiring on-ground surveys of emission activities.
In \cite{GUTTIKUNDA2013101}, Guttikunda et. al. build a detailed GIS-based emissions inventory where they divided their study area into grids of $0.01^o$x$0.01^o$ (approximately 1km x 1km). According to them, the eight major sources of pollution emissions in New Delhi are brick kilns, industries, power plants, domestic emissions, vehicular emissions, road dust, construction activities, and garbage burning. Every cell in their inventory consists of the details of emissions from the eight sources.
Unfortunately, the inventory is not publicly available, so we had to reconstruct an approximate version. In the process, we were unable to find the updated distribution maps for road dust, garbage burning, and construction activities, which according to Guttikunda et. al., account for about 20\% of New Delhi's emissions.

We denote our emissions inventory (E) as a matrix
\begin{equation}
    E[i,j,t] = \{Q_b, Q_i, Q_p, Q_d, Q_v\}
\end{equation}
where, $i$ and $j$ are latitude and longitude indexes, ant $t$ is the time. $Q_b, Q_i, Q_p, Q_d, Q_t$ refer to the intensities of pollution sources corresponding to brick kilns, industries, power plants, domestic emissions, and vehicular emissions.\\

\begin{figure}[t]
    \begin{subfigure}{0.35\columnwidth}
        \centering
        \includegraphics[width=\columnwidth]{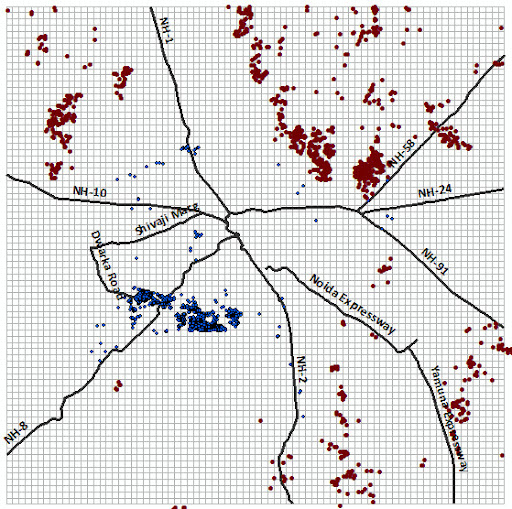}
        \caption{Brick Kilns}
        \label{fig:guttikonda_bricks}
    \end{subfigure}
    ~
    \begin{subfigure}{0.35\columnwidth}
        \centering
        \includegraphics[width=\columnwidth]{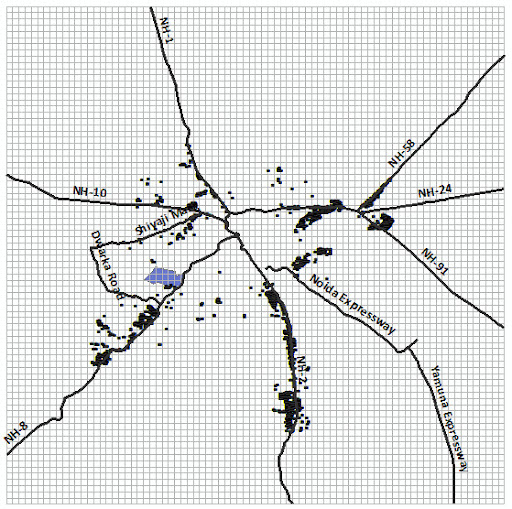}
        \caption{Industries}
        \label{fig:guttikonda_industry}
    \end{subfigure}
    \caption{The maps of brick kilns and major industries around New Delhi \cite{GUTTIKUNDA2013101}. The geographic region  between (28.2$^o$,76.85$^o$) and (29.0$^o$,77.65$^o$) latitude-longitude coordinates has been covered in the map. The figures are taken from \cite{GUTTIKUNDA2013101} and are not our contribution.}
    \label{fig:guttikonda_maps}
\end{figure}

\noindent\textbf{Brick Kilns, Industries, Power Plants and Domestic Emissions:} Using the distribution map of industries and brick kilns from \cite{GUTTIKUNDA2013101}, we reconstructed the positional map of these pollution sources by using the number of colored pixels to approximately infer the number of sources in different cells. Thus
\begin{equation}
    E[i,j][Q_b] = P_r(i,j) 
\end{equation}
where $P_r(i,j)$ is the number of red dots in cell i,j of brick kilns map, and
\begin{equation}
    E[i,j][Q_i] = P_b(i,j) 
\end{equation}
where $P_b(i,j)$ is the number of black dots in cell i,j of industries map from \cite{GUTTIKUNDA2013101}.
Note that we have made the simplifying assumption that all instances of brick kilns and industries emit equal amounts of PM$_{2.5}$ pollutants per unit of time, making the inventory independent of time for $Q_b$ and $Q_i$. 
For the case of power plants, we assumed that the emissions of power plants are proportional to their power capacity. Adjusting for the historical data on the decommissioning and changes in capacities of power plants, we were able to create a similar inventory for power plants.
\begin{equation}
    E[i,j][Q_p] = \sum \text{Capacity}_{pp'}, \text{ for } pp' \in \text{ cell i,j}.
\end{equation}
For domestic sources of pollution, we utilized the Gridded Population of the World (GPWv4) population density data \cite{ColumbiaUniversity2018} as representative of domestic pollution emission intensity. This data was adjusted to fit the $0.01^o$x$0.01^o$ positional map. The same assumption of spatiotemporal uniformity was applied to convert this map into emission intensities. 
\begin{equation}
    E[i,j][Q_d] = \text{Population}_{i,j}
\end{equation}
where Population$_{i,j}$ is the population density in cell i,j.\\

\begin{figure}[t]
    \begin{subfigure}{0.35\columnwidth}
        \centering
        \includegraphics[width=\columnwidth]{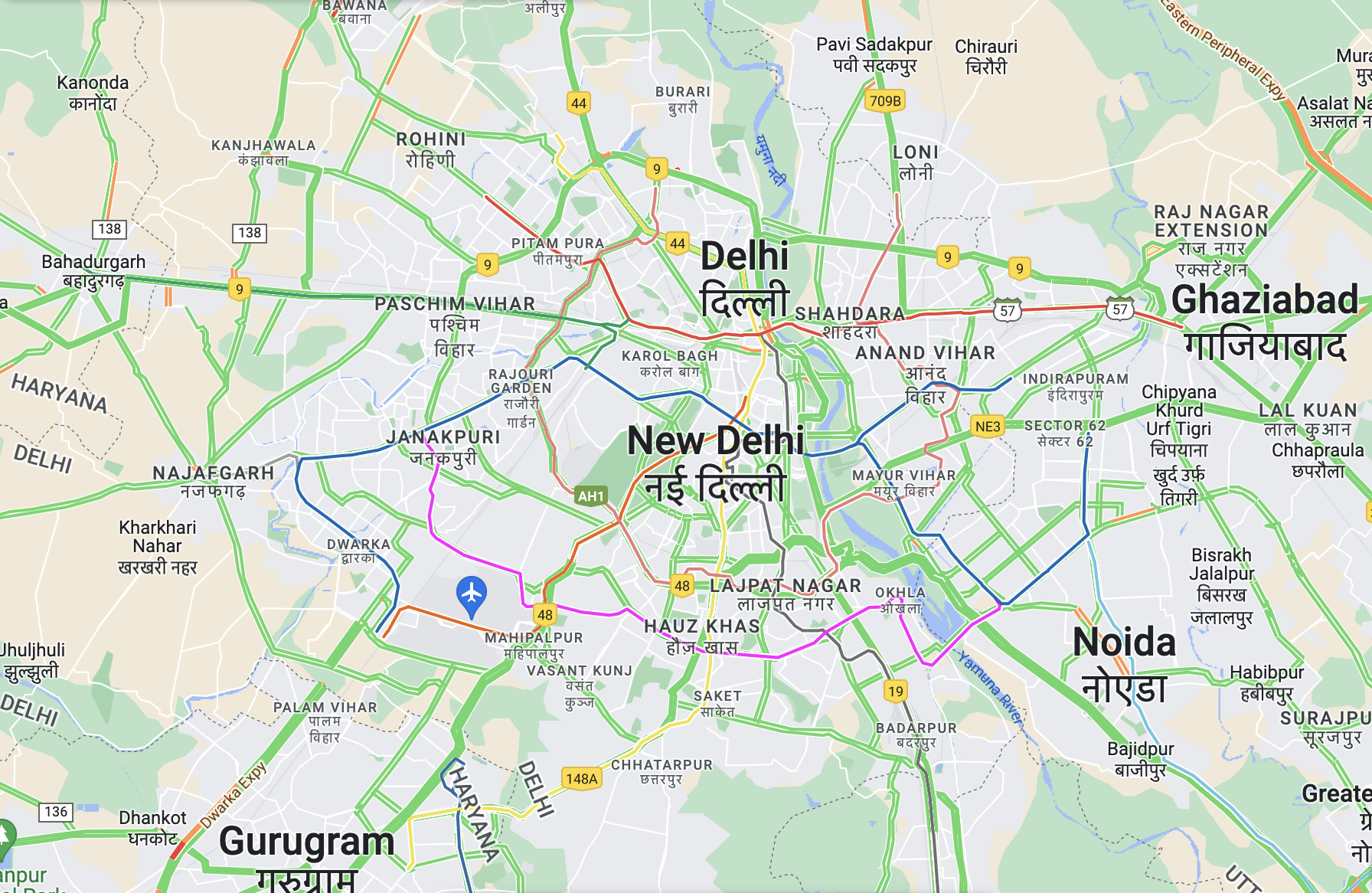}
        \caption{6 AM}
        \label{fig:6am}
    \end{subfigure}
    ~
    \begin{subfigure}{0.35\columnwidth}
        \centering
        \includegraphics[width=\columnwidth]{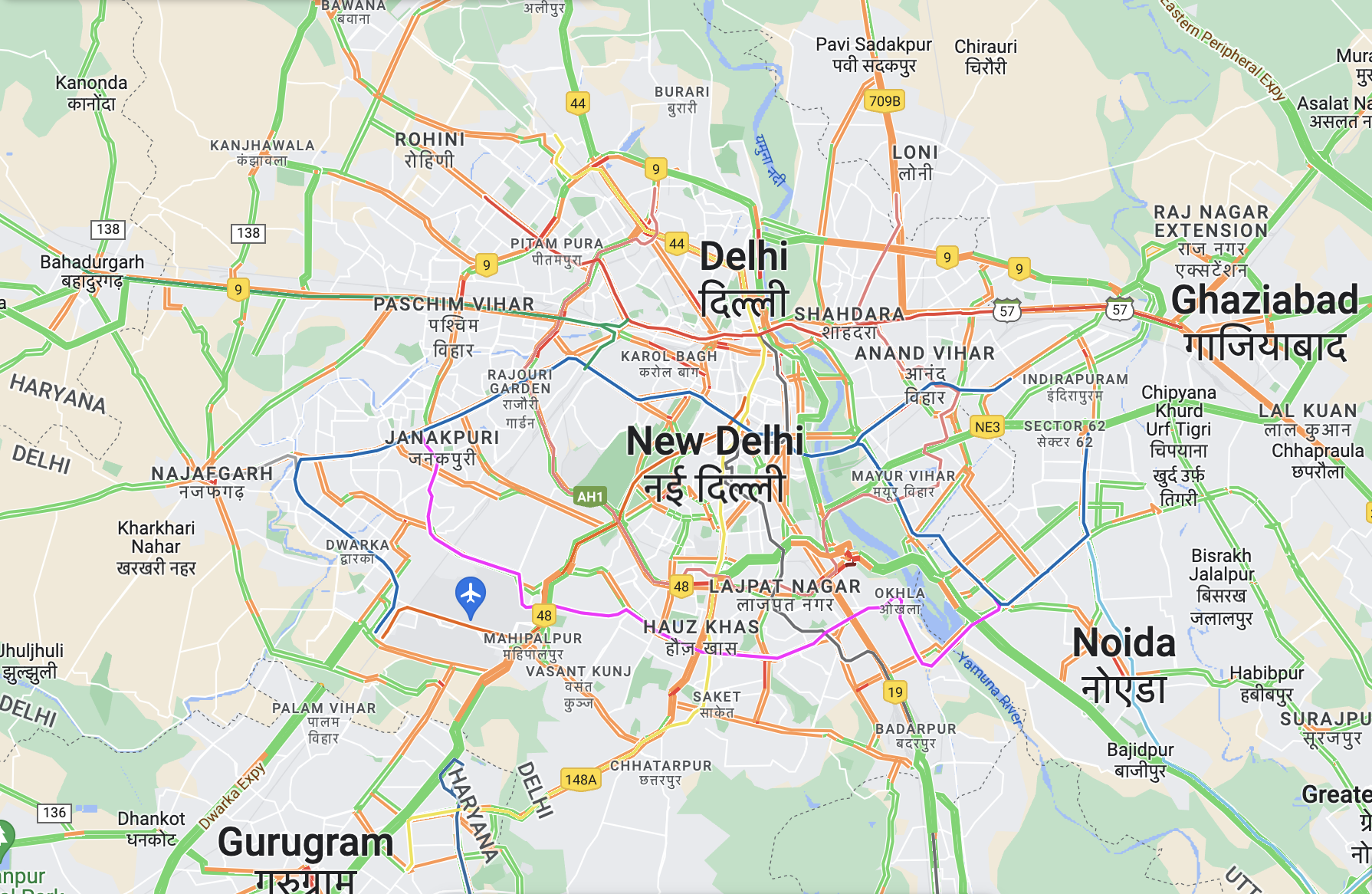}
        \caption{12 PM}
        \label{fig:12pm}
    \end{subfigure}

    \begin{subfigure}{0.35\columnwidth}
        \centering
        \includegraphics[width=\columnwidth]{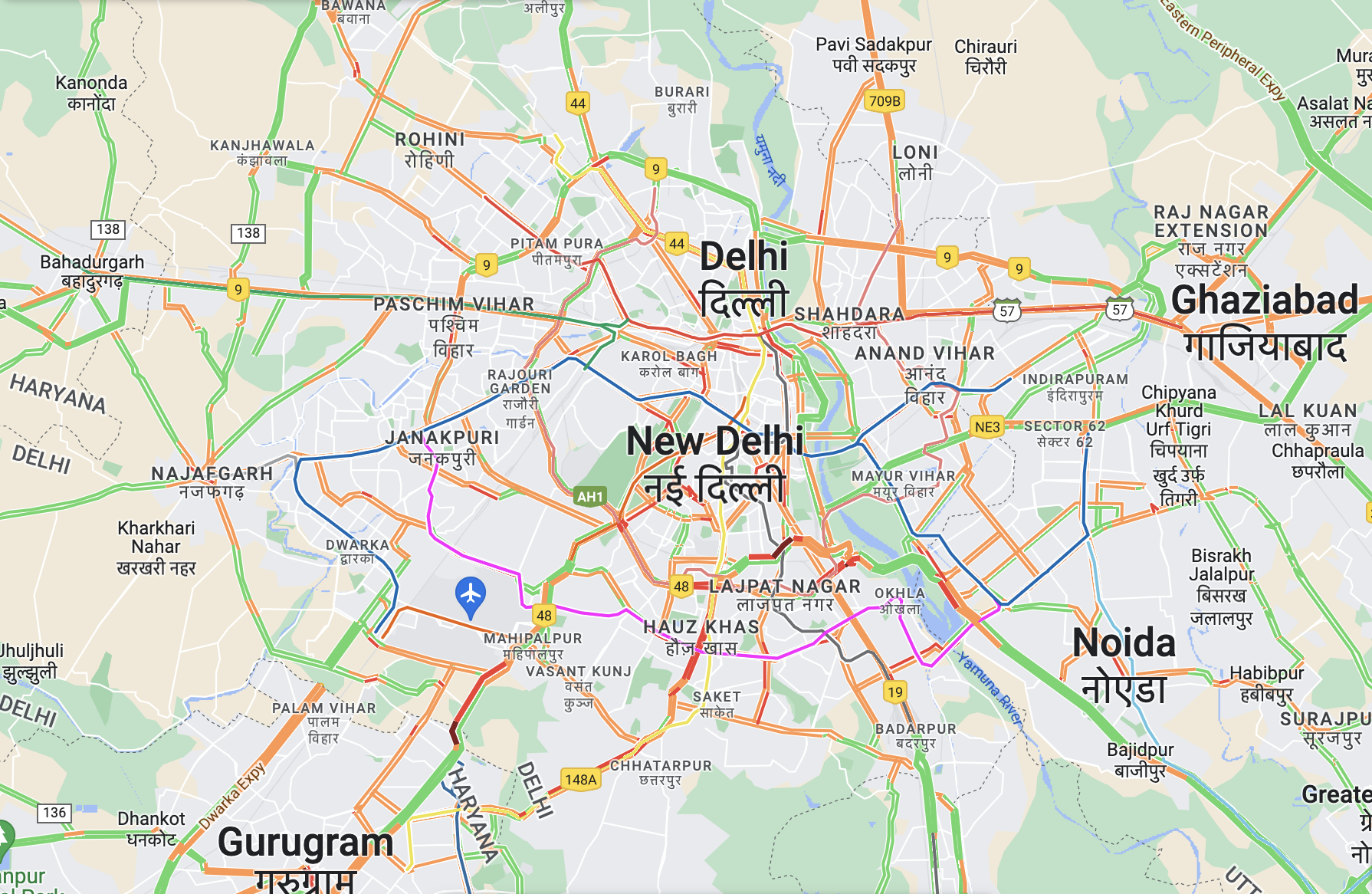}
        \caption{6 PM}
        \label{fig:6pm}
    \end{subfigure}
    ~
    \begin{subfigure}{0.35\columnwidth}
        \centering
        \includegraphics[width=\columnwidth]{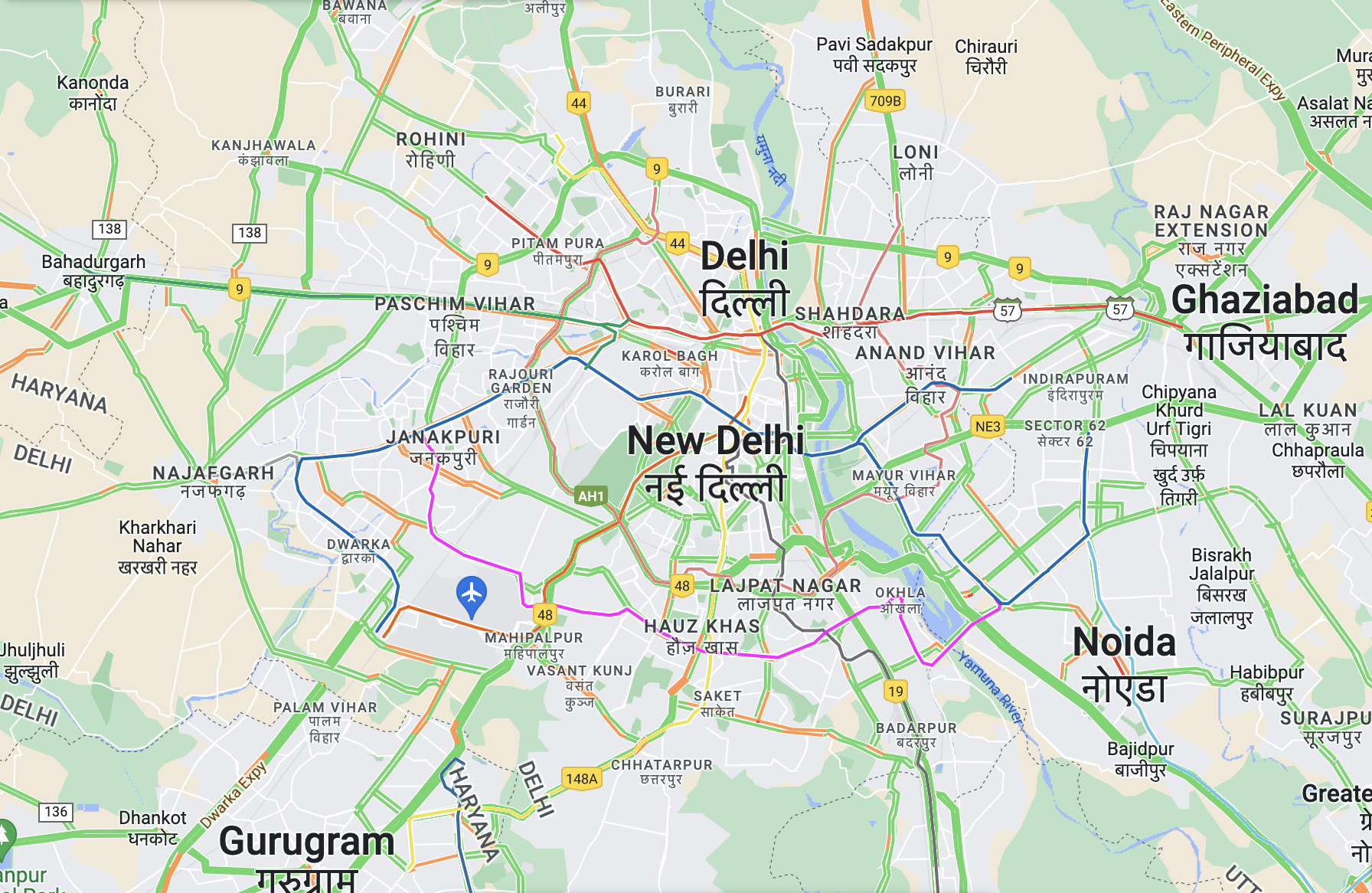}
        \caption{10 PM}
        \label{fig:10pm}
    \end{subfigure}
    \caption{Typical traffic on Mondays in New Delhi at different times as inferred by Google maps.}
    \label{fig:traffic_gmaps}
\end{figure}

\noindent\textbf{Vehicular Emissions:} We could not find open-source detailed traffic measurement data for New Delhi. Therefore, we used Google maps to obtain a representation of New Delhi's traffic at 4 different time snapshots, 6 AM, 12 PM, 6 PM, and 10 PM (see Figure \ref{fig:traffic_gmaps}) for a section of the study area. The typical traffic patterns on different days were only slightly different, hence, we assume that these maps are an average representation of the traffic in the city on all days. From Google's explanation, we know that green roads signify no delays in traffic, which is equivalent to allowing vehicles to move at the speed limit. Evoking prior work on traffic curves \cite{BHARDWAJ2023100105}, we can reasonably bound the maximum possible link density (the density of vehicles on the road) in such a case to less than 0.1. Thus, we assume that pollution caused due to traffic on green-marked roads is negligible. As seen in Figure \ref{fig:6am}, at 6 AM, we have near-zero traffic-based emissions. Therefore,
\begin{equation}
    E[i,j,6am][Q_v] = 0.
\end{equation}

Using the 6 AM map as the baseline, and using image differencing, we computed the additional traffic at other times. 
In the differenced images, we found roads marked `orange', `red', and `maroon' corresponding to progressively increasing levels of traffic delay. Based on the structure of the traffic curves in \cite{BHARDWAJ2023100105}, we posit that `orange' roads corresponding to mild delay be in the free-flow traffic regime, `red' roads corresponding to significant congestion would be in the spiraling regime, and `maroon' roads corresponding to traffic jams would be in the jam regime. For traffic curves derived for typical segments in other cities in developing countries like Nairobi and Sao Paulo, the link densities for the three regions would approximately be in the ratio 1:2:3. Thus, we use this ratio for our analysis.
\begin{equation}
    E[i,j,t][Q_v] = P_o(I_{t}-I_{6am},i,j) + 2*P_r(I_{t}-I_{6am},i,j) + 3*P_m(I_{t}-I_{6am},i,j)
\end{equation}
where $I_t$ refers to traffic snapshot at time $t$, and $P_o, P_r, P_m$ are pixel-counting functions.

Note that here we have assumed the traffic emissions to be proportional to link density (number of vehicles). The resulting emissions inventory was sum-normalized by the total emission statistics in \cite{GUTTIKUNDA2013101}
\begin{equation}
    E[i,j,t] = \left \{\frac{Q_bT_b}{\sum_{i,j,t}Q_b}, \frac{Q_iT_i}{\sum_{i,j,t}Q_i},\frac{Q_pT_p}{\sum_{i,j,t}Q_p},\frac{Q_dT_d}{\sum_{i,j,t}Q_d},\frac{Q_vT_v}{\sum_{i,j,t}Q_v}\right \}
\end{equation}
where, $T_b,T_i,T_p,T_d,T_v$ are corresponding total emission statistics.

\subsection{Gaussian Plume Dispersion Model} 

\begin{table}[t]
    \centering
    \begin{tabular}{c|c|c}
        \textbf{Parameter} & \textbf{Value Range} & \textbf{Sources} \\
        \hline
        Effective Stack Height for Industries ($H_i$) & 30-60 meters & \cite{industry} \\
        Effective Stack Height for Brick Kilns ($H_b$) & 22-60 meters & \cite{brick_kilns}\\
        Effective Stack Height for Power Plants ($H_p$) & 200-400 meters & \cite{power_plant}\\
        Effective Stack Height for Domestic Emissions ($H_d$) & 0-20 meters & \cite{domestic}\\
        Effective Stack Height for Vehicular Emissions ($H_v$) & 0-3 meters & \cite{gp_lecture}\\
        Atmospheric Stability ($\rho$) & A, B, C, D, E, F & \citep{sigma,DO_Martin}\\
        \hline
    \end{tabular}
    \caption{Summary of model parameters, considered ranges, and sources informing the ranges.}
    \label{tab:params}
\end{table}

\noindent\textbf{Model Description:} For modeling atmospheric dispersion, we use the Gaussian-Plume dispersion model for point sources \citep{gaussian_plume,gp_lecture}. The model is the solution of the convection-diffusion differential equation as described by
\begin{equation}
    \frac{\partial C}{\partial t} = -\frac{\partial (CU)}{\partial x} + \frac{\partial}{\partial y} \left(\frac{\partial (D_yC)}{\partial y}\right) + \frac{\partial}{\partial z}\left(\frac{\partial (D_zC)}{\partial z}\right)
\end{equation}
under steady-state state conditions where $\frac{\partial C}{\partial t} = 0$. The solution obtained is expressed as
\begin{equation}
\label{eqn:gp}
    C(x,y,z) = \frac{Q}{2\pi U \sigma_y \sigma_z} e^{\frac{-y^2}{2\sigma_y^2}}\left (e^{\frac{-(z-H)^2}{2\sigma_z^2}} + e^{\frac{-(z+H)^2}{2\sigma_z^2}}\right );
    \sigma_z = cx^d+f, \text{ }\sigma_y = ax^b
\end{equation}

Equation \ref{eqn:gp} describes the formula for computing the steady-state concentration ($C(x,y,z)$) at a point (x,y,z) produced by a source placed at the origin. It is assumed that the wind travels in the positive x-direction. Q refers to source intensity, U refers to wind speed, H refers to effective stack height (height of chimney or exhaust pipe + height smoke plume rises before dispersion), and $\sigma_y$ and $\sigma_z$ are lateral and vertical dispersion coefficients that depend on the distance in direction of the wind as shown in equation \ref{eqn:gp}. The coefficients of the equation are determined by atmospheric stability. The Gaussian-Plume dispersion model is a standard atmospheric dispersion model that is simple but accurate for short distances (<15 km). 
Another important feature of the model is that it is additive in nature, so line sources and area sources of pollution can be represented as a collection of point sources. \\

\noindent\textbf{Modeling Setup:} We model the pollution data using the Gaussian-plume model by assuming multiple point sources coinciding at the center of every grid cell, corresponding to brick kilns, industries, power plants, domestic emissions, and traffic emissions. Thus, the contributions for a cell are computed as
\begin{equation}
    \label{eq:gpdm}
    C_{i,j,t}(x,y,z) = \frac{e^{\frac{-y^2}{2\sigma_y^2}}}{2\pi U(t) \sigma_y \sigma_z}\left (\sum_s E[i,j,t][Q_s]\left (e^{\frac{-(z-H_s)^2}{2\sigma_z^2}} + e^{\frac{-(z+H_s)^2}{2\sigma_z^2}}\right )\right )
\end{equation}
where $E$ is our emissions inventory, and $s$ is the type of source. For the purpose of modeling, we considered sources from a region of 15x15 cells (approximately 15km x 15km) around the hotspot, referred to as source boundary $B$. Thus,
\begin{equation}
    C_{sensor,t} = \sum_{B}C_{i,j,t}(x,y,z)
\end{equation}
where $C_{sensor,t}$ gives us the computed value at the sensor at time t.

For wind speed, we used data at $1^o$x$1^o$ granularity at 10 meters height from ground level from the NCEP FNL Operational Model Global Tropospheric Analyses, continuing from July 1999 dataset \cite{cisl_rda_ds083.2}. The wind speeds for the six nearest data points were averaged to obtain the wind speed in the region of interest. It was further interpolated to match the temporal resolution of our measurement data. The ranges for the parameters of the model were taken from relevant literature \citep{brick_kilns,power_plant,industry,domestic,gp_lecture_17,sensor_height}, and further approximations were made to simplify the model to decrease the computation requirement. \\

\noindent\textbf{Model Approximation and Acceleration:} In Table \ref{tab:params}, we present the range of values for the different parameters required in the Gaussian Plume Dispersion model as expressed in equation \ref{eq:gpdm}.
The effective stack heights for industries, brick kilns, power plants, domestic emissions, and traffic were bound between 30-60, 22-60, 200-400, 0-20, and 0-3 meters based on \citep{brick_kilns,power_plant,industry,domestic,gp_lecture_17}. The value of z was kept at 5 meters, which is within the recommended bounds of 3-10 meters based on \cite{sensor_height}. The value of $\sigma_y$ and $\sigma_z$ was derived from the choice of atmospheric stability ($\rho$), which was bound to the six cases between [A, B, C, D, E, F] documented in \citep{sigma,DO_Martin}. Now, we consider some simplifying approximations of the above expression.
Inputting the corresponding distances between cells in our grid, for atmospheric stability choice A, we find that $\sigma_y$ takes values close to 200 for a downwind distance of 1 km. The exponential term with y in the equation, for the same value of y (1 km) goes to zero. Thus, we can ascertain that any sources that do not lie in cells in the upwind direction of wind from the sensor cell do not contribute any value to the sensor concentration.
On the other hand, the value of $\sigma_z$ at the neighboring cell for the same assumptions will be close to 400. For all sources except power plants, the effective stack height $H << 400$, making the terms 
$$e^{\frac{-(z-H_s)^2}{2\sigma_z^2}} + e^{\frac{-(z+H_s)^2}{2\sigma_z^2}} = 2$$
For power plants, there is a somewhat smaller contribution from the exponential terms only for adjoining cells, which we assume to be 2 for simplicity. Under this simplification, and substitution of the expressions of $\sigma_y$ and $\sigma_z$, we can rewrite the expression as
\begin{equation}
\label{eqn:simple_gpdm}
    C_{i,j,t}(x,y) = \frac{C}{U(t)*x^3}\sum_s E[i,j,t][Q_s]*\delta_{y,0}
\end{equation}
where $\delta_{y,0}$ is the Kronecker Delta function, which is 1 only if y=0. This two-dimensional simplified formulation was defined and computed on the grid of the study area. These simplifications simplify the amount of computations that we need to perform, accelerating the model's implementation.

It should be noted that the cell positions $i,j$ and the distances $x,y$ are measured in different coordinate systems, where the wind-speed is always assumed to be in the x-direction. In the model's implementation, we first computed the wind-velocity vector in the $i,j$ coordinate system. Followed by this, we computed 2-dimensional GPDM\_filters of size (2*source\_radius+1) x (2*source\_radius+1) as a function of wind velocity vectors, where the source\_radius is the distance up to which a source is assumed to have a contribution at any given location. The GPDM\_filter is mathematically equivalent to the concentration contribution of a unit source field of size (2*source\_radius+1) x (2*source\_radius+1) square centered at the location of interest. The computed GPDM\_filters were stacked together for all timesteps, and the entire model was expressed as a convolution operation of the stacked GPDM\_filters on the emissions inventory matrix. The reason we could express the model as such was that the wind-velocity was assumed to be constant over the entire field, making the source contributions at a destination translationally invariant. Thus, we converted the simplified model in equation \ref{eqn:simple_gpdm} into a convolutional operation, further accelerating of the implementation by running it on GPU, and allowing us to use backpropagation to tune the parameters in the model.

\subsection{Evaluation}
\label{sec:explaining}

\begin{figure}
    \centering
    \begin{subfigure}{0.35\columnwidth}
        \includegraphics[width=\columnwidth]{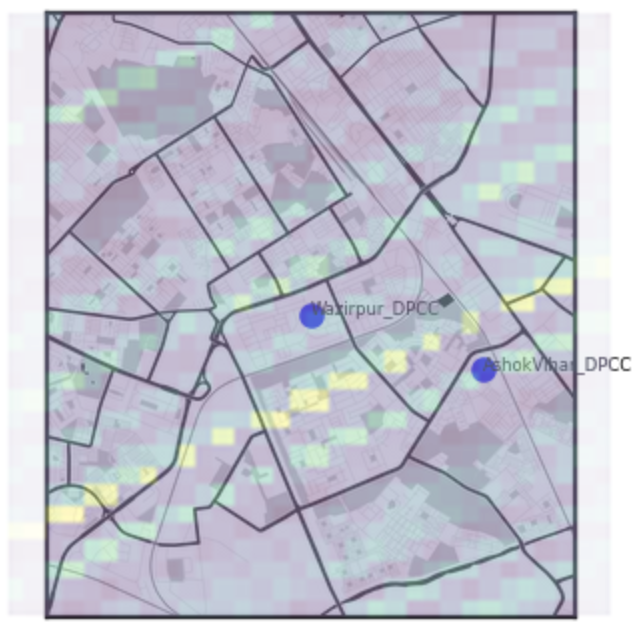}
        \caption{}
        \label{fig:explained}
    \end{subfigure}
    ~
    \begin{subfigure}{0.33\columnwidth}
        \includegraphics[width=\columnwidth]{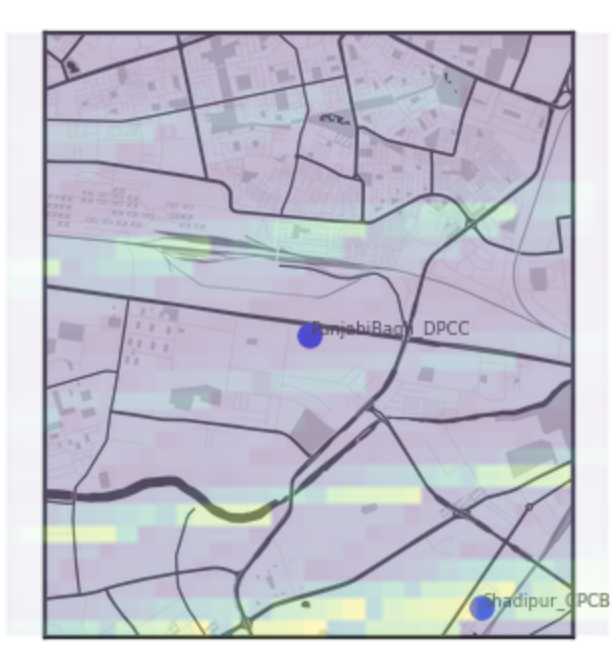}
        \caption{}
        \label{fig:unexplained}
    \end{subfigure}
    \caption{The figure shows the heatmaps derived from dispersion modeling overlayed on top of the geographical maps for two transient hotspots. In Figure \ref{fig:explained}, the heatmap explains the observed measurements well but in Figure \ref{fig:unexplained}, it does not.}
    \label{fig:explaining}
\end{figure}

On the computed error metrics, the dispersion model performs poorly with MAPE of 104.52\% $\pm$ 6.00\%. This error number might seem to be relatively large, it is a well-known fact that dispersion models generally contain large errors due to the uncertainty in the input data taken by these models (wind and emissions inventory data). \cite{GUTTIKUNDA2013101} and \cite{10.1007/978-94-007-2540-9_11} both note very high uncertainties in the input data used in dispersion modeling ranging from 20-30\%. In \cite{GUTTIKUNDA2013101}, even on a monthly averaged scale, the dispersion modeling error at individual monitoring stations can be estimated to range from 20\% to 200\% (based on eyeballing the results bar chart). In fact, in the traditional pollution modeling community, and in the larger scientific community, it is well-known that mechanistic models generally have significant differences from real-world measurements, especially in complex systems, because of several simplifying assumptions and unaccounted externalities in the system. The purpose of mechanistic modeling is to develop a better understanding the formation and causes of hotspots. Since our mechanistic model is based on steady-state assumption, we need to look at hotspots at much smaller time scales, and define them on hourly snapshots instead of monthly averages. To do this, we define a new type of hotspot - \textit{transient hotspots}, which are areas of relatively high pollution value at any hourly snapshot. We employ the algorithms presented in Appendix \ref{sec:transient} to capture these hotspots. Then, the mechanistic model was used to investigate these hotspots.\\

\noindent\textbf{Transient Hotspots Formation and Causes:} Using the Gaussian-Plume dispersion model applied to our approximate emissions inventory after assuming point sources at the center of $0.01^ox0.01^o$ grid cells, we are able to develop heatmaps of the small regions around the transient hotspots. These heatmaps show the distribution of the calculated PM$_{2.5}$ values around the transient hotspot regions. Figure \ref{fig:explaining} shows these heatmaps overlayed on top of the geographic maps of regions near two transient hotspots. It should be noted that the direction of the wind is easily noticed in the heatmaps. This is because the dispersion model used mainly disperses pollutants in the direction of the wind. In Figure \ref{fig:explained}, we see the distribution of one particular transient hotspot that occurs around the Ashok Vihar and Wazipur localities in New Delhi, which are part of high-pollution regions in the city. According to our dispersion model, this transient hotspot is mainly linked equally to nearby industries and domestic emissions, while the contributions of brick kilns, power plants, and traffic are almost zero. We consider the model to have explained this transient hotspot because the calculated pollution values are in agreement with the measured values at the sensors. By this, we mean that the calculated values at the locations of sensors broadly follow the same trends, and not that the predictions are accurate. On the other hand, we also find transient hotspots like those shown in Figure \ref{fig:unexplained} where the computed values are not in agreement with the measured values. In total, \textbf{our model can explain 65\% of the observed transient hotspots}. It should be noted that we have only accounted for 80\% of sources and have used deviated wind speed data (with a temporal deviation of 30 minutes and a spatial deviation in the order of 10 km). For improving the explanations for the remaining 35\% transient hotspots, we would need more accurate and granular wind speed data. Unfortunately, such data is extremely difficult to obtain. We would also need a detailed emissions inventory, but we did not find such an open-source emissions inventory.

An interesting observation from individual transient hotspot modeling is that the contribution of power plants to the formation of transient hotspots is almost zero. This is because power plants boast a stack height of >200 meters, which disperses their pollution in the upper atmosphere. They are macro contributors and the released concentration affects the city as a whole rather than small areas. Another important insight is that localized transient hotspots are largely connected to localized sources and the faraway sources mostly act as macro contributors to the entire region increasing the baseline pollution levels. Therefore, transient hotspots in south Delhi are hardly contributed to by brick kilns mostly distributed in the north and southeast. We also noted that the wind speed and direction can alter the source profile of transient hotspots drastically even if they are created in the same geographical area, indicating their dynamic nature based on meteorological complexities.

\section{Discussion}
\label{sec:discussion}

\noindent\textbf{Hotspot-Centric Mitigation:} Air pollution is one of the key challenges in sustainable urban development. Most city governments are budget-constrained, and hence their machinery for air pollution monitoring and regulation is coarse-grained. As shown in the results, the sensor network in New Delhi, being the densest in any Indian city, is still sparse for effective monitoring of air pollution hotspots. Similarly, most city governments tend to focus on more heavily polluted areas of the city, the so-called hotspots. For the context of New Delhi, one can find overwhelming direct evidence of hotspot-centric mitigation strategy of the city \cite{dpcc_hotspot,PriyangiAgarwal_2020,Shrangi_Pillai_2019,toi2024,doedelhi,it2023}. In the official policy document \cite{dpcc_hotspot,doedelhi}, the mitigation steps mentioned include identification and closure of unauthorized pollution sources (biomass combustion, construction dust), decongestion of road traffic for decreasing vehicular exhaust, road sweeping and water sprinkling for reducing dust, and development of green areas. But, these steps are to be applied only to the 13 identified hotspots. According to \cite{it2023}, the government has also deployed smog guns in these hotspots to reduce the dust levels. The government has, in fact, formed separate teams for each hotspot to address the local pollution sources. Just recently, the government announced another 8 hotspots in addition to the previous 13 \cite{ht2024}. The new identified hotspots are Shadipur, ITO, Mandir Marg, Nehru Nagar, Patparganj, Sonia Vihar, Major Dhyan Chand National Stadium, and Moti Bagh. In this paper, we have hypothesized and then shown that public monitoring networks do not have sufficient resolution to detect hotspots effectively. Thus, the actual problem of hotspots extends beyond what is visible with the government machinery, as shown by our deployed low-cost sensor network. This further increases the importance of our presented algorithms for predicting these hidden hotspots on top of a sparse sensor network using appropriate modeling strategies of spatial interpolation and dispersion modeling. Our algorithm solves the next pertinent problem for the city, which is finding the hidden hotspots. As an added benefit, the algorithm also improves the robustness of the sensor network against random failure. Having framed our entire work in context of hotspots, we believe that our predicted hotspots can be easily adopted in the existing government policy, leading to direct mitigation impact.\\

\noindent\textbf{Health Impacts of Air Pollution:} Pollution is a serious health concern in the urban centers of developing countries. High concentrations of air pollutants such as coarse (PM$_{10}$) and fine (PM$_{2.5}$) particulate matter, nitrogen dioxide (NO$_{2}$), and sulfur dioxide (SO$_{2}$) have directly triggered numerous health problems including lung cancer, chronic obstructive pulmonary disease (COPD), cerebrovascular disease, and higher mortality rates \citep{Gauderman2002,Cropper2010,Sunyer2015,Braun-Fahrlaender1997,world2021air}. Children and elderly populations are the most vulnerable to high pollution \citep{Cropper2010,Braun-Fahrlaender1997,world2021air,Lei2019ShortTerm}. Pollution also has a negative impact on labor productivity and human capital accumulation \citep{Zivin2018}.  The Air Quality Life Index, derived from prior research \citep{chen2013evidence,ebenstein2017new,greenstone2023air}, suggests that a $10 \pmunit$ increase in PM$_{10}$ reduces life expectancy by 0.64 years, and a $10 \pmunit$ increase in PM$_{2.5}$ reduces life expectancy by 0.98 years permanently. In the heavily polluted city of New Delhi, where the PM$_{2.5}$ levels go beyond the recommended level ($5 \pmunit$) \cite{world2021air} by $100 \pmunit$ for a majority of the locations, the expected loss of life expectancy is close to 10 years. In a study conducted in the city, it was shown that an increase of  $25 \pmunit$ in daily PM$_{2.5}$ exposure was associated with a 0.8\% (95\% CI: 0.3, 1.3\%) increase in daily non-accidental mortality and a 1.5\% (95\% CI: 0.8, 2.2\%) increase in mortality among those aged 60 or over \cite{krishna2021air}. In another study, a total of 17,000 premature deaths were attributed to PM$_{2.5}$ and ozone in New Delhi in the year 2015 \cite{miller2019health}. Based on an older world bank report \cite{cropper1997health}, the association between PM$_{2.5}$ levels and mortality in New Delhi, while lower than cities in the U.S.A, results in higher loss of life-years due to affecting younger populations predominately. Another study in the context of New Delhi showed positive correlation between increased pollution and incidence of COPD \cite{agarwal2006assessing}. The 2013 study by Rizwan et. al. \cite{rizwan2013air} provides further details of New Delhi's air pollution's impact on health and points to other studies done for the same study area. In recent years, Dutta et. al. \cite{dutta2022air} have shown robust association between New Delhi's air pollution and respiratory illnesses. 

While it is easy to associate uncovering hidden hotspots to beneficial impact on population health, quantifying this impact accurately is challenging, primarily because there exist very few studies, with the notable exception of \cite{foster2011health}, that have assessed the health impacts of government mitigation policies. The last few years have seen a downward trend in the average air pollution levels in New Delhi \cite{guttikunda2023polluting}, but we could not find a more localized hotspot-based study evaluating the effects of government's mitigation steps.\\

\section{Other Related Work}
\label{sec:related}

There is a rich body of literature on the use of distributed sensor networks to gather information on air pollution and other meteorological variables in urban contexts \citep{LIU2018, Chamblisse2109249118, Liange2106478118, Ferraroe2106489118, Ludeschere1922872118}. Readers can refer to Clements et al. {\cite{clements2017low}} for a comprehensive review. One important direction in the field studies deployment of low-cost sensor networks \citep{jiao2016community, lin2015evaluation, shusterman2016berkeley, moltchanov2015feasibility, sun2016development, tsujita2005gas, gao2015distributed, 10.1145/2668332.2668346, kumar_rise_2015, kortoci_air_2022, qin_fine-grained_2022,ijcai2022}. Except for Gao et al. \citep{gao2015distributed}, who examine the performance of fine particulate sensors in Xi'an in China, most of these deployments have occurred in cities with significantly lower air pollution than New Delhi. Also, prior deployments have been done on smaller sizes and time scales compared to ours, with 28 sensors functional over 2 years. As we have seen in section \ref{sec:proof}, some hotspot definitions accepted in the community and government circles are based on averaging over large time windows. Challenging existing public monitoring systems requires a comparable deployment time and network size scale.

Coming to air pollution hotspots, we find that most prior works use the terminology with their independent understanding of it \cite{querol2006atmospheric, goyal2019assessment, bathory2019hotspot, pant2015characterization}. In 2021, Goyal et al. \citep{APH_paper} noted this and proposed formalization of the notion of air pollution hotspots by combining different such approaches. Their definition is based on the frequency of days of exceedance, the scale of exceedance, and consecutive days of exceedance and they identified hotspots as part of their longitudinal study. Given that our work shares the same study area, we will be using their definitions of hotspots for first establishing that the public monitoring network is not sufficiently dense to discover all hotspots in section \ref{sec:proof}, and then evaluating different field estimation models in section \ref{sec:methods}. Another definition that we will be considering is also based on the scale of exceedance over a threshold, but over the period of a year, as considered by the New Delhi government in their current policy \cite{dpcc_hotspot}.

\section{Conclusion and Future Work}
\label{sec:conclusion}
\noindent\textbf{Conclusion:} In this paper, we have shown the insufficiency of public sensor network in New Delhi, and have presented predictive and mechanistic modeling strategies that can be used to complement the sparse sensor network to effectively and comprehensively monitor air pollution hotspots. Our predictive models based on Space-Time Kriging uncovered hidden hotspots accurately with 95\% precision and 88\% recall with increased robustness and generalization with minimal dependence on additional data collection. Our city-scale Gaussian-Plume Dispersion based on an approximate 0.01$^o$x0.01$^o$ emissions inventory that was collated using alternative online data sources, and implemented in an approximate and accelerated setup, was able to mechanistically explain 65\% of observed transient hotspots. Based on our modeling approaches, we also presented policy recommendations for mitigating pollution hotspots. While our work is centered in New Delhi, the general takeaways are applicable to other urban environments, and a similar setup can be adapted in other environments.

\noindent\textbf{Future Work:} The identification of pollution hotspots and corresponding sources is critical for policymakers for drafting policies aimed at reducing pollution levels. Our approach, accompanied by public investment, would not only allow us to find hidden hotspots but would also provide important directions required for the proactive investigation of unknown emission sources. This is because hidden hotspots are likely to be hiding such sources, which can be inferred by mechanistic modeling. From the results presented in Section \ref{sec:recommendations}, we can see that the health and life-expectancy of at about 16.5 million people can be directly improved by this targeted approach. With one-fifth the cost of existing government sensors, the city of New Delhi can deploy a sensor network a hundred times the current size, increasing the spatial resolution by a factor of 10. At this scale, we can even infer and measure "localized hotspots". This framework could be used for evaluating the effects of policy decisions, differential exposure analysis of different sections of the population, and personal pollution exposure.

\section{Data and Code Availability}
\label{sec:availability}
The main data used in this paper was collected by the authors by deploying low-cost sensors. This data can be made available for peer-review and can be obtained privately by contacting the authors but can not be open-sourced. All further data used in the analysis and modeling are appropriately cited in the manuscript.

The anonymized code to run the analysis is available on \url{https://github.com/ankitbha/hidden_hotspots}.

\section{Author Contributions}
Ankit Bhardwaj (A.Bh), Ananth Balashankar (A.Ba), Shiva Iyer (SI), Anant Sudarshan (AS), Rohini Pande (RP) and Lakshminarayanan Subramanian (LS) collectively helped in problem conceptualization and formalization. A.Bh, A.Ba and LS contributed to the mathematical modeling of the problem, analysis of results, and the writing of the paper. The data collection and sensor network deployment was made possible through a joint collaboration between Yale University, New York University and Kai Air Monitoring Pvt Ltd. with contributions from SI and Nita Soans and the Inclusion Economics India Center (formerly EPoD India) under the guidance of RP. AS, RP and LS provided guidance and mentorship on project execution, analysis of results and writing of the paper.

\section{Acknowledgements}
Ankit Bhardwaj and Lakshminarayanan Subramanian were supported by the NSF Grant (Award Number 2335773) titled "EAGER: Scalable Climate Modeling using Message-Passing Recurrent Neural Networks". Lakshminarayanan Subramanian was also funded in part by the NSF Grant (award number OAC-2004572) titled  "A Data-informed Framework for the Representation of Sub-grid Scale Gravity Waves to Improve Climate Prediction".

\section{Disclosures}
Prof. Subramanian declares the following: Prof. Subramanian is a co-founder of Entrupy Inc, Velai Inc, Velai Health Analytics Pvt. Ltd., and Gaius Networks Inc and has served as a consultant for the World Bank and the Governance Lab. All other authors declare no competing interests.

\newpage

\bibliographystyle{abbrv}
{\footnotesize
\bibliography{main}}

\begin{thebibliography}{100}

\bibitem{power_plant}
Coal-based thermal power plants, discounting the effects of sulphur dioxide emissions on air quality.
\newblock \url{https://www.cseindia.org/content/downloadreports/10807}.

\bibitem{cpcb_portal}
Cpcb data portal.
\newblock \url{https://app.cpcbccr.com/ccr/\#/caaqm-dashboard-all/caaqm-landing/caaqm-comparison-data}.

\bibitem{geoiq}
Geoiq.
\newblock \url{https://geoiq.ai/in}.

\bibitem{sigma}
Plume dispersion coefficients, sigmay sy and sigmaz sz.
\newblock \url{http://courses.washington.edu/cee490/DISPCOEF4WP.htm}.

\bibitem{purpleair}
Purple air.
\newblock \url{https://www2.purpleair.com/}.

\bibitem{brick_kilns}
Brick kilns in india - j. s. kamyotra director, central pollution control board.
\newblock Technical report, India, Feb 2016.
\newblock Pages: 57.

\bibitem{domestic}
Residential building heights might be raised, according to ndmc.
\newblock \url{https://www.propertypistol.com/blog/residential-building-heights-might-be-raised-according-to-ndmc/}, 2022.

\bibitem{PriyangiAgarwal_2020}
P.~Agarwal.
\newblock Night patrol at 13 pollution hotspots in delhi.
\newblock \url{https://timesofindia.indiatimes.com/city/delhi/night-patrol-at-13-pollution-hotspots/articleshow/78264271.cms}, Sep 2020.

\bibitem{agarwal2006assessing}
R.~Agarwal, G.~Jayaraman, S.~Anand, and P.~Marimuthu.
\newblock Assessing respiratory morbidity through pollution status and meteorological conditions for delhi.
\newblock {\em Environmental Monitoring and Assessment}, 114:489--504, 2006.

\bibitem{industry}
S.~ALAM.
\newblock Noida: Industries get 2-hour daily window to use gensets.
\newblock \url{https://timesofindia.indiatimes.com/city/noida/noida-industries-get-2-hour-daily-window-to-use-gensets/articleshow/94234824.cms}, 2022.

\bibitem{Alpert2012}
P.~Alpert, O.~Shvainshtein, and P.~Kishcha.
\newblock {AOD} trends over megacities based on space monitoring using {MODIS} and {MISR}.
\newblock {\em American Journal of Climate Change}, 01(03):117--131, 2012.

\bibitem{Apte2017High}
J.~S. Apte, K.~P. Messier, S.~Gani, M.~Brauer, T.~W. Kirchstetter, M.~M. Lunden, J.~D. Marshall, C.~J. Portier, R.~C. Vermeulen, and S.~P. Hamburg.
\newblock High-{Resolution} {Air} {Pollution} {Mapping} with {Google} {Street} {View} {Cars}: {Exploiting} {Big} {Data}.
\newblock {\em Environmental Science \& Technology}, 51(12):6999--7008, June 2017.
\newblock Publisher: American Chemical Society.

\bibitem{bathory2019hotspot}
C.~B{\'a}thory and A.~B. Palotas.
\newblock Hotspot identification with portable low-cost particulate matter sensor.
\newblock {\em International Journal of Energy Water Food Nexus}, 1(1):13--17, 2019.

\bibitem{ijcai2022}
A.~Bhardwaj, S.~Iyer, Y.~Jalan, and L.~Subramanian.
\newblock Learning pollution maps from mobile phone images.
\newblock In L.~D. Raedt, editor, {\em Proceedings of the Thirty-First International Joint Conference on Artificial Intelligence, {IJCAI-22}}, pages 5024--5030. International Joint Conferences on Artificial Intelligence Organization, 7 2022.
\newblock AI for Good.

\bibitem{BHARDWAJ2023100105}
A.~Bhardwaj, S.~R. Iyer, S.~Ramesh, J.~White, and L.~Subramanian.
\newblock Understanding sudden traffic jams: From emergence to impact.
\newblock {\em Development Engineering}, 8:100105, 2023.

\bibitem{Bikkina2019Air}
S.~Bikkina, A.~Andersson, E.~N. Kirillova, H.~Holmstrand, S.~Tiwari, A.~K. Srivastava, D.~S. Bisht, and O.~Gustafsson.
\newblock Air quality in megacity {Delhi} affected by countryside biomass burning.
\newblock {\em Nature Sustainability}, 2(3):200--205, Mar. 2019.
\newblock Number: 3.

\bibitem{Braun-Fahrlaender1997}
C.~Braun-Fahrl{\"a}nder, J.~C. Vuille, F.~H. Sennhauser, U.~Neu, T.~K{\"u}nzle, L.~Grize, M.~Gassner, C.~Minder, C.~Schindler, H.~S. Varonier, and B.~Wüthrich.
\newblock Respiratory health and long-term exposure to air pollutants in {Swiss} schoolchildren. {SCARPOL} team. {Swiss} study on childhood allergy and respiratory symptoms with respect to air pollution, climate and pollen.
\newblock {\em American Journal of Respiratory and Critical Care Medicine}, 155(3):1042--1049, mar 1997.

\bibitem{cpcb_specs}
{Central Pollution Control Board}.
\newblock Technical specifications for continuous ambient air quality monitoring (caaqm) station (real time).
\newblock \url{https://erc.mp.gov.in/Documents/doc/Guidelines/CAAQMS_Specs_new.pdf}.
\newblock [Online; accessed 17-July-2024].

\bibitem{Chamblisse2109249118}
S.~E. Chambliss, C.~P. Pinon, K.~P. Messier, B.~LaFranchi, C.~R. Upperman, M.~M. Lunden, A.~L. Robinson, J.~D. Marshall, and J.~S. Apte.
\newblock Local- and regional-scale racial and ethnic disparities in air pollution determined by long-term mobile monitoring.
\newblock {\em Proceedings of the National Academy of Sciences}, 118(37), 2021.

\bibitem{chen2013evidence}
Y.~Chen, A.~Ebenstein, M.~Greenstone, and H.~Li.
\newblock Evidence on the impact of sustained exposure to air pollution on life expectancy from {China's Huai} river policy.
\newblock {\em Proceedings of the National Academy of Sciences}, 110(32):12936--12941, 2013.

\bibitem{10.1145/2668332.2668346}
Y.~Cheng, X.~Li, Z.~Li, S.~Jiang, Y.~Li, J.~Jia, and X.~Jiang.
\newblock Aircloud: A cloud-based air-quality monitoring system for everyone.
\newblock In {\em Proceedings of the 12th ACM Conference on Embedded Network Sensor Systems}, SenSys '14, page 251–265, New York, NY, USA, 2014. Association for Computing Machinery.

\bibitem{clements2017low}
A.~L. Clements, W.~G. Griswold, J.~E. Johnston, M.~M. Herting, J.~Thorson, A.~Collier-Oxandale, and M.~Hannigan.
\newblock Low-cost air quality monitoring tools: From research to practice (a workshop summary).
\newblock {\em Sensors}, 17(11):2478, 2017.

\bibitem{ColumbiaUniversity2018}
C.~f. I. E. S. I. N. .~C. Columbia~University.
\newblock Gridded population of the world, version 4 (gpwv4): Population density, revision 11, 20230129 2018.

\bibitem{caqm2022}
{Commission for Air Quality Management in NCR and Adjoining Areas}.
\newblock Policy to curb air pollution in the national capital region.
\newblock \url{https://caqm.nic.in/WriteReadData/LINKS/0031dcb806e-8af7-4b38-a9bc-65b91f2704cd.pdf}, 2022.
\newblock [Online; accessed 17-July-2024].

\bibitem{cressie2011}
N.~Cressie and C.~K. Wikle.
\newblock {\em Statistics for spatio-temporal data}.
\newblock John Wiley \& Sons, Hoboken, {NJ}, 2011.

\bibitem{Cropper2010}
M.~Cropper.
\newblock What are the health effects of air pollution in {China}?
\newblock In {\em Is Economic Growth Sustainable?}, pages 10--46. Palgrave Macmillan {UK}, 2010.

\bibitem{cropper1997health}
M.~Cropper, N.~B. Simon, A.~Alberini, and P.~Sharma.
\newblock The health effects of air pollution in delhi, india.
\newblock {\em India (December 1997)}, 1997.

\bibitem{Cusworth2018Quantifying}
D.~H. Cusworth, L.~J. Mickley, M.~P. Sulprizio, T.~Liu, M.~E. Marlier, R.~S. DeFries, S.~K. Guttikunda, and P.~Gupta.
\newblock Quantifying the influence of agricultural fires in northwest india on urban air pollution in delhi, india.
\newblock {\em Environmental Research Letters}, 13(4):044018, mar 2018.

\bibitem{dpcc_hotspot}
{D}elhi {P}ollution~{C}ontrol {C}ommittee.
\newblock Pollution control at source.
\newblock \url{https://www.dpcc.delhigovt.nic.in/uploads/sitedata/hotspot.pdf}.

\bibitem{doedelhi}
{Department of Environment, Government of NCT of Delhi}.
\newblock Air pollution control at pollution hotspots.
\newblock \url{https://environment.delhi.gov.in/environment/air-pollution-control-pollution-hotspots}.
\newblock [Online; accessed 17-July-2024].

\bibitem{Dinh2022}
T.-V. Dinh, B.-G. Park, S.-W. Lee, J.-H. Park, D.-H. Baek, I.-Y. Choi, Y.-B. Seo, and J.-C. Kim.
\newblock Comparison of pm2.5 monitoring data using light scattering and beta attenuation methods: A case study in seoul metro subway.
\newblock {\em Asian Journal of Atmospheric Environment}, 16(4):2022116, Dec 2022.

\bibitem{downtoearth2014}
{Down to earth}.
\newblock All about effective air quality monitoring.
\newblock \url{https://www.downtoearth.org.in/environment/all-about-effective-air-quality-monitoring-46494}, 2014.
\newblock [Online; accessed 17-July-2024].

\bibitem{dutta2022air}
A.~Dutta and W.~Jinsart.
\newblock Air pollution in delhi, india: It’s status and association with respiratory diseases.
\newblock {\em Plos one}, 17(9):e0274444, 2022.

\bibitem{ebenstein2017new}
A.~Ebenstein, M.~Fan, M.~Greenstone, G.~He, and M.~Zhou.
\newblock New evidence on the impact of sustained exposure to air pollution on life expectancy from {China's Huai} river policy.
\newblock {\em Proceedings of the National Academy of Sciences}, 114(39):10384--10389, 2017.

\bibitem{chem_transport}
U.~Emissions.
\newblock Air pollution monitoring 101.
\newblock \url{https://urbanemissions.info/blog-pieces/air-monitoring-101/}.

\bibitem{Ferraroe2106489118}
P.~J. Ferraro and A.~Agrawal.
\newblock Synthesizing evidence in sustainability science through harmonized experiments: Community monitoring in common pool resources.
\newblock {\em Proceedings of the National Academy of Sciences}, 118(29), 2021.

\bibitem{foster2011health}
A.~Foster and N.~Kumar.
\newblock Health effects of air quality regulations in delhi, india.
\newblock {\em Atmospheric Environment}, 45(9):1675--1683, 2011.

\bibitem{gao2015distributed}
M.~Gao, J.~Cao, and E.~Seto.
\newblock A distributed network of low-cost continuous reading sensors to measure spatiotemporal variations of pm2. 5 in xi'an, china.
\newblock {\em Environmental pollution}, 199:56--65, 2015.

\bibitem{Gauderman2002}
W.~J. Gauderman, G.~F. Gilliland, H.~Vora, E.~Avol, D.~Stram, R.~McConnell, D.~Thomas, F.~Lurmann, H.~G. Margolis, E.~B. Rappaport, K.~Berhane, and J.~M. Peters.
\newblock Association between air pollution and lung function growth in southern {California} children.
\newblock {\em American Journal of Respiratory and Critical Care Medicine}, 166(1):76--84, jul 2002.
\newblock PMID: 12091175.

\bibitem{ge2021multi}
L.~Ge, K.~Wu, Y.~Zeng, F.~Chang, Y.~Wang, and S.~Li.
\newblock Multi-scale spatiotemporal graph convolution network for air quality prediction.
\newblock {\em Applied Intelligence}, 51:3491--3505, 2021.

\bibitem{APH_paper}
P.~Goyal, S.~Gulia, and S.~Goyal.
\newblock Identification of air pollution hotspots in urban areas - an innovative approach using monitored concentrations data.
\newblock {\em Science of The Total Environment}, 798:149143, 2021.

\bibitem{goyal2019assessment}
P.~Goyal, S.~Gulia, S.~K. Goyal, and R.~Kumar.
\newblock Assessment of the effectiveness of policy interventions for air quality control regions in delhi city.
\newblock {\em Environmental Science and Pollution Research}, 26:30967--30979, 2019.

\bibitem{graler2012spatio}
B.~Gr{\"a}ler, M.~Rehr, L.~Gerharz, and E.~Pebesma.
\newblock Spatio-temporal analysis and interpolation of pm10 measurements in europe for 2009.
\newblock {\em ETC/ACM Technical Paper}, 8:1--29, 2012.

\bibitem{greenstone2023air}
M.~Greenstone and C.~Hasenkopf.
\newblock Air quality life index.
\newblock \url{https://aqli.epic.uchicago.edu/}, 2023.
\newblock Accessed: 2023-11-20.

\bibitem{gp_lecture}
B.~R. Gurjar.
\newblock Lecture 16: Gaussian dispersion model for point source.
\newblock Air Pollution and Control, IIT Roorkee July 2018 Youtube Channel, 2018.

\bibitem{gp_lecture_17}
B.~R. Gurjar.
\newblock Lecture 17: Gaussian dispersion model for line source and area source.
\newblock Air Pollution and Control, IIT Roorkee July 2018 Youtube Channel, 2018.

\bibitem{src_apportionment_primer}
S.~Guttikonda.
\newblock A primer on source apportionment.
\newblock Urbanemissions.info, 2011.

\bibitem{sensor_height}
S.~Guttikonda.
\newblock Air pollution monitoring 101.
\newblock Urbanemissions.info, 2018.

\bibitem{GUTTIKUNDA2013101}
S.~K. Guttikunda and G.~Calori.
\newblock A gis based emissions inventory at 1 km × 1 km spatial resolution for air pollution analysis in delhi, india.
\newblock {\em Atmospheric Environment}, 67:101--111, 2013.

\bibitem{guttikunda2023polluting}
S.~K. Guttikunda, S.~K. Dammalapati, G.~Pradhan, B.~Krishna, H.~T. Jethva, and P.~Jawahar.
\newblock What is polluting delhi’s air? a review from 1990 to 2022.
\newblock {\em Sustainability}, 15(5):4209, 2023.

\bibitem{hindustantimes2017}
{Hindustan Times}.
\newblock Delhi to get 10 more pollution monitoring stations by next winter.
\newblock \url{https://www.hindustantimes.com/delhi-news/delhi-to-get-10-more-pollution-monitoring-stations-by-next-winter/story-Ipal3yRk71AMNrPCtW86iL.html}, 2017.
\newblock [Online; accessed 17-July-2024].

\bibitem{hindustantimes2020}
{Hindustan Times}.
\newblock Mumbai gets sensor-based monitors as low cost air monitoring feasibility study begins.
\newblock \url{https://www.hindustantimes.com/india-news/mumbai-gets-sensor-based-monitors-as-low-cost-air-monitoring-feasibility-study-begins/story-LjSYWRqLsC7Q27B6WQDrEN.html}, 2020.
\newblock [Online; accessed 17-July-2024].

\bibitem{ht2024}
{Hindustan Times}.
\newblock Govt marks 8 new pollution hot spots in delhi.
\newblock \url{https://www.hindustantimes.com/cities/delhi-news/govt-marks-8-new-pollution-hot-spots-in-delhi-101698085833613.html}, 2023.
\newblock [Online; accessed 17-July-2024].

\bibitem{hofman2021spatiotemporal}
J.~Hofman, T.~H. Do, X.~Qin, E.~Rodrigo, M.~E. Nikolaou, W.~Philips, N.~Deligiannis, and V.~P.~L. Manna.
\newblock Spatiotemporal air quality inference of low-cost sensor data; application on a cycling monitoring network.
\newblock In {\em Pattern Recognition. ICPR International Workshops and Challenges: Virtual Event, January 10--15, 2021, Proceedings, Part VI}, pages 139--147. Springer, 2021.

\bibitem{it2023}
{India Today}.
\newblock Delhi government forms teams to tackle air pollution at 13 hotspots.
\newblock \url{https://www.indiatoday.in/india/story/delhi-government-forms-teams-to-tackle-air-pollution-at-13-hotspots-2447777-2023-10-12}, 2023.
\newblock [Online; accessed 17-July-2024].

\bibitem{our_npj}
S.~R. Iyer, A.~Balashankar, W.~H. Aeberhard, S.~Bhattacharyya, G.~Rusconi, L.~Jose, N.~Soans, A.~Sudarshan, R.~Pande, and L.~Subramanian.
\newblock Modeling fine-grained spatio-temporal pollution maps with low-cost sensors.
\newblock {\em npj Climate and Atmospheric Science}, 5(1):76, Oct 2022.

\bibitem{jiao2016community}
W.~Jiao, G.~Hagler, R.~Williams, R.~Sharpe, R.~Brown, D.~Garver, R.~Judge, M.~Caudill, J.~Rickard, M.~Davis, et~al.
\newblock Community air sensor network (cairsense) project: evaluation of low-cost sensor performance in a suburban environment in the southeastern united states.
\newblock {\em Atmospheric Measurement Techniques}, 9(11):5281, 2016.

\bibitem{kortoci_air_2022}
P.~Kortoçi, N.~H. Motlagh, M.~A. Zaidan, P.~L. Fung, S.~Varjonen, A.~Rebeiro-Hargrave, J.~V. Niemi, P.~Nurmi, T.~Hussein, T.~Petäjä, M.~Kulmala, and S.~Tarkoma.
\newblock Air pollution exposure monitoring using portable low-cost air quality sensors.
\newblock {\em Smart Health}, 23:100241, Mar. 2022.

\bibitem{krishna2021air}
B.~Krishna.
\newblock {\em Air Pollution and Health in Delhi, India: Health Effects of Short-Term Exposures and Policy Implications}.
\newblock PhD thesis, Harvard University, 2021.

\bibitem{kulshreshta2009}
P.~Kulshreshtha, M.~Khare, and P.~Seetharaman.
\newblock Indoor air quality assessment in and around urban slums of delhi city, india.
\newblock {\em Indoor air}, 18:488--98, 01 2009.

\bibitem{kumar_rise_2015}
P.~Kumar, L.~Morawska, C.~Martani, G.~Biskos, M.~Neophytou, S.~Di~Sabatino, M.~Bell, L.~Norford, and R.~Britter.
\newblock The rise of low-cost sensing for managing air pollution in cities.
\newblock {\em Environment International}, 75:199--205, Feb. 2015.

\bibitem{Lei2019ShortTerm}
R.~Lei, F.~Zhu, H.~Cheng, J.~Liu, C.~Shen, C.~Zhang, Y.~Xu, C.~Xiao, X.~Li, J.~Zhang, R.~Ding, and J.~Cao.
\newblock Short-term effect of {PM2}.5/{O3} on non-accidental and respiratory deaths in highly polluted area of {China}.
\newblock {\em Atmospheric Pollution Research}, 10(5):1412--1419, Sept. 2019.

\bibitem{Liange2106478118}
Y.~Liang, D.~Sengupta, M.~J. Campmier, D.~M. Lunderberg, J.~S. Apte, and A.~H. Goldstein.
\newblock Wildfire smoke impacts on indoor air quality assessed using crowdsourced data in california.
\newblock {\em Proceedings of the National Academy of Sciences}, 118(36), 2021.

\bibitem{lin2015evaluation}
C.~Lin, J.~Gillespie, M.~Schuder, W.~Duberstein, I.~Beverland, and M.~Heal.
\newblock Evaluation and calibration of aeroqual series 500 portable gas sensors for accurate measurement of ambient ozone and nitrogen dioxide.
\newblock {\em Atmospheric Environment}, 100:111--116, 2015.

\bibitem{LIU2018}
T.~Liu, M.~E. Marlier, R.~S. DeFries, D.~M. Westervelt, K.~R. Xia, A.~M. Fiore, L.~J. Mickley, D.~H. Cusworth, and G.~Milly.
\newblock Seasonal impact of regional outdoor biomass burning on air pollution in three indian cities: Delhi, bengaluru, and pune.
\newblock {\em Atmospheric Environment}, 172:83--92, 2018.

\bibitem{Ludeschere1922872118}
J.~Ludescher, M.~Martin, N.~Boers, A.~Bunde, C.~Ciemer, J.~Fan, S.~Havlin, M.~Kretschmer, J.~Kurths, J.~Runge, V.~Stolbova, E.~Surovyatkina, and H.~J. Schellnhuber.
\newblock Network-based forecasting of climate phenomena.
\newblock {\em Proceedings of the National Academy of Sciences}, 118(47), 2021.

\bibitem{lueker2020indoor}
J.~Lueker, R.~Bardhan, A.~Sarkar, and L.~Norford.
\newblock Indoor air quality among mumbai's resettled populations: Comparing dharavi slum to nearby rehabilitation sites.
\newblock {\em Building and Environment}, 167:106419, 2020.

\bibitem{MAHATO2020109835}
S.~Mahato and K.~G. Ghosh.
\newblock Short-term exposure to ambient air quality of the most polluted indian cities due to lockdown amid sars-cov-2.
\newblock {\em Environmental Research}, 188:109835, 2020.

\bibitem{Mardia1998}
K.~V. Mardia, C.~Goodall, E.~J. Redfern, and F.~J. Alonso.
\newblock The kriged kalman filter.
\newblock {\em Test}, 7(2):217--282, Dec 1998.

\bibitem{DO_Martin}
D.~O. Martin.
\newblock Comment on"the change of concentration standard deviations with distance".
\newblock {\em Journal of the Air Pollution Control Association}, 26(2):145--147, 1976.

\bibitem{miller2019health}
J.~Miller, A.~Bandivadekar, and B.~Sathiamoorthy.
\newblock Health impacts of air pollution from transportation sources in delhi.
\newblock 2019.

\bibitem{ncap18}
{MOEF}.
\newblock National clean air programme india, 2018.

\bibitem{moltchanov2015feasibility}
S.~Moltchanov, I.~Levy, Y.~Etzion, U.~Lerner, D.~M. Broday, and B.~Fishbain.
\newblock On the feasibility of measuring urban air pollution by wireless distributed sensor networks.
\newblock {\em Science of The Total Environment}, 502:537--547, 2015.

\bibitem{10.1007/978-94-007-2540-9_11}
A.~Namdeo, I.~Sohel, J.~Cairns, M.~Bell, M.~Khare, and S.~Nagendra.
\newblock Performance evaluation of air quality dispersion models in delhi, india.
\newblock In S.~Rauch and G.~M. Morrison, editors, {\em Urban Environment}, pages 121--130, Dordrecht, 2012. Springer Netherlands.

\bibitem{cisl_rda_ds083.2}
{National Centers for Environmental Prediction, National Weather Service, NOAA, U.S. Department of Commerce}.
\newblock Ncep fnl operational model global tropospheric analyses, continuing from july 1999, 2000.

\bibitem{6588606}
I.~Nevat, G.~W. Peters, and I.~B. Collings.
\newblock Random field reconstruction with quantization in wireless sensor networks.
\newblock {\em IEEE Transactions on Signal Processing}, 61(23):6020--6033, 2013.

\bibitem{Pant2017PM25}
P.~Pant, G.~Habib, J.~D. Marshall, and R.~E. Peltier.
\newblock {PM2}.5 exposure in highly polluted cities: {A} case study from {New} {Delhi}, {India}.
\newblock {\em Environmental Research}, 156:167--174, July 2017.

\bibitem{pant2015characterization}
P.~Pant, A.~Shukla, S.~D. Kohl, J.~C. Chow, J.~G. Watson, and R.~M. Harrison.
\newblock Characterization of ambient {PM2}.5 at a pollution hotspot in {New} {Delhi}, {India} and inference of sources.
\newblock {\em Atmospheric Environment}, 109:178--189, May 2015.

\bibitem{pib_funds}
{Press Information Bureau, Government of India}.
\newblock Allocation of funds to 131 cities under national clean air programme to combat air pollution.
\newblock \url{https://pib.gov.in/PressReleseDetailm.aspx?PRID=1989207}, 2023.
\newblock [Online; accessed 17-July-2024].

\bibitem{qin_fine-grained_2022}
X.~Qin, T.~H. Do, J.~Hofman, E.~R. Bonet, V.~P. La~Manna, N.~Deligiannis, and W.~Philips.
\newblock Fine-{Grained} {Urban} {Air} {Quality} {Mapping} from {Sparse} {Mobile} {Air} {Pollution} {Measurements} and {Dense} {Traffic} {Density}.
\newblock {\em Remote Sensing}, 14(11):2613, Jan. 2022.
\newblock Number: 11 Publisher: Multidisciplinary Digital Publishing Institute.

\bibitem{querol2006atmospheric}
X.~Querol, A.~Alastuey, T.~Moreno, M.~Viana, S.~Castillo, J.~Pey, M.~Escudero, S.~Rodr{\'\i}guez, A.~Crist{\'o}bal, A.~Gonz{\'a}lez, et~al.
\newblock Atmospheric particulate matter in spain: Levels, composition and source origin.
\newblock {\em CSIC and Ministerio de Medioambiente}, 2006.

\bibitem{rizwan2013air}
S.~Rizwan, B.~Nongkynrih, and S.~K. Gupta.
\newblock Air pollution in delhi: its magnitude and effects on health.
\newblock {\em Indian Journal of Community Medicine}, 38(1):4--8, 2013.

\bibitem{seabold2010statsmodels}
S.~Seabold and J.~Perktold.
\newblock statsmodels: Econometric and statistical modeling with python.
\newblock In {\em 9th Python in Science Conference}, 2010.

\bibitem{iitkanpur}
M.~Sharma and O.~Dikshit.
\newblock Comprehensive study on air pollution and green house gases ({GHG}s) in {D}elhi.
\newblock Technical report, {IIT Kanpur}, Jan 2016.

\bibitem{sharma1998indoor}
S.~Sharma, G.~R. Sethi, A.~Rohtagi, A.~Chaudhary, R.~Shankar, J.~S. Bapna, V.~Joshi, and D.~G. Sapir.
\newblock Indoor air quality and acute lower respiratory infection in indian urban slums.
\newblock {\em Environmental health perspectives}, 106(5):291--297, 1998.

\bibitem{Shrangi_Pillai_2019}
V.~Shrangi and S.~Pillai.
\newblock Delhi’s 13 hot spots where air is always polluted. do you live in one of them?
\newblock \url{https://www.hindustantimes.com/cities/delhi-s-13-hot-spots-where-aqi-is-always-in-danger-zone/story-yPneyCnH6gA4bdftKfmkyN.html}, Nov 2019.

\bibitem{shusterman2016berkeley}
A.~A. Shusterman, V.~E. Teige, A.~J. Turner, C.~Newman, J.~Kim, and R.~C. Cohen.
\newblock The berkeley atmospheric co 2 observation network: initial evaluation.
\newblock {\em Atmospheric Chemistry and Physics}, 16(21):13449--13463, 2016.

\bibitem{stein1999interpolation}
M.~L. Stein.
\newblock {\em Interpolation of spatial data}.
\newblock Springer Series in Statistics. Springer-Verlag, New York, 1999.
\newblock Some theory for Kriging.

\bibitem{sun2016development}
L.~Sun, K.~C. Wong, P.~Wei, S.~Ye, H.~Huang, F.~Yang, D.~Westerdahl, P.~K. Louie, C.~W. Luk, and Z.~Ning.
\newblock Development and application of a next generation air sensor network for the hong kong marathon 2015 air quality monitoring.
\newblock {\em Sensors}, 16(2):211, 2016.

\bibitem{Sunyer2015}
J.~Sunyer, M.~Esnaola, M.~Alvarez-Pedrerol, J.~Forns, I.~Rivas, M.~L{\'{o}}pez-Vicente, E.~Suades-Gonz{\'{a}}lez, M.~Foraster, R.~Garcia-Esteban, X.~Basaga{\~{n}}a, M.~Viana, M.~Cirach, T.~Moreno, A.~Alastuey, N.~Sebastian-Galles, M.~Nieuwenhuijsen, and X.~Querol.
\newblock Association between traffic-related air pollution in schools and cognitive development in primary school children: A prospective cohort study.
\newblock {\em {PLOS} Medicine}, 12(3):e1001792, mar 2015.

\bibitem{kaiterra}
K.~Technologies.
\newblock Laser egg.
\newblock \url{kaiterra.com}.

\bibitem{toi2024}
{The Times of India}.
\newblock Delhi: Inventory of pollution sources at bad air hotspots to determine what more must be done.
\newblock \url{https://timesofindia.indiatimes.com/city/delhi/inventory-of-pollution-sources-at-delhis-hotspots/articleshow/111738706.cms}, 2024.
\newblock [Online; accessed 17-July-2024].

\bibitem{thunis2019source}
P.~Thunis, A.~Clappier, L.~Tarras{\'o}n, C.~Cuvelier, A.~Monteiro, E.~Pisoni, J.~Wesseling, C.~Belis, G.~Pirovano, S.~Janssen, et~al.
\newblock Source apportionment to support air quality planning: Strengths and weaknesses of existing approaches.
\newblock {\em Environment international}, 130:104825, 2019.

\bibitem{tsujita2005gas}
W.~Tsujita, A.~Yoshino, H.~Ishida, and T.~Moriizumi.
\newblock Gas sensor network for air-pollution monitoring.
\newblock {\em Sensors and Actuators B: Chemical}, 110(2):304--311, 2005.

\bibitem{urbanem_winter}
{U}rban {E}missions.
\newblock Delhi’s air quality – is it the emissions or meteorology?
\newblock \url{https://urbanemissions.info/blog-pieces/delhis-air-quality-emissions-or-meteorology/}.

\bibitem{gaussian_plume}
A.~Venkatram and J.~Thé.
\newblock Introduction to gaussian plume models.
\newblock In P.~Zannetti, editor, {\em AIR QUALITY MODELING - Theories, Methodologies, Computational Techniques, and Available Databases and Software. Vol. I - Fundamentals}, chapter~7A. Air and Waste Management Association, 2003.

\bibitem{2020SciPy-NMeth}
P.~Virtanen, R.~Gommers, T.~E. Oliphant, M.~Haberland, T.~Reddy, D.~Cournapeau, E.~Burovski, P.~Peterson, W.~Weckesser, J.~Bright, S.~J. {van der Walt}, M.~Brett, J.~Wilson, K.~J. Millman, N.~Mayorov, A.~R.~J. Nelson, E.~Jones, R.~Kern, E.~Larson, C.~J. Carey, {\.I}.~Polat, Y.~Feng, E.~W. Moore, J.~{VanderPlas}, D.~Laxalde, J.~Perktold, R.~Cimrman, I.~Henriksen, E.~A. Quintero, C.~R. Harris, A.~M. Archibald, A.~H. Ribeiro, F.~Pedregosa, P.~{van Mulbregt}, and {SciPy 1.0 Contributors}.
\newblock {{SciPy} 1.0: Fundamental Algorithms for Scientific Computing in Python}.
\newblock {\em Nature Methods}, 17:261--272, 2020.

\bibitem{world2021air}
{WHO}, {UNAIDS}, et~al.
\newblock {\em Air quality guidelines: global update 2021}.
\newblock World Health Organization, 2021.

\bibitem{adftest}
{Wikipedia}.
\newblock Augmented dickey–fuller test.
\newblock \url{https://en.wikipedia.org/wiki/Augmented_Dickey%E2%80%93Fuller_test}, 2024.
\newblock [Online; accessed 17-July-2024].

\bibitem{zareba2023big}
M.~Zareba, H.~Dlugosz, T.~Danek, and E.~Weglinska.
\newblock Big-data-driven machine learning for enhancing spatiotemporal air pollution pattern analysis.
\newblock {\em Atmosphere}, 14(4):760, 2023.

\bibitem{zheng2013uair}
Y.~Zheng, F.~Liu, and H.-P. Hsieh.
\newblock U-air: when urban air quality inference meets big data.
\newblock In {\em Proceedings of the 19th ACM SIGKDD International Conference on Knowledge Discovery and Data Mining}, KDD '13, page 1436–1444, New York, NY, USA, 2013. Association for Computing Machinery.

\bibitem{Zivin2018}
J.~G. Zivin and M.~Neidell.
\newblock Air pollution{\textquotesingle}s hidden impacts.
\newblock {\em Science}, 359(6371):39--40, jan 2018.

\end{thebibliography}

\newpage
\pagenumbering{arabic}
\renewcommand*{\thepage}{A\arabic{page}}
\appendix

\section{Hotspot Boundary}
\label{sec:bounds}
One important extension of interpolating pollution field heatmaps and the definition of hotspots is to delineate the geometric boundary of the hotspot. This can allow policymakers to mark out the neighborhoods and administrative zones that should be focused on for pollution control and management. One method of doing so has been discussed in \cite{6588606}, where the authors introduce a novel algorithm based on the Saddle-point approach to address spatial field reconstruction and exceedance level estimation (hotspot identification) in wireless sensor networks (WSN). Their approach involves predictive distribution calculations that require solving complex integrals, which are approximated using a novel Gaussian basis series expansion of the Saddle-point type. Inspired by this methodology, we also estimate the predictive distribution of the spatial field generated by spatial interpolation by using Gaussian Kernal Density estimation using the Scipy library \cite{2020SciPy-NMeth}. Let $\Phi(x,y)$ be the predictive probability of the pollution field $P$ at point $(x,y)$. Then,
\begin{equation}
    \Phi(x,y) = \text{Gaussian KDE}_{G,P}(P(x,y))
\end{equation}
where G is the grid of points sampled from the study area which is used to fit the Gaussian kernel. Further, the spatial exceedance level $f$ of point $(x,y)$ can be computed as
\begin{equation}
    f(x,y) = 1 - \Phi(x,y)
\end{equation}
Then, appropriate thresholding of the spatial exceedance level can be used for computing the boundary of the hotspots.

\newpage

\section{Observations on Diwali}
\label{sec:diwali}
\begin{figure}[t]
  \centering
  \begin{subfigure}{0.32\columnwidth}
    \centering
    \includegraphics[width=\columnwidth]{figures/d1.jpeg}
    \caption{}
    \label{fig:d_nn}
  \end{subfigure}
  ~
  \begin{subfigure}{0.32\columnwidth}
    \centering
    \includegraphics[width=\columnwidth]{figures/d2.jpeg}
    \caption{}
    \label{fig:d_wz}
  \end{subfigure}
  ~
  \begin{subfigure}{0.32\columnwidth}
    \centering
    \includegraphics[width=\columnwidth]{figures/d3.jpeg}
    \caption{}
    \label{fig:d_jh}
  \end{subfigure}

  \begin{subfigure}{0.32\columnwidth}
    \centering
    \includegraphics[width=\columnwidth]{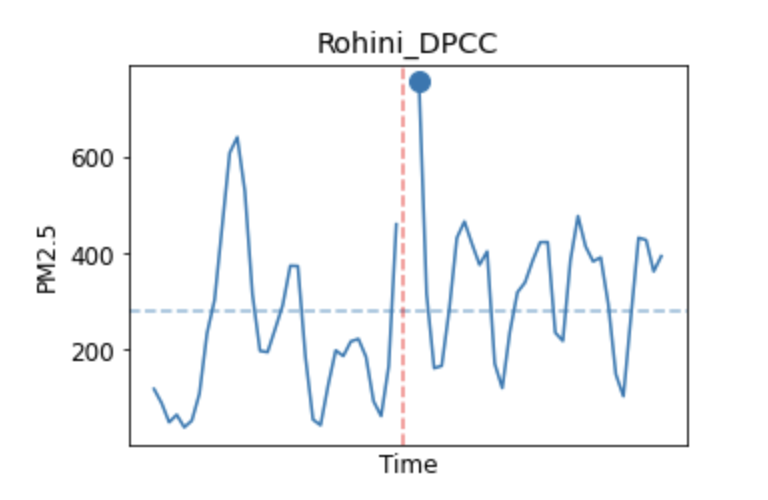}
    \caption{}
    \label{fig:d_ro}
  \end{subfigure}
  ~
  \begin{subfigure}{0.32\columnwidth}
    \centering
    \includegraphics[width=\columnwidth]{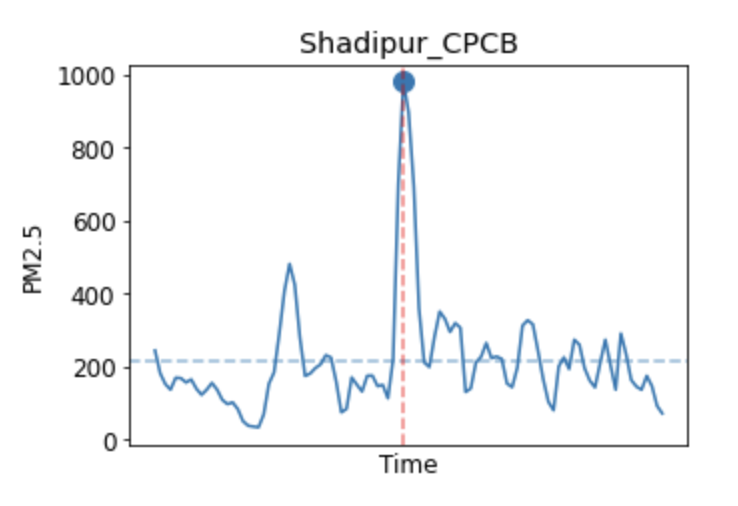}
    \caption{}
    \label{fig:d_sh}
  \end{subfigure}
  ~
  \begin{subfigure}{0.32\columnwidth}
    \centering
    \includegraphics[width=\columnwidth]{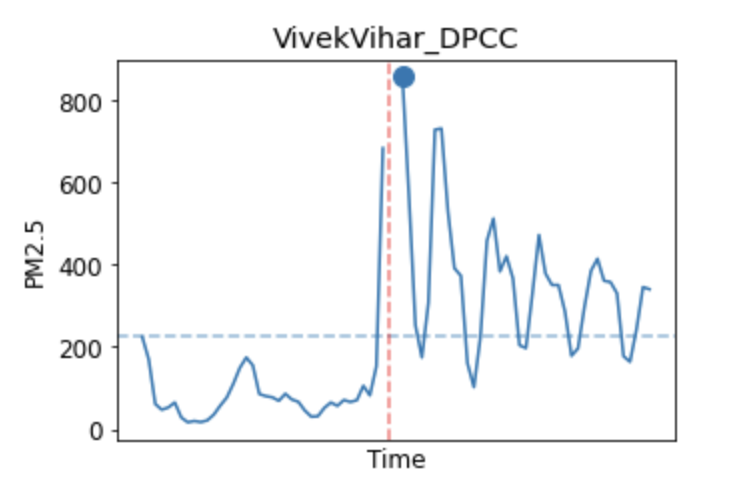}
    \caption{}
    \label{fig:d_vv}
  \end{subfigure}
    
  \begin{subfigure}{0.32\columnwidth}
    \centering
    \includegraphics[width=\columnwidth]{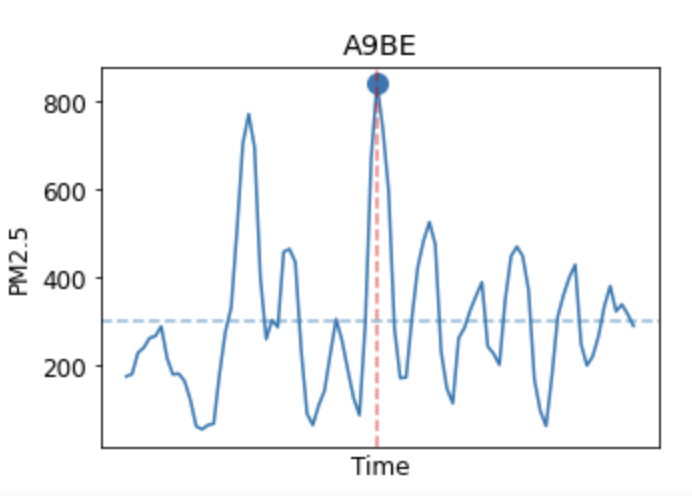}
    \caption{}
    \label{fig:d_k1}
  \end{subfigure}
  ~
  \begin{subfigure}{0.32\columnwidth}
    \centering
    \includegraphics[width=\columnwidth]{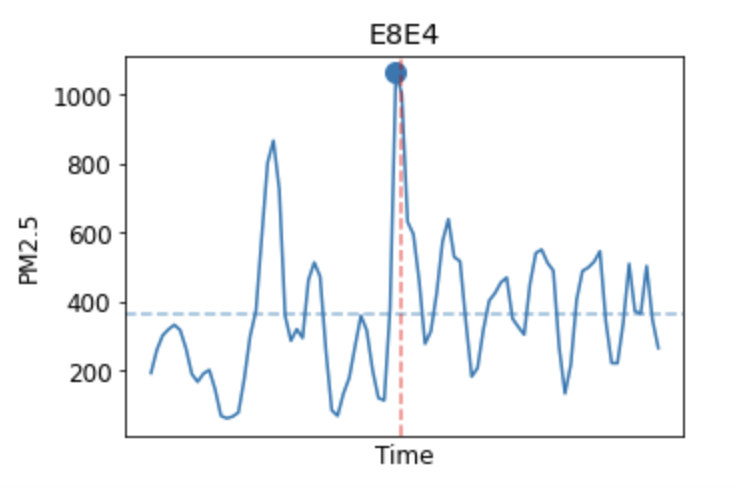}
    \caption{}
    \label{fig:d_k2}
  \end{subfigure}
  ~
  \begin{subfigure}{0.32\columnwidth}
    \centering
    \includegraphics[width=\columnwidth]{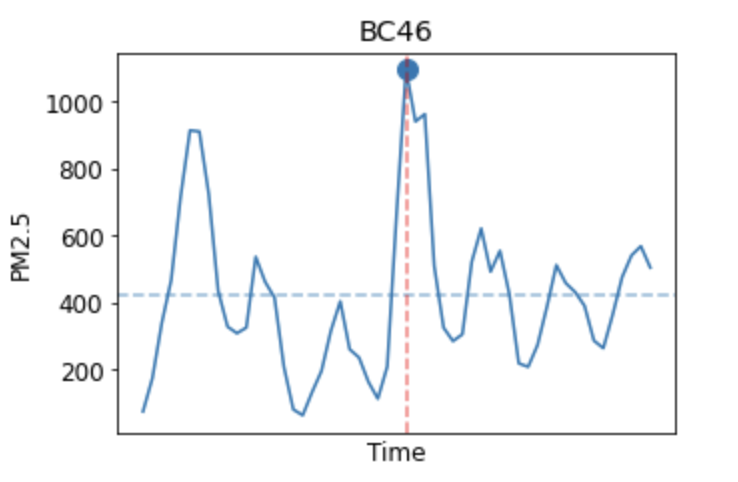}
    \caption{}
    \label{fig:d_k3}    
  \end{subfigure}
  \caption{PM$_{2.5}$ data for some monitoring stations at the time of Diwali. The vertical red line represents Diwali midnight. Note how most stations are not able to record readings as the PM$_{2.5}$ values shoot beyond the measurement limit of 1000 $\pmunit$. When the value is actually recorded, as in figure (c), it is very close to the measurement limit. Our deployed sensors presented in figures (d) at Sadiq Nagar, (e) at Safdarganj Enclave, and (f) at Preet Vihar also show the same pattern.}
  \label{fig:diwali}
\end{figure}

On specific days of Diwali, we find another interesting behavior. Most sensors on Diwali do not register pollution readings. In case when some sensors do register readings, the values recorded are close to their maximum limit, i.e. PM$_{2.5}$ values of $\sim 1000 \pmunit$. Thus, we hypothesize that on the days of Diwali, pollution sensors are not able to register readings since the pollution level goes beyond their limit of measurement. We present the temporal hotspot profiles on the day of Diwali at 3H resolution for some monitoring stations in Figure \ref{fig:diwali}.

\newpage

\section{Purple Air Experiments}
\label{sec:purpleair}
\begin{figure}[h!]
    \centering
    \includegraphics[width=0.9\columnwidth]{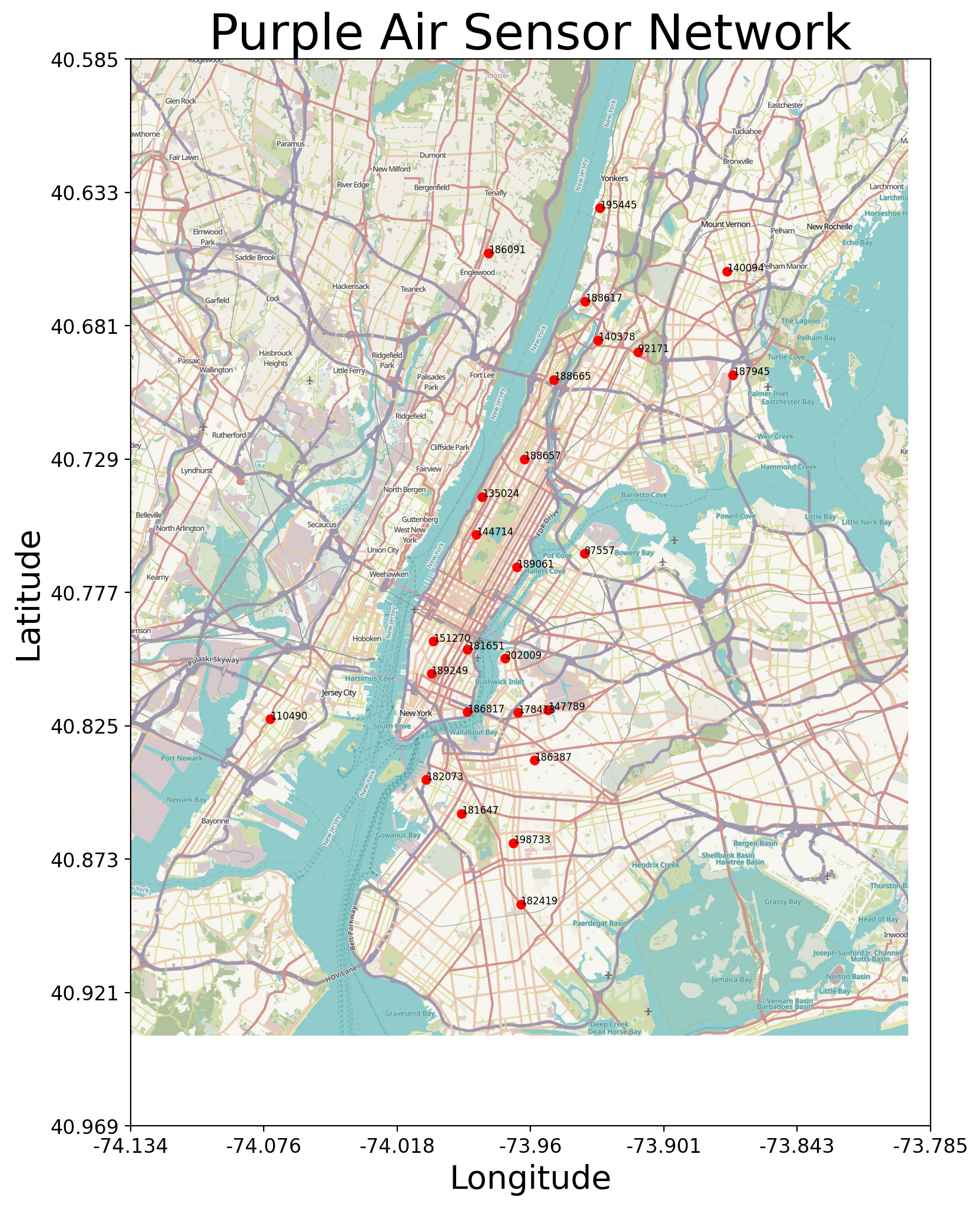}
    \caption{Purple Air sensor network used for generalization experiments. The sensor ids can be used to download the data using Purple Air's data download tool.}
    \label{fig:purpleairnyc}
\end{figure}

For the experiments on the Purple Air dataset, we took a subset of the sensors deployed in and around NYC. We present the map of the sensor network in Figure \ref{fig:purpleairnyc}. The dataset was downloaded for the entire year of 2023 for 60 minute resolution. After running the Space-Time Kriging model on the data, the RMSE obtained was normalized by the following approach
\begin{align*}
    NRMSE &= RMSE * \frac{\Bar{\sigma_{New Delhi}}}{\Bar{\sigma_{NYC}}}\\
    \Bar{\sigma_{i}} &= \frac{\sum_{t=t_0}^{t_f}\sigma_{i}(t)}{\sum_{t=t_0}^{t_f}1}, i \in \{New Delhi, NYC\}
\end{align*}
where, $t_0$ and $t_f$ refer to the first timestamp and last timestep for the data. $\sigma(t)$ refers to the spatial variation observed for timestep t.

We chose to normalize based on the average spatial variation because it makes sense in the context of the modeling strategy. In our Space-Time Kriging methodology, we are essentially learning separate models for each time step, thus the data variation actually affecting the performance is along the spatial axis.

\newpage

\section{Derivation of Ordinary Kriging}
\label{sec:kriging_proof}
We have \( n \) known spatial data points \( \{ (u_i, z_i) \}_{i=1}^n \), where \( u_i \) represents the location and \( z_i \) the observed value at that location. Our goal is to estimate the value \( \hat{Z}(u_0) \) at an unobserved location \( u_0 \).

The ordinary kriging estimator for the unknown value at \( u_0 \) is a weighted sum of the known values:

\[
\hat{Z}(u_0) = \sum_{i=1}^n \lambda_i z_i
\]

where \( \lambda_i \) are the kriging weights that need to be determined.

To ensure that the estimator is unbiased, the weights must sum to 1:

\[
\sum_{i=1}^n \lambda_i = 1
\]

The spatial dependence between the points is characterized by the semivariogram \( \gamma(h) \), where \( h \) is the distance between points. The semivariogram is defined as:

\[
\gamma(h) = \frac{1}{2} \text{Var}[Z(u) - Z(u+h)]
\]

The key to ordinary kriging is to minimize the estimation variance (also called kriging variance). The kriging variance is defined as the variance of the estimation error:

\[
\sigma^2_{\hat{Z}(u_0)} = \text{Var}[\hat{Z}(u_0) - Z(u_0)]
\]

We express this as:

\[
\sigma^2_{\hat{Z}(u_0)} = \text{Var}\left[\sum_{i=1}^n \lambda_i z_i - Z(u_0)\right]
\]

Using the properties of variances and covariances, this can be expanded as:

\[
\sigma^2_{\hat{Z}(u_0)} = \text{Var}\left[\sum_{i=1}^n \lambda_i z_i\right] + \text{Var}[Z(u_0)] - 2\text{Cov}\left[\sum_{i=1}^n \lambda_i z_i, Z(u_0)\right]
\]

Since \( z_i \) are observed values, we know their semivariogram values \( \gamma_{ij} \). We can rewrite the terms using the semivariograms. The semivariogram values are related to the covariance values by the following relationship:

\[
\gamma_{ij} = \gamma(h_{ij}) = C(0) - C(h_{ij})
\]

where \( h_{ij} \) is the distance between points \( u_i \) and \( u_j \), and \( C(h_{ij}) \) is the covariance between points \( u_i \) and \( u_j \). Using this relationship, we rewrite the variances and covariances in terms of the semivariogram:

\[
\text{Var}\left[\sum_{i=1}^n \lambda_i z_i\right] = \sum_{i=1}^n \sum_{j=1}^n \lambda_i \lambda_j (C(0) - \gamma_{ij})
\]

\[
\text{Var}[Z(u_0)] = C(0)
\]

\[
\text{Cov}\left[\sum_{i=1}^n \lambda_i z_i, Z(u_0)\right] = \sum_{i=1}^n \lambda_i (C(0) - \gamma_{0i})
\]

So, the kriging variance becomes:

\[
\sigma^2_{\hat{Z}(u_0)} = \sum_{i=1}^n \sum_{j=1}^n \lambda_i \lambda_j (C(0) - \gamma_{ij}) + C(0) - 2\sum_{i=1}^n \lambda_i (C(0) - \gamma_{0i})
\]

Simplifying:

\[
\sigma^2_{\hat{Z}(u_0)} = C(0) \left( 1 + \sum_{i=1}^n \sum_{j=1}^n \lambda_i \lambda_j - 2\sum_{i=1}^n \lambda_i \right) - \sum_{i=1}^n \sum_{j=1}^n \lambda_i \lambda_j \gamma_{ij} + 2\sum_{i=1}^n \lambda_i \gamma_{0i}
\]

Given the unbiasedness constraint \( \sum_{i=1}^n \lambda_i = 1 \), the term involving \( C(0) \) simplifies to zero:

\[
\sigma^2_{\hat{Z}(u_0)} = - \sum_{i=1}^n \sum_{j=1}^n \lambda_i \lambda_j \gamma_{ij} + 2\sum_{i=1}^n \lambda_i \gamma_{0i}
\]

To minimize this variance subject to the unbiasedness constraint, we use the method of Lagrange multipliers. Define the Lagrangian:

\[
\mathcal{L} = - \sum_{i=1}^n \sum_{j=1}^n \lambda_i \lambda_j \gamma_{ij} + 2\sum_{i=1}^n \lambda_i \gamma_{0i} + \mu \left(1 - \sum_{i=1}^n \lambda_i \right)
\]

where \( \mu \) is the Lagrange multiplier.

To find the minimum, we take the partial derivatives of \( \mathcal{L} \) with respect to \( \lambda_i \) and set them to zero:

\[
\frac{\partial \mathcal{L}}{\partial \lambda_i} = -2 \sum_{j=1}^n \lambda_j \gamma_{ij} + 2 \gamma_{0i} - \mu = 0 \quad \forall i
\]

\[
\frac{\partial \mathcal{L}}{\partial \mu} = 1 - \sum_{i=1}^n \lambda_i = 0
\]

Rearranging the first equation:

\[
\sum_{j=1}^n \lambda_j \gamma_{ij} + \frac{\mu}{2} = \gamma_{0i} \quad \forall i
\]

These equations can be written in matrix form as:

\[
\begin{pmatrix}
\gamma_{11} & \gamma_{12} & \cdots & \gamma_{1n} & 1 \\
\gamma_{21} & \gamma_{22} & \cdots & \gamma_{2n} & 1 \\
\vdots & \vdots & \ddots & \vdots & \vdots \\
\gamma_{n1} & \gamma_{n2} & \cdots & \gamma_{nn} & 1 \\
1 & 1 & \cdots & 1 & 0
\end{pmatrix}
\begin{pmatrix}
\lambda_1 \\
\lambda_2 \\
\vdots \\
\lambda_n \\
\mu
\end{pmatrix}
=
\begin{pmatrix}
\gamma_{01} \\
\gamma_{02} \\
\vdots \\
\gamma_{0n} \\
1
\end{pmatrix}
\]

This system can be solved for the \( \lambda_i \) and \( \mu \). The weights \( \lambda_i \) are then used in the kriging estimator:

\[
\hat{Z}(u_0) = \sum_{i=1}^n \lambda_i z_i
\]

Thus, we've derived the kriging weights \( \lambda_i \) through the minimization of the estimation variance under the unbiasedness constraint, resulting in the ordinary kriging estimator.

\section{Algorithms for Transient Hotspots}
\label{sec:transient}
\begin{algorithm}[hbt!]
\caption{Enforce Monotonic Decrease}\label{alg:mon_dec}
\begin{algorithmic}
\For{h in H}
    \State Consider the triangle formed by s,h,n.
    \State Assert: $\forall x \in H-\{h\}$, value at x > value at h and 
    \State value at x > value at n.
\EndFor
\end{algorithmic}
\end{algorithm}

\begin{algorithm}[hbt!]
\caption{Transient Hotspot Detection}\label{alg:sp_hsp}
\begin{algorithmic}
\For{t in timestamps}
    \For{s in sensors}
        \State Consider radial neighborhood of $rad\_max$ around s.
        \State For the radial neighborhood of size $rad\_max$*$sp\_nh$,
        \State get average readings.
        \State Arrange neighbors in increasing distance from s.
        \State Set hotspot set H to be empty.
        \If{value at s > average value + $abs\_th$}
            \For{n in neighbors}
                \If{Value at s < value at n + $bnd\_th$}
                    \If{Enforce Monotonic Decrease(s,H,n)}
                        \State Add n to H.
                        \State Remove s,n from sensors.
                    \EndIf
                \EndIf
            \EndFor
            \If{H is not empty}
                \State Report Hotspot.
            \EndIf
        \EndIf
    \EndFor
\EndFor
\end{algorithmic}
\end{algorithm}

For these algorithms, we can tune the parameters $abs\_th$, $bnd\_th$, $rad\_max$, and $sp\_nh$. We have used the values 150, 75, 5 km, and 2 in our analysis. Note that we have chosen parameter values erring towards not qualifying hotspots. Clearly, $abs\_th = 150$, and $sp\_nh=5$km will only qualify hotspots when the pollution value at a point is higher than its surroundings by a very large amount. We have preferred absolute thresholds over relative thresholds because it enforces that hotspots indicate a significant jump in pollution readings that cannot be attributed to noise.

\end{document}